\documentclass[11pt,preprint,nofootinbib,aps,superscriptaddress,eqsecnum]{revtex4}
 \pdfoutput=1
 \usepackage{graphicx}
 \usepackage{amsmath}
\usepackage{amsfonts}
\usepackage{amssymb}
\usepackage{hyperref}
\usepackage{slashed}
\usepackage{float}
\usepackage{subfigure}
\usepackage{multirow}
\usepackage{color}
\usepackage{caption}
\usepackage{cancel}
\newcommand{\newc}{\newcommand}
\newc{\wt}{\widetilde}
\newc{\ra}{\rightarrow}

\def\({\left(}
\def\){\right)}
\def\[{\left[}
\def\]{\right]}

\newcommand{\ba}{\begin{array}}
\newcommand{\ea}{\end{array}}
\newcommand{\bd}{\begin{displaymath}}
\newcommand{\ed}{\end{displaymath}}
\newcommand{\be}{\begin{equation}}
\makeatletter
\newcommand*{\rom}[1]{\expandafter\@slowromancap\romannumeral #1@}
\makeatother
\newcommand{\ee}{\end{equation}}
\def\bt{\begin{table}}
\def\et{\end{table}}
\def\bc{\begin{center}}
\def\ec{\end{center}}
\def\bi{\begin{itemize}}
\def\ei{\end{itemize}}
\def\bw{\begin{widetext}}
\def\ew{\end{widetext}}

\usepackage{array}
\newcolumntype{L}[1]{>{\raggedright\let\newline\\\arraybackslash\hspace{0pt}}m{#1}}
\newcolumntype{C}[1]{>{\centering\let\newline\\\arraybackslash\hspace{0pt}}m{#1}}
\newcolumntype{R}[1]{>{\raggedleft\let\newline\\\arraybackslash\hspace{0pt}}m{#1}}

\def\bea{\begin{eqnarray}}
\def\eea{\end{eqnarray}}
\def\beas{\begin{eqnarray*}}
\def\eeas{\end{eqnarray*}}

\def\N0{\widetilde{\chi}^0}

\def \met{\rm E{\!\!\!/}_T}

\def \charginopm{{\chi}^{\pm}}
\def \mcharginopm{m_{\charginopm}}
\def \chonep {{\chi_1^+}}
\def \chone {{\chi_1}}
\def \ch2p {{\chi_2^+}}
\def \chonem {{\chi_1^-}}
\def \ch2m {{\chi_2^-}}
\def \chplus {{\chi^+}}
\def \chminus {{\chi^-}}
\def \chip{{\chi_i}^{+}}
\def \chim{{\chi_i}^{-}}
\def \chipm{{\chi_i}^{\pm}}
\def \chjp{{\chi_j}^{+}}
\def \chjm{{\chi_j}^{-}}
\def \chjpm{{\chi_j}^{\pm}}
\def \chonepm{\chi_1^{\pm}}
\def \chonemp{\chi_1^{\mp}}
\def \mchonepm{m_{\chonepm}}
\def \mchonemp{m_{\chonemp}}
\def \mchtwopm{m_{\chtwopm}}
\def \chtwop {{\chi_2^+}}
\def \chtwom {{\chi_2^-}}
\def \chtwopm{{\chi_{2}^{\pm}}}
\def \chtwomp{{\chi_2}^{\mp}}

\def \chkpm{{\wt\\def \charginopm{{\wt\chi}^{\pm}}
\def \mcharginopm{m_{\charginopm}}
\def \chonep {{\wt\chi_1^+}}
\def \chone {{\wt\chi_1}}
\def \ch2p {{\wt\chi_2^+}}
\def \chonem {{\wt\chi_1^-}}
\def \ch2m {{\wt\chi_2^-}}
\def \chplus {{\wt\chi^+}}
\def \chminus {{\wt\chi^-}}
\def \chip{{\wt\chi_i}^{+}}
\def \chim{{\wt\chi_i}^{-}}
\def \chipm{{\wt\chi_i}^{\pm}}
\def \chjp{{\wt\chi_j}^{+}}
\def \chjm{{\wt\chi_j}^{-}}
\def \chjpm{{\wt\chi_j}^{\pm}}
\def \chonepm{{\wt\chi_1}^{\pm}}
\def \chonemp{{\wt\chi_1}^{\mp}}
\def \mchonepm{m_{\chonepm}}
\def \mchonemp{m_{\chonemp}}
\def \mchtwopm{m_{\chtwopm}}
\def \chtwop {{\chi_2^+}}
\def \chtwom {{\chi_2^-}}
\def \chtwopm{{\chi_2}^{\pm}}
\def \chtwomp{{\chi_2}^{\mp}}

\def \chkpm{{\chi_k}^{\pm}}
\def \chkmp{{\chi_k}^{\mp}}
\def \chlpm{{\chi_\ell}^{\pm}}
\def \chlmp{{\chi_\ell}^{\mp}}chi_k}^{\pm}}
\def \chkmp{{\chi_k}^{\mp}}
\def \chlpm{{\chi_\ell}^{\pm}}
\def \chlmp{{\chi_\ell}^{\mp}}

\def \lspi{\chi_i^0}

\def \lspj{\chi_j^0}

\def \lspone{\chi_1^0}

\def \lsptwo{\chi_2^0}

\def \lspthree{\chi_3^0}

\def \lspfour{\chi_4^0}

\begin{document}

\title{\bf Electroweakino searches at the HL-LHC in the baryon number violating MSSM}

\author{Rahool Kumar Barman	}
\email{rahoolbarman@iisc.ac.in}
\affiliation{Centre for High Energy Physics, Indian Institute of Science, Bangalore 560012, India}
\affiliation{School of Physical Sciences, Indian Association for the Cultivation of Science, Kolkata 700032, India}

\author{Biplob Bhattacherjee}
\email{biplob@iisc.ac.in}
\affiliation{Centre for High Energy Physics, Indian Institute of Science,
Bangalore 560012, India}

\author{Indrani Chakraborty}
\email{indranic@iitk.ac.in}
\affiliation{Department of Physics, Indian Institute of Technology Kanpur, Kanpur, Uttar Pradesh-208016, India}

 \author{Arghya Choudhury}
\email{arghya@iitp.ac.in}
\affiliation{Department of Physics, Indian Institute of Technology Patna, Bihta -801106, India}

\author{Najimuddin Khan}
\email{khanphysics.123@gmail.com }
\affiliation{Centre for High Energy Physics, Indian Institute of Science,
Bangalore 560012, India}
\affiliation{School of Physical Sciences, Indian Association for the Cultivation of Science, Kolkata 700032, India}
  
\begin{abstract}

The projected reach of direct electroweakino searches at the HL-LHC ($\sqrt{s}=14~{\rm TeV}, ~3000~{\rm fb^{-1}}$ LHC) in the framework of simplified models with $R$-parity violating (RPV) operators: $\lambda_{112}^{\prime \prime}u^{c}d^{c}s^{c}$ and $\lambda_{113}^{\prime\prime}u^{c}d^{c}b^{c}$, is studied. 
Four different analysis channels are chosen: $Wh$ mediated $1l+2b+jets+\met$, $Wh$ mediated $1l+2\gamma+jets+\met$, $WZ$ mediated $3l+jets+\met$ and $WZ$ mediated $3l+2b+jets+\met$ and the projected exclusion/discovery reach of direct wino searches in these channels is analyzed by performing a detailed cut-based collider analysis. The projected exclusion contour reaches up to $600-700~{\rm GeV}$ for a massless bino-like $\lspone$ from searches in the $Wh$ mediated $1l+2b+jets+\met$, $Wh$ mediated $1l+2\gamma +jets+\met$ and $WZ$ mediated $3l+jets+\met$ channels, while the $WZ$ mediated $3l+2b+jets+\met$ search channel is found to have a projected exclusion reach up to $600~{\rm GeV}$ for $150~{\rm GeV} < M_{\lspone} < 250~{\rm GeV}$. The baryon number violating simplified scenario considered in this work is found to furnish a weaker projected reach (typically by a factor of $\sim 1/2$) than the $R$-parity conserving (RPC) case. The projected reach at the HL-LHC in these four channels is also recast for realistic benchmark scenarios. 

\end{abstract}

\maketitle

\tableofcontents

\section{Introduction}
\label{sec:intro}

Supersymmetry (SUSY)~\cite{WESS197439,NILLES19841,Haber:1984rc,Nilles:1983ge} has been among the most attractive frameworks for formulating physics beyond the Standard Model~(SM). Numerous studies have reported the plausibility of SUSY in resolving various inadequacies within the SM of particle physics~\cite{GLASHOW1961579,PhysRevLett.19.1264,Salam:1968rm,GellMann:1964nj} viz. the hierarchy problem~\cite{PhysRevD.14.1667,Veltman:1980mj}, gauge coupling unification~\cite{ELLIS1990441,Amaldi:1991cn,ROSS1992571,Giunti:1991ta}, existence of a viable dark matter (DM) candidate~\cite{1933AcHPh...6..110Z,1937ApJ....86..217Z,Sofue:2000jx}, and naturalness of Higgs mass~\cite{PhysRevD.20.2619,Susskind:1982mw}. In adherence to the experimental observations, SUSY has to be broken, and solution to the hierarchy problem implores the SUSY breaking scale to be $\sim O({\rm TeV})$, thus, bringing the SUSY particles within the potential reach of current and future LHC. The minimal supersymmetric extension of the Standard Model (MSSM) (see \cite{Martin:1997ns,aitchison_2007,Djouadi:2005gj,Baer:2006rs,Drees:873465} for a detailed review) has been among the prominent class of nominees considered to address the shortcomings within the SM and in the pursuit of new physics phenomenology. A legion of studies have focused on investigating the current status and future prospects of the MSSM parameter space in light of the LHC Run-I and Run-II results~\cite{Carena:2011aa,Arbey:2011ab,Baer:2012mv,Arbey:2012dq,Altmannshofer:2012ks,Bechtle:2012jw,Djouadi:2013lra,Cheung:2013kla,Ahmadov:2013qga,Chakrabortty:2015ika,Kowalska:2015zja,Chowdhury:2015yja,Bhattacherjee:2015sga,Bechtle:2015pma,Barman:2016jov,Barr:2016sho,Kowalska:2016ent,Han:2016xet,Buckley:2016kvr,Choudhury:2017fuu, Zhao:2017qpe,Athron:2017yua,Bagnaschi:2017tru,Costa2018,Tran:2018kxv, Endo:2020mqz}. Since the advent of the LHC, the ATLAS and CMS Collaborations have performed a multitude of searches to probe the sparticles using the LHC Run-I and Run-II dataset, however, they are yet to observe a clear signature of physics beyond the SM. Robust lower bounds have been derived on the masses of strongly interacting sparticles. Searches by the ATLAS and CMS collaboration using the LHC $\sqrt{s}=13~{\rm TeV}$ data collected at $\sim 137~{\rm fb^{-1}}$ of integrated luminosity ($\mathcal{L}$) have excluded gluinos ($\tilde{g}$) up to $\sim 2.2~{\rm TeV}$ and $\sim 2.3~{\rm TeV}$, respectively, for a lightest SUSY particle~(LSP) neutralino ($\lspone$) with mass up to $\sim 600~{\rm GeV}$~\cite{CMS-PAS-SUS-19-006,CMS-PAS-SUS-19-007,ATLAS-CONF-2019-040} at $95\%$ C.L., however, within the framework of a simplified SUSY scenario. Using the same respective datasets and within a simplified scenario with some specific decay modes and mass hierarchy, ATLAS and CMS have also excluded the stops ($\tilde{t}$) and sbottoms ($\tilde{b}$) up to $\sim 1.2~{\rm TeV}$ for a $M_{\lspone} \sim 400~{\rm GeV}$ at $95\%$ C.L.~\cite{CMS-PAS-SUS-19-005,CMS-PAS-SUS-19-009,ATLAS-CONF-2019-040}. On the other hand, the electroweakly interacting sparticles viz. electroweakinos and sleptons, are rather feebly constrained~\cite{Sirunyan:2017zss,CMS-PAS-SUS-19-002,CMS-PAS-SUS-18-007,Sirunyan:2018ubx,Aad:2019qnd,Aad:2019vvi,Aad:2019vvf}
\footnote{A few phenomenological analyses in this context may be seen in Refs.\cite{Chakraborti:2014gea,Das:2014kwa,Chakraborti:2015mra,Choudhury:2016lku,Chakraborti:2017vxz,Pozzo:2018anw,Datta:2018lup}.}.

The MSSM is endowed with an exact symmetry related to the baryon number ($B$), lepton number ($L$) and the spin of the particle ($S$), referred to as $R$-parity ($R_{p}$)\footnote{$R_{p}$ is defined as $R_{p} = (-1)^{(3B+L+2S)}$. The SM particles and their superpartners are $R_{p}=+1$ and $R_{p}=-1$, respectively.}. $R$-parity conservation (RPC) entails pair production of SUSY particles at colliders and also ensures that the lightest SUSY particle, typically the $\lspone$, is stable and a viable nonbaryonic DM candidate. The presence of a stable LSP DM candidate results in missing transverse energy ($\met$) signatures at the colliders, making the RPC scenarios extensively attractive to be analyzed at the LHC. Consequently, a myriad of studies have addressed the phenomenological implications of $R$-parity conserved scenarios and a nonexhaustive list of such studies can be found in \cite{Hooper:2002nq,Bhattacharyya:2011se, AlbornozVasquez:2011yq,Choudhury:2012tc,Fowlie:2013oua,Roszkowski:2014iqa,Hamaguchi:2015rxa,Han:2014nba, Belanger:2013pna,Ananthanarayan:2013fga,Dreiner:2012ex,Choudhury:2013jpa,Calibbi:2011ug,Belanger:2003wb,Belanger:2001am,Belanger:2000tg,Barman:2017swy,Chakraborti:2017dpu}. 

Although the $R$-parity conserved scenarios display a tempting landscape for collider and astrophysical searches, it must be noted that $R$-parity conservation is not  fundamentally necessary to obtain a viable SUSY framework. Ensuring the stability of the proton was the prime intent behind introducing $R$-parity conservation (relevant discussions can be found in \cite{PhysRevD.47.279,Bhattacharyya:1998dt}). However, several studies have also explored the possibility to stabilize the proton without conserving $R$-parity~\cite{Ibanez:1991hv,Dreiner:2005rd,Dreiner:2006xw,Dreiner:2012ae}. Another strong incentive to consider RPC scenarios is the possibility of a viable DM candidate with a correct DM relic density as discussed previously. In the presence of $R$-parity violation (RPV) (see \cite{Dreiner:1997uz,Barbier:2004ez} for reviews), the LSP would undergo decay and would no longer remain a viable DM candidate. However, results from \cite{Chun:1999cq,Chun:2006ss,Takayama:2000uz} indicate that axinos and gravitinos could generate a correct relic abundance in RPV scenarios. Furthermore, the presence of $R$-parity violation has also been shown to ease the amount of fine-tuning required to obtain a $\sim 125~{\rm GeV}$ Higgs boson in SUSY~\cite{Dreiner:2014lqa} by weakening the bounds on gluino~\cite{Bhattacherjee:2013gr,Asano:2014aka,ATLAS-CONF-2016-057,Sirunyan:2017dhe,Aaboud:2018lpl} and stop masses~\cite{Graham:2014vya,CMS-PAS-EXO-16-029}. Another critical consequence of RPV terms is the successful explanation of the observed pattern of neutrino masses and mixing~\cite{Hall:1983id, Mukhopadhyaya:1999wn,Kong:2002hb,Rakshit:2004rj, Dey_2008,Datta:2009dc,Bose:2014vea}. Within RPV scenarios, lepton number violating couplings can initiate lepton flavor violating processes (viz. the scattering of unoscillated $\nu_{\mu}$ into $\tau$) even in the absence of neutrino oscillation~\cite{Datta:2000ci}. The $R$-parity framework has also been studied in light of offering a plausible explanation for the $(g-2)_{\mu}$ discrepancy~\cite{Chakraborty:2015bsk}. In addition, the $\met$ dependent collider search strategies, which are a trademark of RPC scenarios, would be rendered ineffective in the presence of $R$-parity violating terms, and the collider bounds are expected to alter. Thus, the introduction of RPV terms would result in characteristically distinct final states, a study of which would be extremely relevant in the context of collider searches at the LHC\cite{Bardhan:2016gui,Guo:2018hbv}. Overall, the discussion until now motivates the impulse of probing the sector of RPV MSSM.

The most general, gauge invariant and renormalizable $R$-parity violating terms~\cite{Martin:1997ns,Mohapatra:2015fua} which could be added to the MSSM superpotential ($W_{MSSM}$) are the following (the notation of \cite{Domingo2019} has been followed): 
 \begin{equation}
 W_{RPV} = W_{MSSM} + \frac{1}{2}\lambda_{ijk}L_{i}\cdot L_{j}e^{c}_{k} + \frac{1}{2}\lambda_{ijk}^{\prime}L_{i}\cdot Q_{j}d^{c}_{k} + \frac{1}{2}\lambda^{\prime\prime}_{ijk} \epsilon_{\alpha\beta\gamma}u^{c\,\alpha}_{i}d^{c\,\beta}_{j}d^{c\,\gamma}_{k} + \mu_{i} H_{u}\cdot L_{i}
 \label{Eqn:RPV_term}
 \end{equation}
where $L$ and $Q$ represent the left-handed lepton and quark superfields, respectively, while, $e$, $u$ and $d$ corresponds to the right-handed lepton, up-type quark and down-type quark superfields, respectively. $\lambda$, $\lambda^{\prime}$ and $\lambda^{\prime\prime}$ are the dimensionless Yukawa couplings while $\epsilon$ is the three dimensional Levi-Civita symbol. Here, $i,j,k$ are the generation indices, $\alpha,\beta,\gamma$ are the flavor indices and $c$ represents charge conjugation. The first and second terms in Eq.~\ref{Eqn:RPV_term} violate the lepton number by $1$ unit, while the third term in Eq.~\ref{Eqn:RPV_term} violates the baryon number by $1$ unit. 

The collider implications of the lepton number violating RPV couplings: $\lambda_{12k}L_{1}\cdot L_{2}e^{c}_{k}$ ($k \in 1,2$) and $\lambda_{i33}L_{i}\cdot L_{3}e^{c}_{3}$ ($i \in 1,2$) have been studied by the ATLAS collaboration through an interpretation in simplified scenarios with winolike next-to-lightest supersymmetric particle~(NLSP) pair production ($pp \to \lsptwo/\chonepm + \chonepm$) and higgsinolike NLSP pair production ($pp \to \lspone/\lsptwo/\chonepm + \chonepm$) in the $WZ$ and $Wh$ mediated $4l$ ($l =$ electrons ($e$) and muons ($\mu$)) final state~\cite{PhysRevD.98.032009} using the LHC Run-II data collected at $\mathcal{L} = 36.1~{\rm fb^{-1}}$. Results from \cite{PhysRevD.98.032009} exclude a winolike $\chonepm,\lsptwo$ up to $\sim 1.46~{\rm TeV}$ ($\sim 980~{\rm GeV}$) for a binolike $\lspone$ with mass $M_{\lspone}$ $\sim 500~{\rm GeV}$ ($M_{\lspone} \in [400-700]~{\rm GeV}$) in the presence of $\lambda_{12k}L_{1}\cdot L_{2}e^{c}_{k}$ ($\lambda_{i33}L_{i}\cdot L_{3}e^{c}_{3}$)-type RPV coupling. The ATLAS collaboration has also probed direct wino production in the context of RPC scenarios and has excluded winos up to $\sim 350~{\rm GeV}$ for a $M_{\lspone} \sim 50~{\rm GeV}$ (at $95\%$ C.L.) from searches in the trilepton ($l=e,\mu$) + $\met$ final state~\cite{Aad:2019vvi} using LHC Run-II data ($\sim\rm 139~fb^{-1}$). Similarly, direct wino searches by CMS in three or more charged $l$ final states in a winolike RPC scenario, using the LHC Run-II $36~{\rm fb^{-1}}$ dataset, have excluded winos up to $\sim 650~{\rm GeV}$ ($WZ$ topology) and $\sim 480~{\rm GeV}$ ($Wh$ topology)~\cite{Sirunyan:2018ubx}. Thus, the $\frac{1}{2}\lambda_{ijk}L_{i}\cdot L_{j}e^{c}_{k}$-type RPV scenarios imply a more stringent exclusion on the electroweakino sector compared to the RPC scenarios due to harder leptons in the final state. In \cite{Aad:2014iza} as well, ATLAS has analyzed the four or more lepton final state in the context of RPV simplified scenario containing $\lambda_{ijk}L_{i}\cdot L_{j}e^{c}_{k}$-type couplings using the $\sqrt{s}=8~{\rm TeV}$ LHC data collected at $\sim 20.3~{\rm fb^{-1}}$ integrated luminosity. Results from \cite{Aad:2014iza} exclude a winolike chargino below $\sim 750~{\rm GeV}$, gluino below $\sim 1350~{\rm GeV}$ and left-handed (right-handed) sleptons below $\sim 490~{\rm GeV}$ ($410~{\rm GeV}$), for $M_{\lspone} = 300~{\rm GeV}$ at $95\%$ C.L., within a simplified RPV scenario where the binolike LSP $\lspone$ can decay only into electrons and muons. If the tau-rich decays are also included, the corresponding exclusion limits get weaker: winolike chargino ($\lesssim 450~{\rm GeV}$), gluino ($\lesssim 950~{\rm GeV}$), left-handed sleptons ($\lesssim 300~{\rm GeV}$) and right-handed sleptons ($\lesssim 240~{\rm GeV}$). The CMS collaboration also analyzed the $\sqrt{s}=8~{\rm TeV}$ LHC data ($19.5~{\rm fb^{-1}}$) and excluded stops up to $\lesssim 1100~{\rm GeV}$ and $\lesssim 950~{\rm GeV}$~\cite{Chatrchyan:2013xsw} at $95\%$ C.L. in simplified scenarios containing $\lambda_{122}$ and $\lambda_{233}$-type RPV couplings, respectively, for a binolike $\lspone$ with mass $\sim 400~{\rm GeV}$. The RPV scenario with $\frac{1}{2}\lambda_{233}^{\prime}L_{2}\cdot Q_{3}d^{c}_{3}$-type coupling has also been investigated in ~\cite{Chatrchyan:2013xsw} and has excluded stops with mass between $\sim 550~{\rm GeV}$ and $\sim 700~{\rm GeV}$ for a binolike $\lspone$ with mass $\sim 500~{\rm GeV}$ at $95\%$ C.L. 

Gluino searches in multijet final state~\cite{Aad:2015lea} and the jets plus two same-sign lepton or three lepton final state~\cite{Aad:2014pda} by the ATLAS collaboration using the LHC $\sqrt{s}=8~{\rm TeV}$ data ($\sim 20 ~{\rm fb^{-1}}$) within $\lambda^{\prime\prime}_{ijk} \epsilon_{\alpha\beta\gamma}u^{c\,\alpha}_{i}d^{c\,\beta}_{j}d^{c\,\gamma}_{k}$-type RPV simplified scenario has excluded gluinos up to $\lesssim 1100~{\rm GeV}$ and $\lesssim 1050~{\rm GeV}$, respectively, for $M_{\lspone} \sim 400~{\rm GeV}$, at $95\%$ C.L. CMS has also searched for the gluinos in multijet~\cite{Chatrchyan:2013gia} and same-sign dilepton plus jets final state~\cite{Chatrchyan:2013fea} using LHC $\sqrt{s}=8~{\rm TeV}$ data ($\sim 19.5~{\rm fb^{-1}}$) within the framework of $\lambda^{\prime\prime}_{ijk} \epsilon_{\alpha\beta\gamma}u^{c\,\alpha}_{i}d^{c\,\beta}_{j}d^{c\,\gamma}_{k}$-type RPV simplified scenarios, and have excluded gluinos below $\lesssim 650~{\rm GeV}$ and $\lesssim 900~{\rm GeV}$, respectively, at $95\%$ C.L. The phenomenology of $\lambda^{\prime\prime}_{ijk} u^{c}_{i}d^{c}_{j}d^{c}_{k}$-type of RPV operators has also been analyzed in \cite{Bhattacherjee:2013gr,Dercks:2017lfq} and the distinct collider signatures emerging in consequence to $\lambda^{\prime\prime} u^{c}_{i}d^{c}_{j}d^{c}_{k}$-type of RPV coupling have been analyzed in \cite{Bhattacherjee:2013tha,Graham:2014vya,Li:2018qxr}. At this point, it would be essential to take a look at the analogous exclusion limits in the RPC framework. Searches by the ATLAS and CMS Collaborations within RPC scenarios (using the LHC $\sqrt{s} = 8~{\rm TeV}$, $\sim 20~{\rm fb^{-1}}$ dataset) have excluded gluinos up to $\sim 1400~{\rm GeV}$~\cite{Aad:2014wea} and $\sim 1300~{\rm GeV}$~\cite{Chatrchyan:1630049}, respectively, for a binolike $\lspone$ with mass $\sim 400~{\rm GeV}$ at $95\%$ C.L. Using the same dataset, ATLAS and CMS also set lower limits on the mass of squarks ($\lesssim 900$~\cite{Aad:2014wea}) and stops ($\lesssim 760$~\cite{CMS-PAS-SUS-13-023}), respectively, at $95\%$ C.L. It is worthwhile to note that the RPC scenario and the RPV scenario discussed until now imply a comparable exclusion limit on the masses of gluinos and squarks. However, the electroweakino sector of $\lambda^{\prime\prime}_{ijk} \epsilon_{\alpha\beta\gamma}u^{c\,\alpha}_{i}d^{c\,\beta}_{j}d^{c\,\gamma}_{k}$-type RPV models still remain to be explored, and that is precisely the goal of this work.  

Our aim is to study the collider constraints on electroweakinos in RPV simplified scenarios with $\lambda^{\prime\prime}_{112} u^{c}d^{c}s^{c}$ and $\lambda^{\prime\prime}_{113} u^{c}d^{c}b^{c}$-types of RPV couplings in context of searches at the future HL-LHC ($\sqrt{s} = 14~{\rm TeV}$, $\mathcal{L} = 3000~{\rm fb^{-1}}$). In this analysis, we have assumed the $\lspone$ to decay promptly. The mean decay length of the LSP neutralino decaying via $\lambda_{ijk}^{\prime\prime}$-type RPV coupling is given by~\cite{Dreiner:1991pe, Barbier:2004ez}:

\begin{equation}
L ~(\rm{cm}) = 0.3 (\beta \gamma)_{\lspone} \left(\frac{M_{\tilde{q}}}{100 ~ \rm{GeV}}\right)^4 \left(\frac{1~ \rm{GeV}}{M_{\tilde{\chi_1^0}}}\right)^5 \frac{3}{{\lambda_{ijk}^{\prime \prime}}^2} \,,
\label{decay_length}
\end{equation} 

The consideration of a promptly decaying $\lspone$~($L \lesssim 1~{\rm cm}$) in Eq.~\ref{decay_length} restricts $\lambda_{ijk}^{\prime\prime}$ to values larger than $\gtrsim 10^{-4}$ and $\gtrsim 10^{-3}$ for $M_{\lspone} = 30~{\rm GeV}$ and $ 10~{\rm GeV}$, with $M_{\tilde{q}} = 100~{\rm GeV}$. Additionally, the $\lambda_{ijk}^{\prime\prime}$ couplings are also bounded from above by various experimental measurements. Although, our analysis is not sensitive to the value of the RPV coupling, nevertheless, it is worthwhile to discuss the existing constraints on $\lambda_{ijk}^{\prime\prime}$ from various experimental measurements and theoretical limitations. Indirect upper bounds have been derived on $\lambda_{112}^{\prime\prime}$~($\lesssim 10^{-15} \times {(m_{\tilde{q}}/\tilde{\Lambda}~{\rm GeV})}^{5/2}$, where $\tilde{\Lambda}$ is a hadronic scale and can be varied from 0.003 to 1~GeV, and $\tilde{q}$ is the squark mass) from double neutron beta decay~\cite{Goity:1994dq}. Indirect upper limits have also been placed on $\lambda_{113}^{\prime\prime}$ from neutron oscillations~\cite{Zwirner:1984is}, $\lambda_{113}^{\prime\prime} \lesssim 10^{-4}$ for $m_{\tilde{q}} = 100~{\rm GeV}$. This upper limit weakens to 0.002 and 0.1 for $m_{\tilde{q}}~=$~200 and 600~GeV~\cite{Zwirner:1984is,Allanach:1999ic}, respectively. $\lambda_{3jk}^{\prime\prime}$ has been constrained from $R_{l} = \Gamma(Z \to hadronic)/\Gamma(Z \to l \bar{l})$ to $\lesssim 0.5$ for $m_{\tilde{q}} = 100~{\rm GeV}$~\cite{BHATTACHARYYA1995193,Bhattacharyya:1997vv}. At higher values of $m_{\tilde{q}}$, the theoretical bounds on $\lambda_{3jk}^{\prime\prime}$ from the requirement of perturbative unification at the grand unified theory~(GUT) scale are more stringent. The remaining $\lambda_{ijk}^{\prime\prime}$ couplings are also indirectly constrained from the perturbativity bound~\cite{Barbier:2004ez,Allanach:1999ic}. The $\lambda^{\prime\prime}_{ijk}$ couplings are also constrained from the required lifetime of the LSP in order for it to be a viable candidate for dark matter. An unstable $\lspone$ LSP can also be a DM candidate provided it is abundantly long lived. Such a condition can manifest if the RPV couplings are very small. In such cases, various cosmological constraints can become relevant depending on the lifetime of the $\lspone$. For example, if the lifetime of the long-lived LSP $\lspone$ is slightly larger or comparable to the present age of the Universe, then its decay can lead to an excess of positrons or antiprotons which could lead to an incompatibility with the measurements. Resolution of this discrepancy requires that the lifetime of the $\lspone$ is sufficiently larger than the age of the Universe which in turn can be made possible only if the RPV couplings~(including $\lambda_{ijk}^{\prime\prime}$) are roughly of the order of $\lesssim 10^{-20}$~\cite{Barbier:2004ez}. The $\lambda_{ijk}^{\prime\prime}$ couplings are also strongly constrained from the three-body decay of the long-lived $\lspone$: $\lspone \to e^{+} + 2 f$~($f=~$ fermions) upon comparison with the observed positron flux, $\lesssim 4 \times 10^{-23} N_{1l}^{-1} \left(m_{\tilde{f}}/100~{\rm GeV}\right)^{2} \left(m_{\lspone}/100~{\rm GeV}\right)^{-9/8}\left(1~{\rm GeV}/m_{f}\right)^{1/2}$~(where, $N_1l$~($l=3,4$) is the higgsino admixture in $\lspone$ and $m_{f}$ is the mass of the emitted fermion)~\cite{Barbier:2004ez}. Constraints from nucleosynthesis also impose an upper limit on the value of $\lambda_{ijk}^{\prime\prime}$. In order to remain consistent with the predictions of light nuclei abundance from big bang nucleosynthesis, the lifetime of the LSP $\lspone$, if smaller than the age of the Universe, must be within an upper limit~\cite{Barbier:2004ez}. This restriction, in turn, imposes a lower bound on the RPV couplings, $\gtrsim 10^{-12}$~\cite{Barbier:2004ez}. 

Direct searches to probe $\lambda_{ijk}^{\prime\prime}$ have also been performed in the context of various colliders. The work in Ref.~\cite{Berger:1999zt} had showed that it should be possible to probe $\lambda^{\prime\prime}_{3jk} > 0.02 - 0.06$ for stop mass in the range of 180~-~325~GeV at the Run II of the Tevatron considering the stop decay: $\tilde{t_{1}} \to b \chonepm \to l\nu\lspone$. The data collected by DELPHI has also been analyzed in Ref.~\cite{Abreu:2000ne} where searches for pair produced gauginos and squarks were performed and lower limits on the stop and sbottom masses~($\gtrsim 70~{\rm GeV}$) were derived at $90\%$ C.L. for various benchmark scenarios. The study in Ref.~\cite{Abreu:2000ne} considered a RPV coupling strength of $\gtrsim 10^{-3}$. Considering the stop decaying promptly into two down quarks and a sbottom, followed by a subsequent decay of the sbottom into an up and a down quark (via $\lambda_{ijk}^{\prime \prime}$ coupling), the stop masses lower than 77~GeV have been excluded by OPAL \cite{Acton:1993xj,Abbiendi:1998ff,Abbiendi:1999wm,Abbiendi:1999is,Abbiendi:2003rn}. In light of the single top quark production, various earlier studies had indicated that the LHC will play a better role in probing $\lambda_{ijk}^{\prime \prime}$ among the hadron colliders \cite{Stelzer:1995mi,Datta:1997us,Oakes:1997zg,Chiappetta:1999cd}. In Ref.~\cite{Chiappetta:1999cd}, projections have been derived on the value of $\lambda_{ijk}^{\prime \prime}~\lambda_{lmn}^{\prime \prime}$ for different masses of the exchanged squarks, assuming an integrated luminosity of 30 fb$^{-1}$ at the LHC. The CMS and ATLAS Collaborations have also performed numerous searches to probe the implications of the $\lambda_{lmn}^{\prime \prime}$-type RPV couplings and have derived exclusion limits on sparticle masses. The ATLAS collaboration has performed a search in the multiple $b$-$jet$ final state produced from the cascade decay of direct produced lightest stop pair using the LHC Run-II data collected at $\mathcal{L} \sim 139~{\rm fb^{-1}}$~\cite{Aad:2020uwr}. The light stops are assumed to decay into a higgsinolike $\chonepm$ and a top quark or a $\lspone/\lsptwo$ and a bottom quark. The $\chonepm$, $\lspone$ and the $\lsptwo$ are considered to be mass degenerate and they have been assumed to undergo decay via: $\chonepm \to b b s$ and $\lsptwo/\lspone \to t bs$~(on account of the $\lambda_{323}^{\prime\prime}$ RPV coupling). The analysis in Ref.~\cite{Aad:2020uwr} considers $\lambda_{323}^{\prime\prime}$ in the order of $\sim O~(10^{-2} - 10^{1})$ which ensures promptly decaying charginos and neutralinos, and has excluded light stops with masses up to $\sim 950~{\rm GeV}$ in the $m_{\tilde{t}} \leq m_{t} + m_{\lsptwo,\lspone,\chonepm}$ region at $95\%$ C.L. Ref.~\cite{Aaboud:2017nmi} utilizes the LHC Run-II data collected at $\mathcal{L} \sim 36~{\rm fb^{-1}}$ to search for LSP stops decaying into two $jets$ via the $\lambda^{\prime\prime}_{ijk}$ couplings, and excludes stop mass in the range of $100$-$400$~GeV at $95\%$ C.L. for stop decaying into two light flavored $jets$. The stop mass ranges, 100-470~GeV and 480-610~GeV, are excluded at $95\%$ C.L., when the stop decays into a $b$ $jet$ and a light $jet$. The CMS collaboration has also used the same dataset to search for pair produced stops~\cite{Sirunyan:2018rlj}, each decaying to a pair of quarks via the $\lambda^{\prime\prime}_{ijk}$ couplings. The study in Ref.~\cite{Sirunyan:2018rlj}, in particular, has considered two different RPV coupling scenarios, $\tilde{t} \to q q^{\prime}$~(via $\lambda^{\prime\prime}_{312}$) and $\tilde{t} \to bq^{\prime}$~(via $\lambda^{\prime\prime}_{323}$), and has considered a relatively large RPV coupling such that the stops undergo prompt decay. This search has excluded stop masses in the range of 80-520~GeV with the $\lambda^{\prime\prime}_{312}$-type RPV couplings and 80-270~GeV, 285-340~GeV, 400-525~GeV, with the $\lambda^{\prime\prime}_{323}$-type RPV couplings, at $95\%$ C.L. We conclude the ongoing summarized revisit section by making a comment on the intermediate case, where the decay length of LSP is neither too large nor zero, {\em i.e.} where the decay vertex of the mother particle is displaced. The minimum distance between the production and decay vertex for differentiating them experimentally being $\mathcal{O}(2 \times 10^{-5} \rm{m})$, from Eq.\ref{decay_length} the upper limit on the RPV coupling reads:
\bea
\lambda_{ijk}^{\prime \prime} < \sqrt{3}\times 1.2 \times 10^{-4} \gamma^{\frac{1}{2}} \left(\frac{M_{\tilde{f}}}{100 ~ \rm{GeV}}\right)^2 
\left(\frac{100 ~\rm{GeV}}{M_{\tilde{\chi_1^0}}}\right)^{\frac{5}{2}} \,.
\eea
with $\gamma$ being Lorentz boost factor. Thus the long-lived LSP might be probed through the detection of displaced vertices as well if the couplings lie within the window of $\mathcal{O} (10^{-5}-10^{-4})$ \cite{Barbier:2004ez}.

A $\lambda^{\prime\prime}_{ijk} u^{c}_{i}d^{c}_{j}d^{c}_{k}$-type of RPV scenario, where the LSP would decay into a multijet final state: $\lspone \to j_{u}j_{d}j_{d}$ ($j_{u} = u,c,t$ and $j_{d} = d,s,b$), would be expected to be amply sensitive to search strategies which consider large jet multiplicity in the final state.
However, if the $jets$ produced from the decay of $\lspone$ are highly collimated, then they would evade identification as isolated $jets$, thereby, altering the reach of collider search strategies. 
Within simplified RPC scenarios, direct wino searches in the $Wh$ mediated $1l+2b+\met$ and $WZ$ mediated $3l+\met$ ($l=e,\mu$) final states~\cite{ATL-PHYS-PUB-2018-048} furnish robust bounds on the mass of winos as a function of $M_{\lspone}$. In this work, we analyze these collider searches in the context of $\lambda^{\prime\prime}_{112} u^{c}d^{c}s^{c}$-type RPV simplified scenarios\footnote{The respective final states feature additional light $jets$ produced from the decay of $\lspone$.} (in Section~\ref{Sec:wh_6jbblnu} and Section~\ref{Sec:6j3lmet}, respectively) and contrast them with the results for RPC scenarios in \cite{ATL-PHYS-PUB-2018-048}. Additionally, we also explore the future reach of direct wino searches in the $Wh$ mediated $1l+ 2\gamma + jets +\met$ final state (Section~\ref{Sec:wh_6jgagalnu}) and $WZ$ mediated $3l+2b+jets+\met$ final state (Section~\ref{sec:3l2bjmet}), respectively characterized by $\lambda^{\prime\prime}_{112} u^{c}d^{c}s^{c}$ and $\lambda^{\prime\prime}_{113} u^{c}d^{c}b^{c}$-type of RPV operators. We have considered final states containing leptons/photons in addition to the multiple $jets$ since they are easier to trigger and offer a cleaner signature. A few benchmark scenarios and their collider implications are discussed in Section~\ref{Sec:benchmark}. We conclude in Section~\ref{Sec:Conclusion}.

\section{Collider analysis}
\label{sec-3}

A simplified SUSY scenario~\cite{Alves:2011wf} with mass degenerate wino like $\lsptwo,~\chonepm$, and a binolike $\lspone$ is considered in this analysis. We consider the direct production of winolike $\lsptwo\chonepm$ pair due to its higher production cross section compared to wino-type neutralino pair ($\lsptwo\lsptwo$) or chargino pair ($\chonepm\chonemp$). Furthermore, the wino production cross section is also roughly $\sim 2$ times larger than the higgsino production rates. Correspondingly, the other SUSY particles namely sleptons, squarks, heavy Higgses and the heavier electroweakinos ($\lspthree$, $\lspfour$, $\chtwopm$) have been fixed at a higher mass in order to decouple their effects on our study. 

\begin{figure}[htpb!]{\centering
 \subfigure[]{
 \includegraphics[scale=0.55]{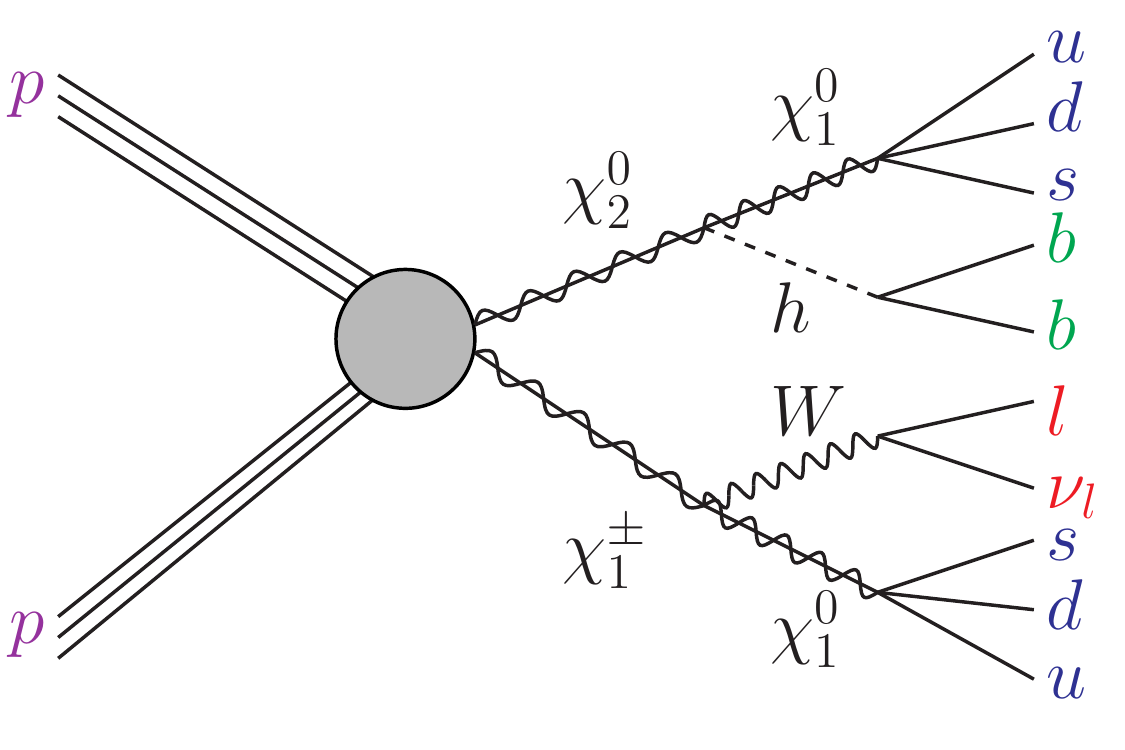}\label{fig:1a}}
 \hskip -5pt
 \subfigure[]{
 \includegraphics[scale=0.55]{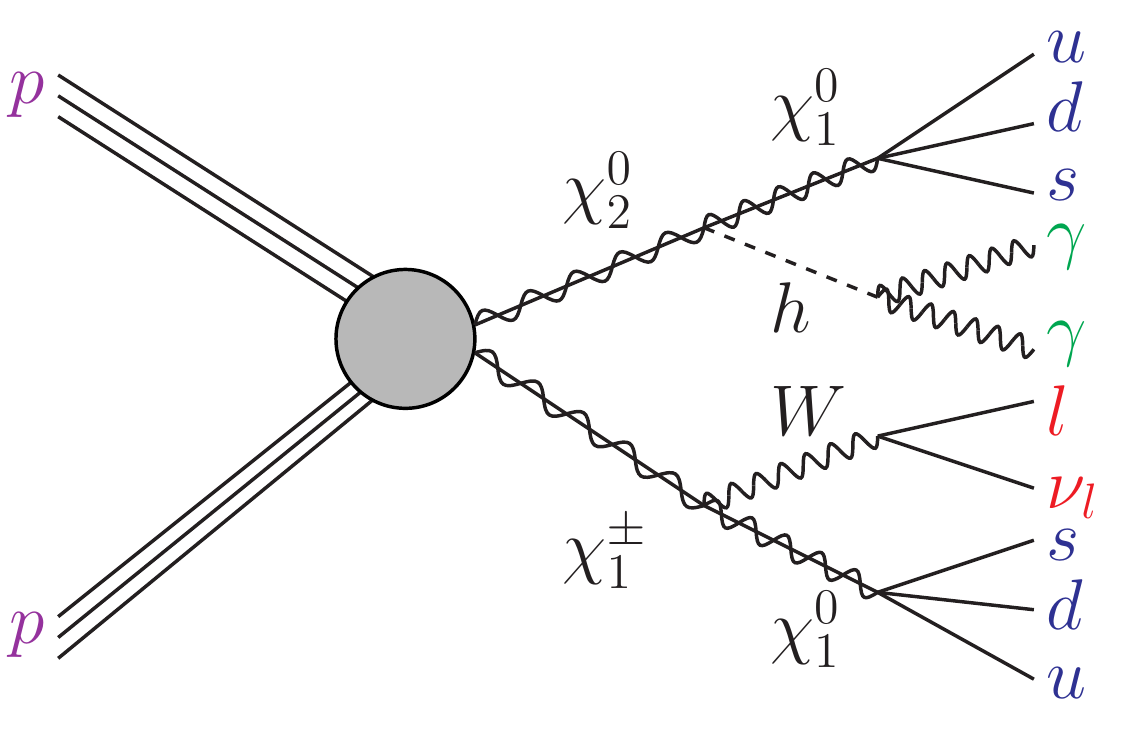}\label{fig:1b}} \\
 \hskip -5pt
 \subfigure[]{
 \includegraphics[scale=0.55]{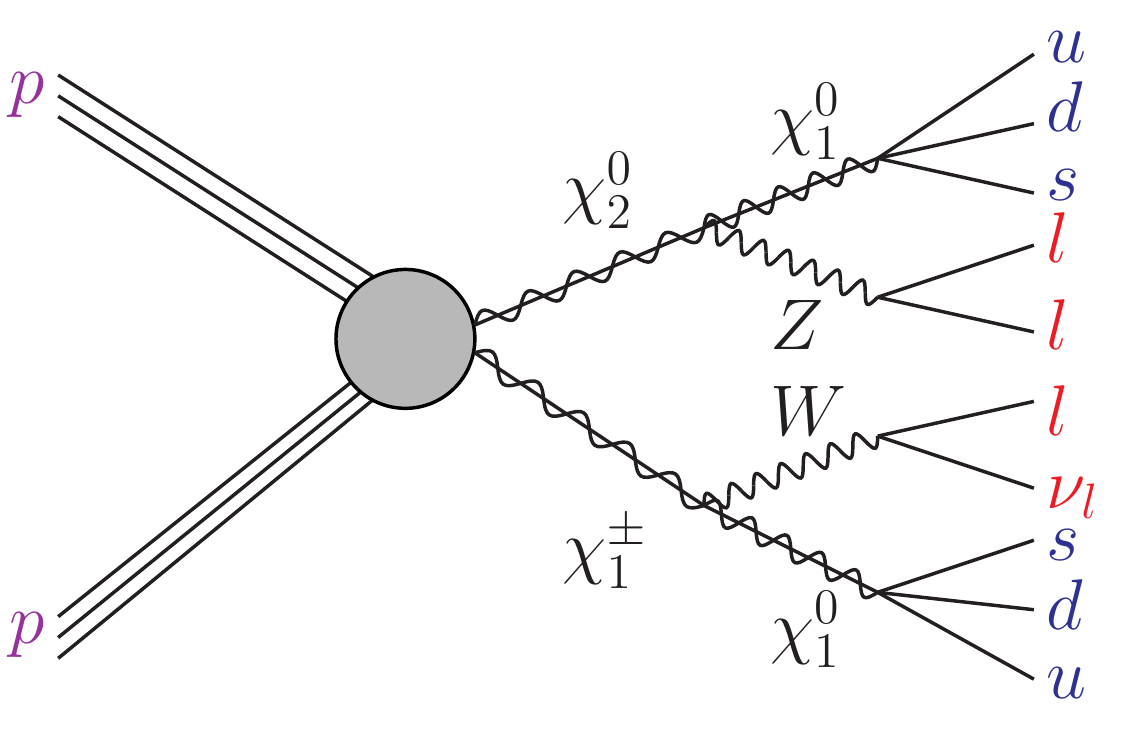}\label{fig:1c}}
  \hskip -5pt
 \subfigure[]{
 \includegraphics[scale=0.55]{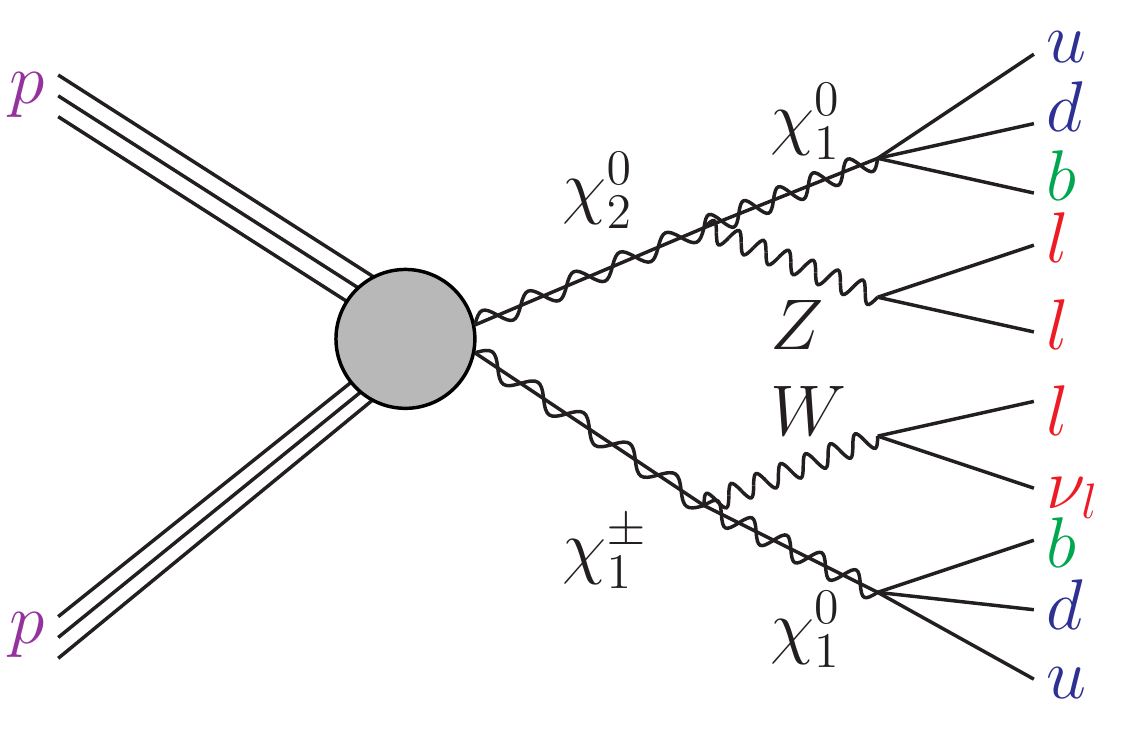}\label{fig:1c}}}
 \caption{\it \footnotesize Feynman diagrams representing the cascade decay chain of $\lsptwo\chonepm$ pair into (a) $Wh$ mediated $1l + 2b + jets +\met$, (b) $Wh$ mediated $1l + 2\gamma + jets +\met$, (c) $WZ$ mediated $3l+ jets + \met$ and (d) $WZ$ mediated $3l+2b+jets+\met$ final state. Here, $jets$ represents the light quark $jets$, while $l$ corresponds to an electron or muon.}\label{fig:proc}
 \end{figure}

Direct wino pair production is considered ($pp \to \lsptwo\chonepm$) and a detailed collider analysis is performed in multifarious final states originating from the cascade decay of the aforesaid $\lsptwo\chonepm$ pair. As stated in Section~\ref{sec:intro}, our aim is to study the collider ramifications of the baryon number violating RPV operator in simplified MSSM. To reiterate the structure of this paper, we study the implications of $\lambda_{112}^{\prime\prime}u^{c}d^{c}s^{c}$-type RPV term in Section~\ref{Sec:wh_6jbblnu}, \ref{Sec:wh_6jgagalnu} and \ref{Sec:6j3lmet} and $\lambda_{113}^{\prime\prime} u^{c}d^{c}b^{c}$-type RPV term in Section~\ref{sec:3l2bjmet}. In light of these terms, the $\lspone$ decays into a $uds$ final state in the initial three cases while the $\lspone$ decays into a $udb$ final state in the latter case, resulting in final states with large $jet$ multiplicity. 
The Feynman diagrams of the signal processes considered in Section~\ref{Sec:wh_6jbblnu}, \ref{Sec:wh_6jgagalnu}, \ref{Sec:6j3lmet} and \ref{sec:3l2bjmet} have been illustrated in Figure~\ref{fig:proc} (a), (b), (c) and (d), respectively.

In the present work, the signal events have been generated using \texttt{Pythia-6.4.28}~\cite{Sjostrand:2001yu,Sjostrand:2014zea}, while the \texttt{$\rm MadGraph\_aMC@NLO$}~\cite{Alwall:2014hca} framework has been used to generate the background events at leading order (LO) parton level in SM. Subsequent showering and hadronization has been performed through \texttt{Pythia-6.4.28}. The background events have been generated by matching up to $3$ $jets$ (the $3~jet$ matched sample of a background process $bkg$ will be represented as $bkg+jets$) except for the $W/Z+jets$ background process for which the $4$ $jet$ matched sample is used. The fast detector response has been simulated using \texttt{Delphes-3.4.1}~\cite{deFavereau:2013fsa}. The default ATLAS configuration card which comes along with \texttt{Delphes-3.4.1} package has been used in the entirety of this analysis\footnote{The $b$ $jet$ tagging efficiency has been assumed to be $70\%$ while the $c \to b$ ($u,d,s \to b$) mistag efficiency has been assumed to be $30\%$ ($1\%$).}. For the background processes, we have considered the leading order (LO) cross sections computed by \texttt{$\rm MadGraph\_aMC@NLO$} unless stated otherwise. The NLO-NLL order cross sections (taken from \cite{Fuks:2012qx,Fuks:2013vua}) have been considered for the signal processes (direct wino production: $\sigma^{wino}_{pp \to \lsptwo\chonepm}$).

In the following subsections, we present a detailed discussion of the collider search strategy employed to maximize the signal significance in the corresponding search channels and present our results on the projected reach of direct wino searches in these respective search channels at the HL-LHC.

\subsection{Searches in $Wh$ mediated $1l$~+~$2b$~+~$jets(N_{j} \geq 2)$~+~$\met$ channel}
\label{Sec:wh_6jbblnu}

The signal process considered in this subsection is illustrated in Figure~\ref{fig:proc}(a). The $\chonepm$ and $\lsptwo$ are assumed to decay into $ W\lspone$ and $h\lspone$, respectively, with a branching ratio of $100\%$, while, the SM branching values for $h \to b\bar{b}$ ($\sim 57\%$~\cite{Tanabashi:2018oca}) and $W \to l^{\prime} \nu$ ($\sim 31.7\%$~\cite{Tanabashi:2018oca}, $l^{\prime} = e,\mu,\tau$) have been considered. The cascade decay process culminates in two $\lspone$ along with other SM particles. The RPV operator: $\lambda^{\prime \prime}_{112} u^{c}d^{c}s^{c}$ implies $\lspone \to uds$, resulting in multiple light quark $jets$ in the final state. The cascade decay chain eventually results in $Wh$ mediated $1l+2b+jets+\met$ final state.

An event is required to have exactly one isolated lepton ($l = e,\mu$), at least two light $jets$ ($N_{j} \geq 2$), and exactly two $b$ $jets$ in the final state. The electron (muon) is considered to be isolated if $\Sigma p_{T}^{others}/p_{T}^{l}$ is $\leq 0.12$ for electrons and $\leq 0.25$ for muons, where, $\Sigma p_{T}^{others}$ is the scalar sum of transverse momenta of charged particles with $p_{T} \geq 0.5~{\rm GeV}$ (excluding the lepton under consideration) within a cone of radius $\Delta R = 0.5$ centred around the direction of lepton momentum and $p_{T}^{l}$ is the transverse momentum of the lepton. Here, $\Delta R$ is defined as: $\Delta R = \sqrt{\Delta \eta^{2} + \Delta \phi^{2}}$, where, $\Delta\eta$ and $\Delta\phi$ are the differences in pseudorapidity and the azimuthal angle, respectively, between the lepton under consideration and the charged particle. The isolated electron (muon) is required to have $p_{T} > 30~{\rm GeV}$, while the lighter $jets$ and the $b$ $jets$ are required to have $p_{T} > 20~{\rm GeV}$. In addition, the final state leptons and quarks must lie within a pseudorapidity range of $\leq |2.5|$.

The most dominant source of background is $t\bar{t}+jets$. Contributions to the background also arise from $WZ+jets$, $WW+jets$ and $ZZ+jets$ when $W,Z$ undergoes decay via leptonic decay modes. Additional contributions arise from $Wh+jets$ and $Zh+jets$ when the $h$ decays to $b\bar{b}$ while the $W/Z$ decays leptonically. Contributions from $Wb\bar{b}+jets$, $Wc\bar{c}+jets$ (here, the $c$ jet get mistagged as a $b$ jet) and $W+jets$ are also considered. Here, we have considered the NLO cross section for $t\bar{t}+jets$, where the NLO cross section has been computed by multiplying the NLO $k$ factor ($k=1.5$) with the LO cross section obtained from \texttt{$\rm MadGraph\_aMC@NLO$}. The cross section of background processes has been listed in the Appendix~(see Table~\ref{tab:Appendix_cs}).

Signal events have been generated for various combinations of $M_{\lsptwo}\left(=M_{\chonepm}\right)$ and $M_{\lspone}$. $M_{\lsptwo}$ has been varied from $200~{\rm GeV}$ to $1~{\rm TeV}$ with a step size of $25~{\rm GeV}$, while $M_{\lspone}$ has been varied between $25~{\rm GeV}$ to $M_{\lsptwo} - 125~{\rm GeV}$ with a step size of $10~{\rm GeV}$. Three different signal regions are chosen, SR1-A, SR1-B and SR1-C, aimed at maximizing the significance of signal events with small, intermediate and large mass difference, respectively, between the NLSP $\lsptwo,\chonepm$ and LSP $\lspone$. The selection cuts for SR1-A, SR1-B and SR1-C have been chosen by performing a cut-based analysis for the three representative benchmark points: BP1-A: $M_{\lsptwo}= 200~{\rm GeV}$, $M_{\lspone} = 55~{\rm GeV}$, BP1-B: $M_{\lsptwo}= 350~{\rm GeV}$, $M_{\lspone} = 165~{\rm GeV}$ and BP1-C: $M_{\lsptwo}= 500~{\rm GeV}$, $M_{\lspone} = 25~{\rm GeV}$, respectively. The values of $\sigma_{pp \to \lsptwo\chonepm}^{wino}$ for BP1-A, BP1-B and BP1-C have been listed in Appendix~\ref{Appendix:cs}. The signal yield ($S$) has been computed as follows:
\begin{eqnarray}
S = \sigma_{pp \to \lsptwo\chonepm}^{wino} \times  \mathcal{L} \times Eff. \times Br(\lsptwo\chonepm \to W\lspone h\lspone \to l \nu b \bar{b}+jets)
\label{Eqn:signal_yield} 
\end{eqnarray}
where, $\mathcal{L}$ is the integrated luminosity ($\mathcal{L}=3000~ {\rm fb^{-1}}$ for HL-LHC) and $Eff.$ represents the efficiency of the signal region\footnote{$Eff.$ is the ratio of the number of signal events which pass through a certain signal region ($\rm IEV$) to the total number of generated signal events ($\rm NEV$); $Eff.$ = $\frac{\rm IEV}{\rm NEV}$.}. 

\begin{figure}[htpb!]{\centering
 \subfigure[]{
 \includegraphics[height=2.3in,width=3.2in]{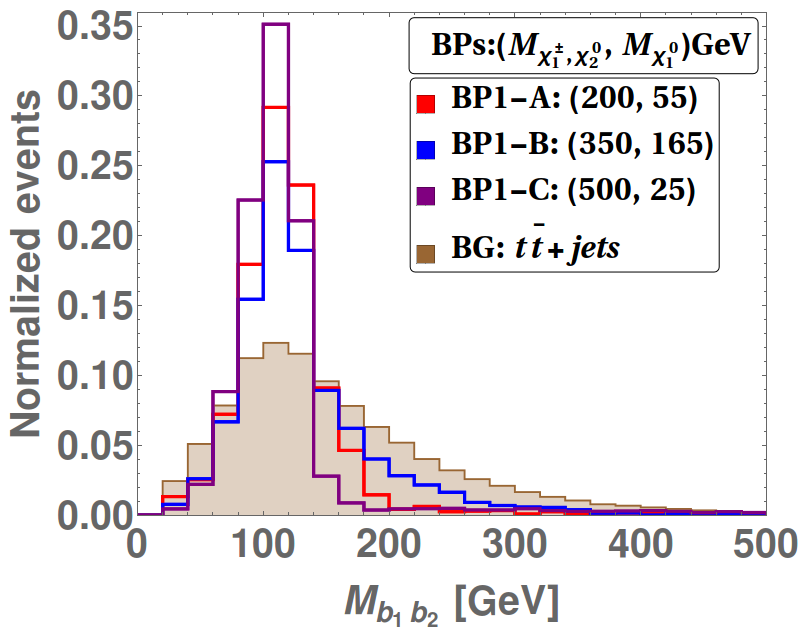}} 
 \subfigure[]{
 \includegraphics[height=2.27in,width=3.2in]{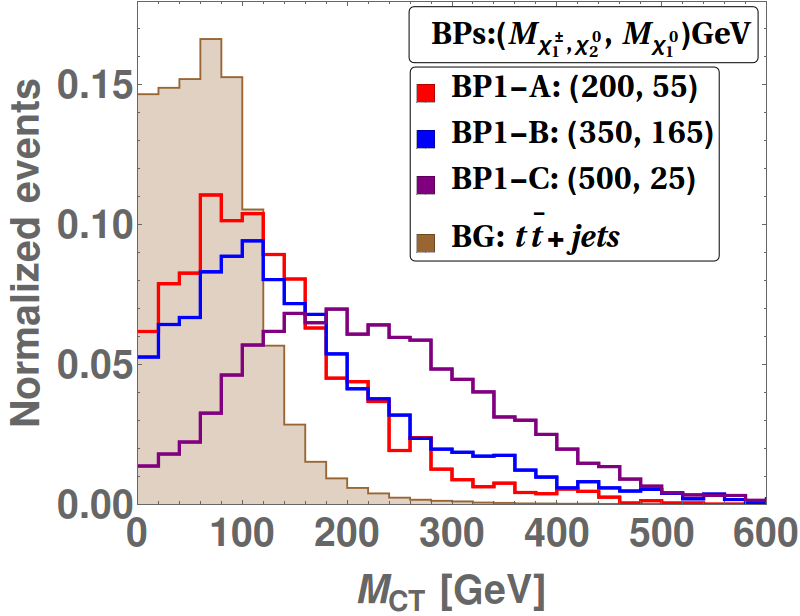}} \\
   \subfigure[]{
 \includegraphics[height=2.3in,width=3.2in]{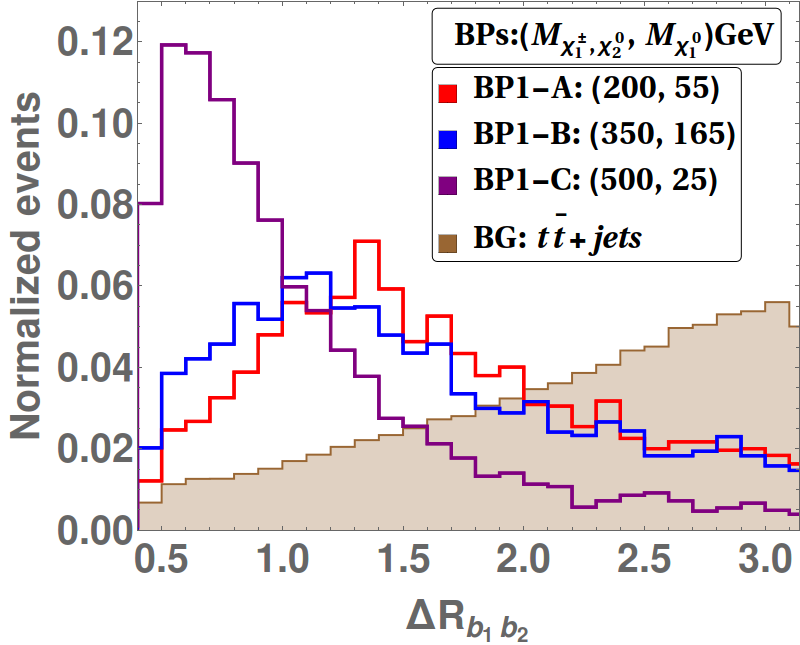}}
  \subfigure[]{
 \includegraphics[height=2.3in,width=3.2in]{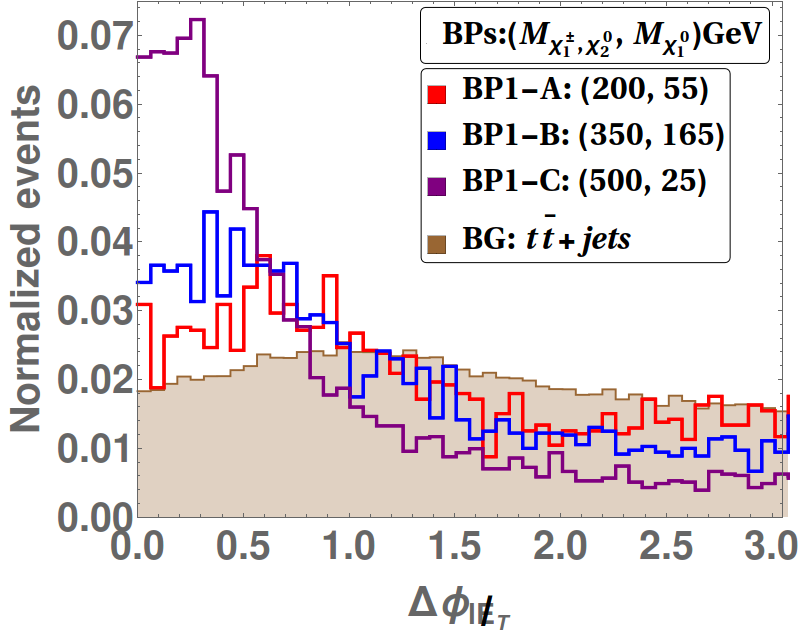}}
}
 \caption{\it \footnotesize Normalized distributions of $M_{b_{1}b_{2}}$ (top left), $M_{CT}$ (top right), $\Delta R_{b_{1}b_{2}}$ (bottom left), $\Delta \Phi_{l\met}$ (bottom right) are shown for BP1-A~(red solid line), BP1-B~(blue solid line), BP1-C~(purple solid line) and the $t \bar{t}+jets$ background~(brown color).}
 \label{fig:DistWHbb}
 \end{figure}

A variety of kinematic variables have been used to design the optimized signal regions. First and foremost, the invariant mass of the two final state $b$ $jets$, $M_{b_{1}b_{2}}$ ($b_{1}$ and $b_{2}$ represents the $p_{T}$ ordered leading and subleading $b$ $jets$ in the final state), is used to discriminate the background. For the signal process, the $b\bar{b}$ pair is produced from the decay of the $h$ and thereby peaks roughly around $\sim 110-115~{\rm GeV}$. On the other hand, the $M_{b_{1}b_{2}}$ distribution for the most dominant $t\bar{t}+jets$ background has a smoothly falling distribution since the two $b$ $jets$ are produced from the decay of two different top quarks. The $M_{b_{1}b_{2}}$ distribution for the signal benchmark points and the $t\bar{t}+jets$ background has been shown in Figure~\ref{fig:DistWHbb}(a)\footnote{The cross section for $t\bar{t}+jets$ process is roughly an order of magnitude higher than the other relevant backgrounds, and, therefore, for the sake of illustrative clarity, we display the kinematic distribution for the $t\bar{t}+jets$ background process only in Figure~\ref{fig:DistWHbb}.}. The distributions for BP1-A, BP1-B and BP1-C in Figure~\ref{fig:DistWHbb} have been illustrated as red, blue and purple solid colors while the $t\bar{t}+jets$ background has been shown in brown color.  

The $t\bar{t}+jets$ background also undergoes a considerable reduction upon the application of a lower bound on the contransverse mass ($M_{CT}$)~\cite{Tovey:2008ui,Polesello:2009rn}, where $M_{CT}$ is defined as, $ M_{\rm{CT}} = \sqrt{(E_T^{b1} + E_T^{b2})^2 - |\vec{p_T}^{b1} - \vec{p_T}^{b2}|^2}$. Here, $\vec{p_T}^{bi}$ and $E_T^{bi}$ are transverse momentum and energy of the $i$-th $b$-$jet$. The normalized $M_{CT}$ distribution for the signal benchmark points and the $t\bar{t}+jets$ background has been shown in Figure~\ref{fig:DistWHbb} (b).

The invariant mass of the first three $p_{T}$ ordered light $jets$, $M_{j_{1}j_{2}j_{3}}$, and the scalar sum of their transverse momenta, $H_{T}$, are also utilized in performing the cut-based analysis. The larger mass difference ($\Delta M$) between ($\lsptwo,\chonepm$) and  $\lspone$ in BP1-C results in the $\lspone$ being produced with a relatively larger boost, thereby, producing more collimated light $jets$ from the decay of $\lspone$. As a result, the kinematic variables constructed by using the momenta of the leading light $jets$ ($H_{T}$ and $M_{j_{1}j_{2}j_{3}}$) peak at a higher value for signal scenarios with large mass difference between the NLSP and the LSP (viz. BP1-C) as compared to the cases where the $\lsptwo$ and $\lspone$ are closer in mass to each other (viz. BP1-B and BP1-A).

\begin{table}[htpb!]
\begin{center}\scalebox{0.65}{
\begin{tabular}{||C{2.5cm}||C{3.0cm}|C{2.0cm}|C{2cm}|C{2cm}|C{2cm}|C{2cm}|C{2cm}|C{3cm}|} 
\hline
\hline
\multicolumn{9}{|c|}{SR1-A} \\ \cline{1-9}
\multicolumn{2}{|c|}{Cut variables} &~$M_{b_{1}b_{2}}$&~$M_{CT}$ &$\Delta R_{b_{1}b_{2}}$&$\Delta \phi_{l,\met}$&~$M_{l\met b}$ & - & - ~\\
\multicolumn{2}{|c|}{•} &(GeV)&(GeV)& & & (GeV) & - & - \\
\hline
\multicolumn{2}{|c|}{Selection cuts for SR1-A} & 70-130 & $ > 200$ & $< 1$ & $ <1.25 $ & $>190$ & - & - \\\hline \hline 
\multirow{7}{*}{\rotatebox[origin=c]{90}{\centering Signal and background yields}} & Signal (BP1-A) & $4.54\times 10^{4}$ & $ 5823$ & $ 4881$ & $ 3311$ & $ 2655$ & - & - \\ \cline{2-9}
& $t\bar{t}+jets$ &$2.78\times 10^7 $ & $ 5.35\times 10^{5}$ & $ 4.80\times 10^{5}$ & $ 3.01\times 10^{5}$ & $ 2.01\times 10^{5}$ & - & -
\\
& $WZ+jets $ & $4.38\times 10^{4}$ & $ 7211$  & 7211 & $ 4635$ & $ 3708$ & - & - \\
& $WW+jets$  &$1.41 \times 10^{4}$ & $ 1393 $ & 1393 & $ 806$ & $ 440$ & - & - \\
& $ZZ+jets$  &$6064$ & $ 683$ & $ 668$ & $ 282$ & $ 222$ & - & - \\
& $Wh+jets$  &$5595$ & $ 859$ & $ 676$ & $ 553$ & $ 417$ & - & -
\\
& $Zh+jets$  &$1249$ & $129$ & $ 51$ & $ 38$ & $ 23$ & - & - \\ \hline
\multicolumn{3}{|c|}{Total background yield: 2.05 $\times 10^{5}$} & \multicolumn{3}{c|}{Total signal yield: $2655$} & \multicolumn{3}{c|}{Signal significance: $5.9$} \\
\hline
\hline
\multicolumn{9}{|c|}{SR1-B} \\ \cline{1-9}
\multicolumn{2}{|c|}{Cut variables} &~$M_{b_{1}b_{2}}$&~$M_{CT}$&$M_{j_1j_2j_3}$&$H_T$&$\Delta R_{b_{1}b_{2}}$&$\Delta \phi_{l,\met}$&~$M_{l\met b}$~\\
\multicolumn{2}{|c|}{•} &(GeV)&(GeV)&(GeV)&(GeV)&& & (GeV)\\
\hline 
\multicolumn{2}{|c|}{Selection cuts for SR1-B} & 70-130 & $>$ 160 & $>$ 140 & $>$ 210 & $<$ 1 & $<$ 1 & $>$ 200 \\ \hline \hline
\multirow{7}{*}{\rotatebox[origin=c]{90}{\centering Signal and background yields}} & Signal (BP1-B) & $6975$ & $ 2317$ & $ 2228$ & $ 1999 $ & $ 1277$ & $ 729$ & $ 572$ \\ \cline{2-9}
& $t\bar{t}+jets$ &$2.79\times10^7 $ & $ 1.22\times10^6 $ & $ 9.37\times 10^{5}$ & $ 6.92\times 10^{5}$ & $ 4.63\times 10^{5}$ & $ 1.88\times 10^{5}$ & $ 7.75\times 10^{4}$
\\
& $WZ+jets$  &$4.38\times 10^{4}$ & $ 1.24\times 10^{4}$ & $ 8447$ & $ 7313$ & $ 6799 $ & $ 3914$ & $ 2575$
\\
& $WW+jets$  &$1.41\times 10^{4}$ & $ 2199$ & $ 1906$ & $ 1686$ & $ 1466$ & $ 586$ & $ 293$\\
& $ZZ+jets$  &$6063$ & $ 1293$ & $ 907$ & $ 743$ & $ 684$ & $ 193$ & $ 134$\\
& $Wh+jets$  &$5595$ & $ 859$ & $ 631$ & $ 602$ & $334$ & $209$ & $129$\\
& $Zh+jets$  &$1249$ & $ 130$ & $ 45$ & $ 43$ & $ 28$ & $21$ & $ 6.4$\\ \hline
\multicolumn{3}{|c|}{Total background yield: 8.07 $\times 10^{4}$} & \multicolumn{3}{c|}{Total signal yield: $572$} & \multicolumn{3}{c|}{Signal significance: $2.0$} \\
\hline\hline 
\multicolumn{9}{|c|}{SR1-C} \\ \cline{1-9}
\hline 
\multicolumn{2}{|c|}{Selection cuts for SR1-C} & 70-130 & $>$ 160 & $>$ 140 & $>$ 240 & $<$ 1 & $<$ 1 & $>$ 360 \\ \hline
\multirow{7}{*}{\rotatebox[origin=c]{90}{\centering Signal and background yields}} & Signal (BP1-C) & $3147$ & $ 2168$ & $ 2061$ & $ 1950 $ & $ 1672 $ & $ 1206 $ & $464$ \\ \cline{2-9}
& $t\bar{t}+jets$ &$2.79\times10^7 $ & $ 1.22\times10^6 $ & $ 9.37\times 10^{5}$ & $ 6.01\times 10^{5}$ & $ 4.03\times 10^{5}$ & $ 1.63 \times 10^{5}$ & $ 1.06\times 10^{4}$\\
 & $WZ+jets$  &$4.39\times 10^{4}$ & $ 1.24\times 10^{4}$ & $ 8447$ & $ 6696$ & $ 6181$ & $ 3708$ & $ 515$\\
& $WW+jets$  &$1.4148\times 10^{4}$ & $ 2199$ & $ 1906$ & $ 1393$ & $ 1173$ & $ 440$ & $ 73.3$\\
& $ZZ+jets$  &$6063$ & $ 1292$ & $ 907$ & $ 624$ & $ 580$ & $ 178$ & $ 0.0$\\
& $Wh+jets $  &$5595$ & $ 859$ & $ 631$ & $ 602$ & $ 334$ & $ 188$ & $ 32$\\
& $Zh+jets $  & $1249$ & $ 130$ & $ 45$ & $ 42$ & $ 28$ & $ 19$ & $ 0.4$\\\hline
\multicolumn{3}{|c|}{Total background yield: 1.13 $\times 10^{4}$} & \multicolumn{3}{c|}{Total signal yield: $464$} & \multicolumn{3}{c|}{Signal significance: $4.4$} \\ \hline
\end{tabular}}
\end{center}
\caption{\it \footnotesize Selection cuts corresponding to SR1-A, SR1-B and SR1-C, optimized to maximize the signal significance of signal processes with small, intermediate and large $\Delta M$ between the NLSP $\chonepm,\lsptwo$ and the LSP $\lspone$, respectively, for searches in the $Wh$ mediated $1l +2b+ jets +\met$ channel at the HL-LHC, are shown. The cut flow table for BP1-A, BP1-B and BP1-C and the relevant backgrounds are also tabulated along with the respective signal significance values.}
\label{table:1_cut_flow_1a}
\end{table}

In addition, the $\Delta R$ between the two final state $b$ $jets$, $\Delta R_{b_{1}b_{2}}$, and the difference between the azimuthal angles of the final state lepton and the $\met$, $\Delta \phi_{\ell,\met}$, are also used in constructing the optimized signal regions. The three benchmark points, BP1-A, BP1-B and BP1-C, feature an on-shell $h$ produced from the decay of $\lsptwo \to \lspone h$. For the case of BP1-C, the relatively larger $\Delta M$ between $\lsptwo$ and $\lspone$ results in a $h$ with relatively larger boost as compared to the case of BP1-A and BP1-B. Thereby, the $b\bar{b}$ pair in the final state of BP1-C is more collimated. As a result, $\Delta R_{b_{1}b_{2}}$ in BP1-C peaks at a smaller value ($\Delta R_{b_{1}b_{2}} \sim 0.7$) than in BP1-B ($\Delta R_{b_{1}b_{2}} \sim 1.4$). Similarly, the $\Delta R_{b_{1}b_{2}}$ distribution for BP1-B peaks at a relatively lower value (at $\Delta R_{b_{1}b_{2}} \sim 1.1$) than for BP1-A. Furthermore, the $b\bar{b}$ pair which originates from the $t\bar{t}+jets$ background is generated from two different mother particles, and therefore, are widely separated in the azimuthal angle. Consequently, $\Delta R_{b_{1}b_{2}}$ for the $t\bar{t}+jets$ background peaks at further larger values ($\Delta R_{b_{1}b_{2}} \sim 3$). The normalized $\Delta R_{b_{1}b_{2}}$ distribution has been illustrated in Figure~\ref{fig:DistWHbb} (c), where the red, blue and purple solid lines represents BP1-A, BP1-B and BP1-C, respectively, while the brown colored region represents the $t\bar{t}+jets$ background. Additionally, we also consider the azimuthal angular separation between the $l$ (produced from $W \to l\nu$) and $\met$, represented as $\Delta \phi_{\ell,\met}$, in performing the cut-based analysis. For BP1-C, $\Delta \phi_{\ell,\met}$ peaks at a lower value than in BP1-B/BP1-A. The normalized distributions of $\Delta \phi_{\ell,\met}$ for BP1-A, BP1-B, BP1-C and $t\bar{t}+jets$ have been shown in Figure~\ref{fig:DistWHbb} (d). 

It is to be noted that the top quark dominantly decays into a $bW$ pair and effectively contributes to the background when one of the $W$ boson undergoes leptonic decay while the other $W$ decays hadronically. One obtains two solutions for the z-component of momentum of the neutrino ($\slashed{p}_{z}^{\nu}$) produced from the leptonically decaying $W$: $\slashed{p}_{z}^{\nu} = (a_1 \,p_{z}^l  \pm \sqrt{ a_3})/a_2$, were, $a_1= p_{x}^l \slashed{E}_{x} + p_{y}^l \slashed{E}_{y} + \frac{M_W^2}{2},$
$a_2=E^{l^2} - p_{z}^{l^2}$ and $a_3=E^{l^2} a_1^2  - a_2 E^{l^2}\, (\slashed{E}_{x}^{2} + \slashed{E}_{y}^{2})$, with $p_{x,y,z}^l$ representing the $x$-, $y$-, $z$- components of momentum of the lepton, $E^{l}$ representing the energy of the lepton, and $\slashed{E}_{x,y}$ represents the x-component and y-component of the missing transverse energy. The signal process considered in the current analysis contains two $b$ $jets$ in the final state, and, coupled with the two possible solutions for $\slashed{p}_{z}^{\nu}$, results in two different values of $M_{l\met^{j} b_{1}}$ ($j=1,2$) and two values of $M_{l\met^{j} b_{2}}$. Here, $M_{l\met b_{i}}~(i=1,2)$ represents the invariant mass of the final state lepton, the missing energy and the $b$ $jets$. The $M_{l\met^{j} b_{i}}$ variable is effective against the $t\bar{t}+jets$ background where the only contribution to $\met$ comes from the $\nu$ produced by the leptonically decaying $W$. In this regard, we compute all four values of $M_{l\met^{j} b_{i}}$ and choose the combination whose value is closest to the top mass. The invariant mass of the aforesaid combination is represented as $M_{l\met b}$ and has been used in performing the cut-based analysis.

\begin{figure}[htpb!]
\centering
\includegraphics[scale=0.5]{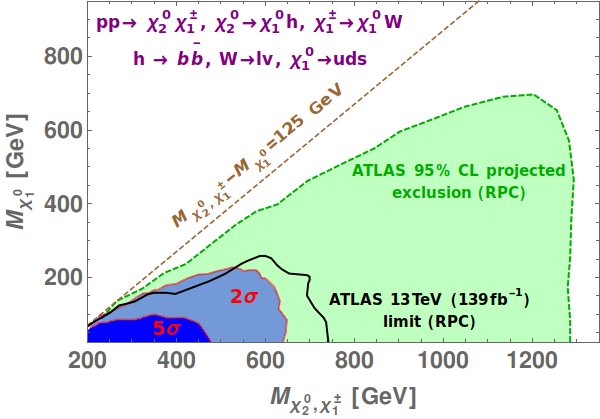}
\caption{\it \footnotesize The projected discovery and exclusion regions are shown in dark blue and light blue colors, respectively, in the $M_{\lspone}$-$M_{\lsptwo,\chonepm}$ plane. The projection contours have been derived from searches in the $1l+2b+jets+\met$ final state resulting from the cascade decay of directly produced winolike $\lsptwo\chonepm$ pair within a simplified model scenario containing $\lambda^{\prime\prime}_{112}u^{c}d^{c}s^{c}$-type RPV term, at the HL-LHC~($\sqrt{s} = 14~{\rm TeV}$, $\mathcal{L} \sim 3~{\rm ab^{-1}}$). The solid black line represents the current observed limit at $95\%$ C.L. from direct wino searches in the $WH$ mediated $1l+2b+\met$ final state, in a RPC simplified scenario, derived by ATLAS using the LHC Run-II dataset collected at $\mathcal{L}\sim 139~{\rm fb^{-1}}$\cite{Aad:2019vvf}. The light green colored region corresponds to the projected exclusion reach (at $95\%$ C.L.) of HL-LHC, derived by ATLAS, in direct wino searches in the $WH$ mediated $1l+2b+\met$ final state within a simplified  RPC scenario \cite{ATL-PHYS-PUB-2018-048}. 
The brown dashed line represents the condition for on-shell Higgs production ($ M_{\lsptwo,\chonepm}-M_{\lspone} =125$~${\rm GeV}$).}
\label{fig:1_excl_contour}
\end{figure}

The optimized selection cuts corresponding to SR1-A, SR1-B and SR1-C have been shown in Table~\ref{table:1_cut_flow_1a}. The signal yields for BP1-A, BP1-B and BP1-C, along with the corresponding background yields obtained after successive application of selection cuts listed in SR1-A, SR1-B and SR1-C, respectively, have also been shown in Table~\ref{table:1_cut_flow_1a}. It should be noted that the signal significances\footnote{The signal significance is computed as $S/\sqrt{B}$, where $S$ and $B$ are the signal and background yields.} tabulated in Table~\ref{table:1_cut_flow_1a} have been obtained without assuming any systematic uncertainty. SR1-A results in a signal significance of $5.9$ for BP1-A, while SR1-B and SR1-C have a signal significance of $2.0$ and $4.4$ for BP1-B and BP1-C, respectively.

We also derive the projected exclusion limits in the $M_{\lspone}-M_{\lsptwo,\chonepm}$ plane from direct wino searches at the HL-LHC in the $Wh$ mediated $1l+2b+jets+\met$ search channel. The value of signal significance is computed for the three optimized signal regions and the maximum among them is considered in deriving the projection regions\footnote{The same strategy has been followed in all the analyses considered in this work.}. 
The projected exclusion and discovery region corresponds to the sector with signal significance $> 2\sigma$ and $> 5\sigma$, respectively. They have been represented in light blue and dark blue colors, respectively, in Figure~\ref{fig:1_excl_contour}. The brown dashed line in the same figure corresponds to the on-shell mass condition for $h$ production and represents the mass correlation: $M_{\lsptwo} - M_{\lspone} = 125~{\rm GeV}$. It can be observed from Figure~\ref{fig:1_excl_contour} that within the framework of a $\lambda^{\prime\prime}_{112}u^{c}d^{c}s^{s}$-type RPV simplified scenario, direct wino searches at the HL-LHC in the $Wh$ mediated $1l+2b+jets+\met$ search channel has a potential exclusion (discovery) reach up to $\sim 630~{\rm GeV}$ ($\sim 450~{\rm GeV}$) for a binolike $M_{\lspone} \sim 0~{\rm GeV}$.


The ATLAS collaboration has also analyzed the current observed limit ($\sqrt{s}=13~{\rm TeV}$, $\mathcal{L}=139~{\rm fb^{-1}}$) as well as the projected reach of direct wino production at the HL-LHC in the analogous channel for the RPC scenario: $Wh$ mediated $1l+2b+\met$ final state in \cite{Aad:2019vvf} and \cite{ATL-PHYS-PUB-2018-048}, respectively. The current observed limit (at $95\%$ C.L.) reaches up to $M_{\lsptwo,\chonepm} \sim 720~{\rm GeV}$ for $M_{\lspone} = 100~{\rm GeV}$. The projected exclusion and discovery contour of ATLAS reaches up to $M_{\lsptwo,\chonepm} \sim 1300~{\rm GeV}$ and $\sim 600~{\rm GeV}$ for a binolike $\lspone$ with mass up to $100~{\rm GeV}$ at $95\%$ C.L. The ATLAS exclusion contour has been shown in light green color in Figure~\ref{fig:1_excl_contour}. Thus, the projected reach of direct wino searches interpreted in a RPC simplified scenario can get significantly weakened in the presence of $\lambda_{112}^{\prime \prime} u^{c}d^{c}s^{c}$-type RPV coupling. In the next three subsections, we further study the collider implications of RPV couplings in different final states.



\subsection{Searches in $Wh$ mediated $1l$~+~$2\gamma$~+~$jets (N_{j} \geq 2)$~+~$\met$ channel}
\label{Sec:wh_6jgagalnu}

In the current subsection, we consider the process: $pp \to \chonepm\lsptwo$ $\to$ $\left(W \lspone\right)$ $\left(h \lspone\right)$ $ \to \left( l \nu uds \right)$ $\left(\gamma \gamma u d s\right)$, which culminates in $1l+2\gamma +jets+\met$ final state (Figure~\ref{fig:proc}(b)). Here, $Br(\chonepm \to W \lspone)$ and $Br(\lsptwo \to h\lspone)$, have been assumed to be $100\%$. The small branching rate of $h \rightarrow \gamma \gamma$ is a significant drawback for this channel, however, a large photon detection efficiency, sharp diphoton invariant mass peak and a smaller background makes it a promising one. Unlike the $Wh$ mediated $1l+2b+jets+\met$ channel (discussed in Section~\ref{Sec:wh_6jbblnu}), no HL-LHC projection study has been performed for the analogous RPC scenario channel ($Wh$ mediated $1l+2\gamma+\met$). 

The event selection criteria requires the presence of exactly one isolated lepton ($l = e,\mu$), two photons, and at least two light $jets$ ($N_{j} > 2$) in the final state. The lepton isolation criteria specified in Section~\ref{Sec:wh_6jbblnu} is implemented here as well. 
The final state lepton, jets and photons satisfy the criteria : $|\eta^{l,~jet,~\gamma}| < 2.5$ and $p_T^{l,~jet,~\gamma} > 30, 30, 20~ \rm{GeV}$ respectively.
In addition, we demand that no pair of final state particles must be within $\Delta R < 0.5$ of each other. 
 Furthermore, a $b$-$jet$ veto is applied. 

 \begin{figure}[!htb]{\centering
   \subfigure[]{
 \includegraphics[height=2.3in,width=3.2in]{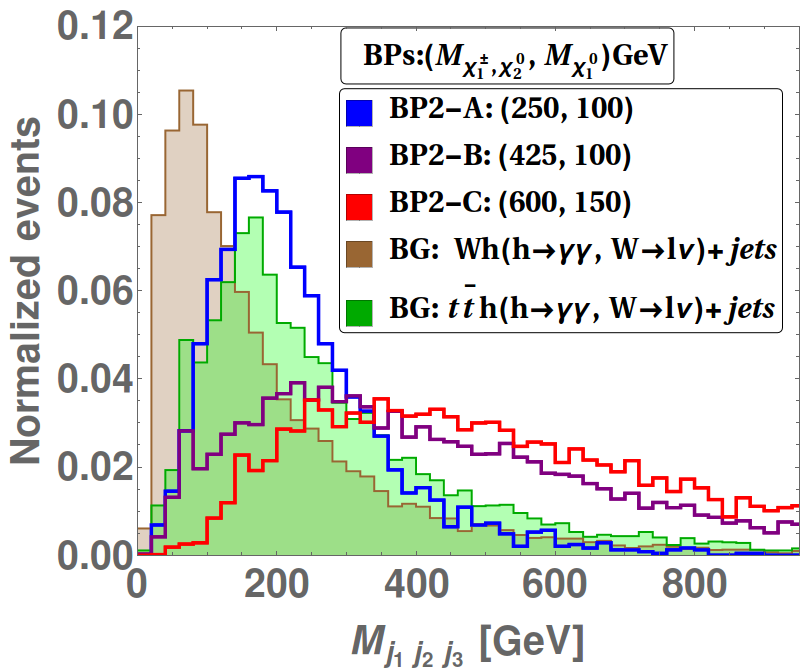}}
  \subfigure[]{
 \includegraphics[height=2.24in,width=3.2in]{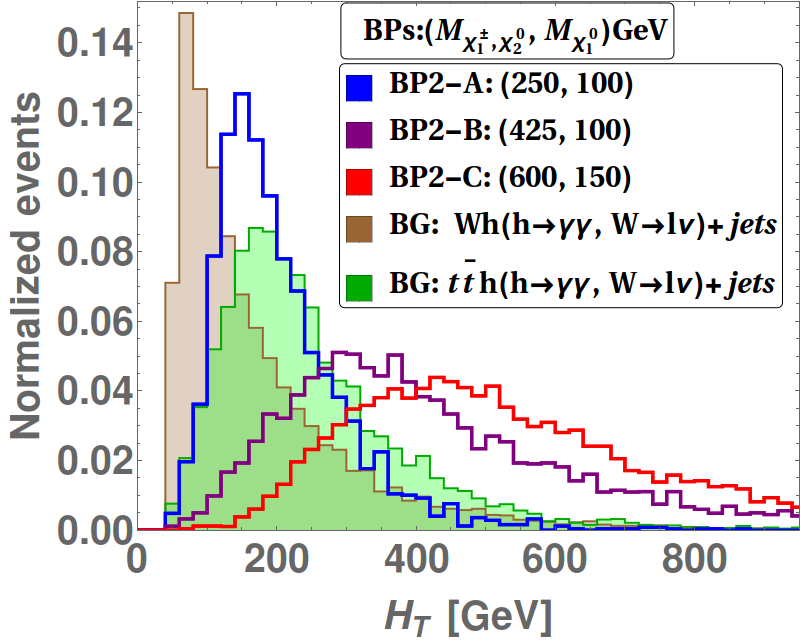}} \\
 \subfigure[]{
 \includegraphics[height=2.3in,width=3.2in]{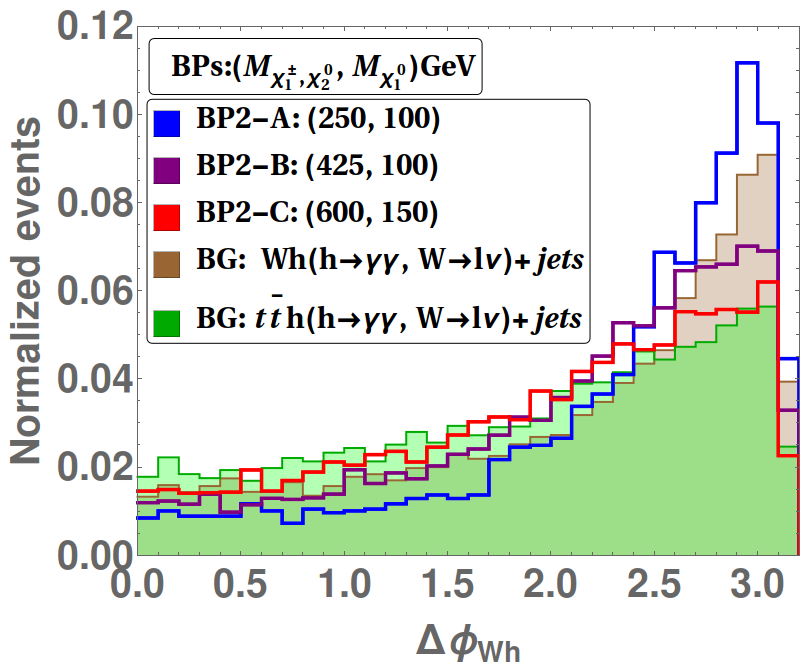}}
  \subfigure[]{
 \includegraphics[height=2.25in,width=3.2in]{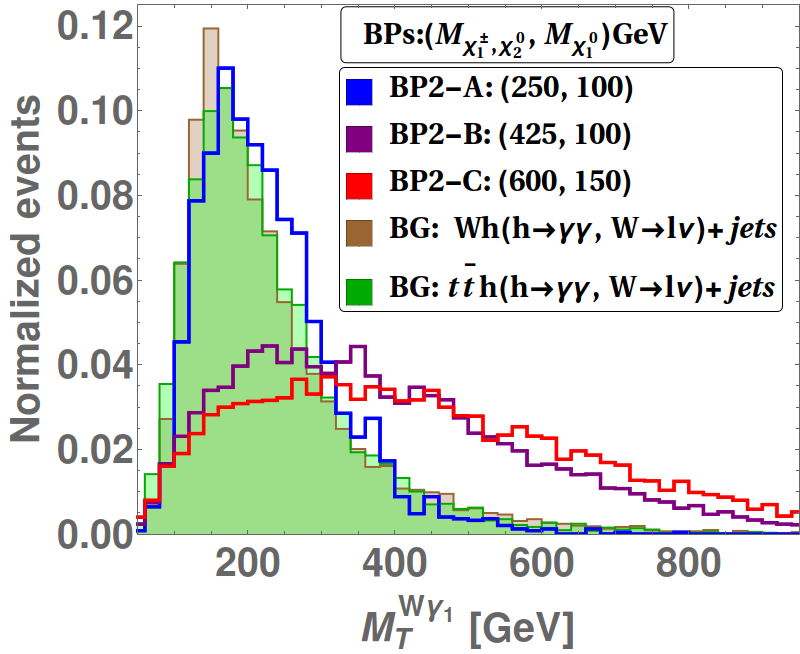}}}
 \caption{\it \footnotesize Normalized distribution of $M_{j_1 j_2 j_3}$ (top left), $H_{T}$ (top right), $\Delta \phi_{Wh}$ (bottom left) and $M_T^{W\gamma_1}$ (bottom right) are shown for BP2-A (blue solid line), BP2-B (purple solid line), BP2-C (red solid line). The corresponding distributions for $t\bar{t}h+jets$ and $Wh+jets$ are also shown in brown and green colored regions.}
 \label{fig:DistWHaa}
 \end{figure}


\begin{table}[htbp!]
\begin{center}\scalebox{0.65}{
\begin{tabular}{||C{1.25cm} C{1.25cm}||C{3.0cm}|C{2.0cm}|C{2cm}|C{2cm}|C{2cm}|C{2cm}|} 
\hline
\hline
\multicolumn{8}{|c|}{SR2-A} \\ \cline{1-8}
\multicolumn{3}{|c|}{Cut variables} &~$M_{\gamma\gamma}$ &~$M_{j_1j_2j_3}$ &~$H_T$ &~$\Delta \phi_{Wh}$~&$M_T^{W\gamma_1}$ \\
\multicolumn{3}{|c|}{•} &(GeV)&(GeV)&(GeV)& & (GeV)\\
\hline 
\multicolumn{3}{|c|}{Selection cuts for SR2-A} & 122-128 & $>$ 280 & $>$ 170 & $>$ 1.4 & $>$ 170 \\ \hline \hline
\multirow{6}{*}{\rotatebox[origin=c]{90}{\centering Signal and}} & \multirow{6}{*}{\rotatebox[origin=c]{90}{\centering background yields}} & Signal (BP2-A) & $115$ & $ 33$ & $ 31$ & $ 27 $ & $ 22 $ \\ \cline{3-8}
& & $t\bar{t}h+jets$ &$23$ & $ 9.0 $ & $8.6$ & $ 5.7$ & $ 4.8$ \\
& & $Wh+jets$  & $19$ & $ 5.5$ & $4.7$ & $2.9$ & $ 2.4$  \\
& & $Zh+jets$  & $4.5$ & $1.1$ & $1.0$ & $0.6$ & $0.4$ \\
& & $W+jets$  &$3.7$ & $ 0.8$ & $ 0.6$ & $0.4$ & $0.1$ \\
& & $Z+jets $ &$1.2$ & $ 0.3$ & $ 0.1$ & 0.1 & $ 0.03 $ \\ \hline 
\multicolumn{4}{|c|}{Total background yield: 7.6} & \multicolumn{2}{c|}{Total signal yield: $22$} & \multicolumn{2}{c|}{Signal significance: $8.0$} \\
\hline
\hline
\multicolumn{8}{|c|}{SR2-B} \\ \cline{1-8}
\multicolumn{3}{|c|}{Selection cuts for SR2-B} & 122-128 & $>$ 300 & $>$ 120 & $>$ 1.6 & $>$ 220 \\ \hline\hline
\multirow{6}{*}{\rotatebox[origin=c]{90}{\centering Signal and}} & \multirow{6}{*}{\rotatebox[origin=c]{90}{\centering background yields}} & Signal (BP2-B) & $45$ & $28$ & 28 & $ 21 $ & $19$ \\ \cline{3-8}
& & $t\bar{t}h+jets$ &$23$ & $ 8.2 $ & $8.1$ & $ 5.0$ & $ 3.4$ \\
& & $Wh+jets$  & $19$ & $ 4.9$ & $4.8$ & $2.7$ & $ 1.8$  \\
& & $Zh+jets$  & $6.6$ & $1.1$ & $1.0$ & $0.6$ & $0.3$ \\
& & $W+jets$  &$3.7$ & $ 0.6$ & $ 0.5$ & $0.3$ & $0$ \\
& & $Z+jets $ &$1.2$ & $ 0.19$ & 0.19 & $0.17$ & $0$ \\ \hline 
\multicolumn{4}{|c|}{Total background yield: 5.5} & \multicolumn{2}{c|}{Total signal yield: $20$} & \multicolumn{2}{c|}{Signal significance: $8.5$} \\
\hline
\hline
\multicolumn{8}{|c|}{SR2-C} \\ \cline{1-8}
\multicolumn{3}{|c|}{Selection cuts for SR2-C} & 122-128 & $>$ 330 & $>$ 100 & $>$ 1.5 & $>$ 220 \\ \hline
\multirow{6}{*}{\rotatebox[origin=c]{90}{\centering Signal and}} & \multirow{6}{*}{\rotatebox[origin=c]{90}{\centering background yields}} & Signal (BP2-C) & $10$ & $ 7.8$ & 7.8 & $ 5.5 $ & $ 5.3 $ \\ \cline{3-8}
& & $t\bar{t}h+jets$ & $23$ & $ 7.0 $ & 7.0 & $ 4.4$ & $ 2.7$ \\
& & $Wh+jets$  & $19$ & $ 4.2$ & $4.1$ & $2.5$ & $ 1.5$  \\
& & $Zh+jets$  & $4.5$ & $0.84$ & 0.84 & $0.50$ & $0.22$ \\
& & $W+jets$  &$3.7$ & $ 0.46$ & 0.46 & $0.18$ & $0$ \\
& & $Z+jets $ &$1.2$ & $ 0.08$ & 0.08 & 0.08 & $0$ \\ \hline 
\multicolumn{4}{|c|}{Total background yield: 4.5} & \multicolumn{2}{c|}{Total signal yield: $5.3$} & \multicolumn{2}{c|}{Signal significance: $2.5$} \\
\hline
\hline
\end{tabular}}
\end{center}
\caption{\it \footnotesize The selection cuts corresponding to the signal regions: SR2-A, SR2-B and SR2-C designed to maximize the signal significance of BP2-A, BP2-B and BP2-C, respectively, for searches in the $Wh$ mediated $ 1l+2\gamma + jets +\met$ channel, are listed. The cut flow table for BP2-A, BP2-B, BP2-C and other relevant backgrounds are also tabulated. The signal significance values have also been listed.}
\label{table:2_cut_flow_1a}
\end{table}

The most dominant contribution to the background comes from the $t\bar{t}h +jets $ and $Wh+jets$ processes. Subdominant contribution to the background arises from $Zh+jets$ and $W/Z + jets$\footnote{The $W/Z+jets$ process contributes to the background of the $1l+2\gamma+jets+\met$ signal when the $W/Z$ decays leptonically and $jets$ get faked as photons. In the present analysis, we have assumed a $jet \to \gamma$ fake rate of $0.05\%$~\cite{ATL-PHYS-PUB-2017-001}.}.

The signal events have been generated by varying $M_{\lsptwo} (=M_{\chonepm})$ in between $200~{\rm GeV}$ to $1000~{\rm GeV}$ with a step size of $25~{\rm GeV}$, while $M_{\lspone}$ has been varied between $25~{\rm GeV}$ to $M_{\lsptwo,\chonepm} - 125~{\rm GeV}$, with a step size of $25~{\rm GeV}$. We choose three representative benchmark points, BP2-A: $M_{\lsptwo} =  250~{\rm GeV}$, $M_{\lspone} = 100~{\rm GeV}$ (small $\Delta M$), BP2-B: $M_{\lsptwo} = 425~{\rm GeV}$, $M_{\lspone} = 100~{\rm GeV}$ (intermediate $\Delta M$), and BP2-C: $M_{\lsptwo} = 600~{\rm GeV}$, $M_{\lspone} = 150~{\rm GeV}$ (large $\Delta M$), and perform a cut-based collider analysis. The cross section values for the background processes and the signal benchmark points have been listed in Appendix~\ref{Appendix:cs}.


The kinematic variables used to perform the cut-based analysis are : 
$M_{\gamma\gamma}$ (invariant mass of the diphoton pair), $M_{j_1 j_2 j_3}$ (invariant mass of the three leading $jets$), $H_T$ (scalar sum of transverse momenta of the three leading $jets$), $\Delta \phi_{Wh}$ (difference in the azimuthal angle of the lepton-$\met$ system and the $\gamma\gamma$ system (originated from $h$)) and the transverse mass of the $W\gamma_i$ system, $M_T^{W\gamma_i} (i=1,2)$.
Here $M_T^{W\gamma_i} (i=1,2)$ is defined as $M_T^{W\gamma_i} = \sqrt{(M_T^{W})^2 + 2 E_T^W E_T^{\gamma_i} -2 \vec{p}_T^{\,\,W} \vec{p}_T^{\,\, \gamma_i}} $. $M_T^{W}$, $E_T^W$ (=$\sqrt{M_W^2+|\vec{p}_T^{\,\,W}|^2}$) and $\vec{p}_T^{\,\,W}$ are the transverse mass, energy and momentum of the leptonically decaying $W$ boson, respectively. $E_T^{\gamma_i}$ and $\vec{p}_T^{\,\, \gamma_i}$ are the transverse energy and momentum of the $i^{\rm th}$ photon. 

The normalized distributions of $M_{j_1 j_2 j_3}$, $H_{T}$, $\Delta \phi_{Wh}$ and $M_T^{W\gamma_1}$ for BP2-A (blue solid line), BP2-B (purple solid line) and BP2-C (red solid line), are represented in Figure~\ref{fig:DistWHaa} (a), (b), (c) and (d), respectively. The normalized distributions of the dominant background processes: $t\bar{t}h+jets$ and $Wh+jets$, are also shown as brown and green colored regions, respectively.
Considering the high reconstruction and identification capability of the photons at the LHC, we restrict $M_{\gamma\gamma}$ to a narrow range around $m_{h}$: 122~GeV~$\leq M_{\gamma\gamma} \leq$~128~GeV. The cut on $M_{\gamma\gamma}$ effectively diminishes the contribution from the $W/Z + jets$ background. The other kinematic variables: $M_{j_1j_2j_3}$, $H_{T}$ and $M_T^{W}$, help in suppressing the contributions from the other background processes, all of which contain a $h$. The $M_{j_1j_2j_3}$ distribution for the most dominant $t\bar{t}h$ and $Wh$ background peaks roughly at $200~{\rm GeV}$ and $75~{\rm GeV}$, respectively, while the $M_{j_1j_2j_3}$ peak for the signals heavily depend upon the mass difference between the NLSPs and the $\lspone$ and $m_{\lspone}$. For BP2-A, where both $\Delta M$ and $M_{\lspone}$ are relatively smaller~(150~{\rm GeV} and 100~{\rm GeV}, respectively), the $M_{j_1j_2j_3}$ peak mostly overlaps with the $t\bar{t}h$ background. As $\Delta M$ and/or $M_{\lspone}$ are increased, as in the case of BP2-B and BP2-C, the peak of the $M_{j_1j_2j_3}$ distribution shifts towards higher values and becomes relatively flatter. The $H_{T}$ variable also displays a similar behavior. The variations in $H_{T}$ upon changing $\Delta M$ or $M_{\lspone}$ are more pronounced as can be seen from Figure~\ref{fig:DistWHaa} (b). The variable: $M_T^{W\gamma_{i}}~(i=1,2)$, is another important discriminator between the backgrounds and the signal benchmarks with a relatively larger $\Delta M$. In the signal, the mass difference between the NLPs and $\lspone$ determines the boost carried by the $W$ boson and the Higgs boson. Consequently, the transverse mass of the $W$ system and the photons~(produced from $h$) increases with an increase in $\Delta M$. For the $t\bar{t}h$ and $Wh$ background, $M_T^{W\gamma_1}$ peaks at $\sim 165$ and $ \sim 150~{\rm GeV}$, respectively. The normalized distribution of BP2-A also peaks roughly at 165~GeV, while BP2-B and BP2-C peak at $\sim 250$ and $300~{\rm GeV}$, respectively. The distributions for the latter two also become flatter compared to the distributions for BP2-A and the backgrounds. We also use $\Delta \phi_{Wh}$ while performing the cut-based analysis. We impose a lower limit on $\Delta \phi_{Wh}$ for the three signal benchmarks, details of which can be found in Table~\ref{table:2_cut_flow_1a}.


\begin{figure}[htpb!]
\centering
\includegraphics[scale=0.5]{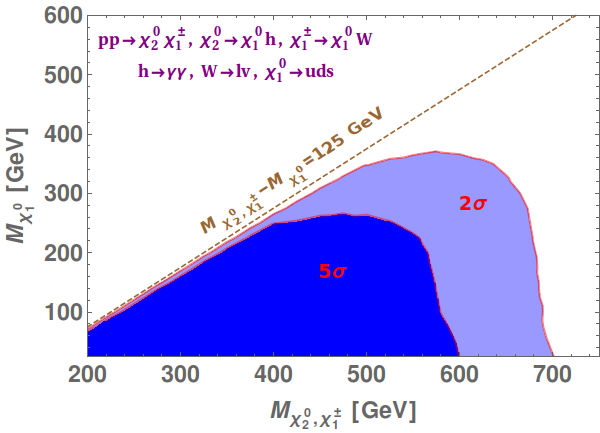}
\caption{\it \footnotesize The projected exclusion (light blue) and discovery (dark blue) regions in the mass plane of $M_{\lspone}$ vs. $M_{\lsptwo,\chonepm}$ in the $Wh$ simplified model with mass degenerate wino like $\chonepm,\lsptwo$ and binolike $\lspone$, and with one RPV term ($\lambda_{112}^{\prime\prime} u^{c}d^{c}s^{c}$). The brown line denotes the line $M_{\lsptwo,\chonepm}-M_{\lspone}=125$ GeV.}
\label{fig:2_exclusion_contour}
\end{figure}

The list of selection cuts for these three signal regions along with the cut flow for the three benchmark points have also been tabulated in Table~\ref{table:2_cut_flow_1a}. The total background yield corresponding to the three signal regions, the corresponding signal yields for BP2-A, BP2-B and BP2-C and the signal significances obtained from the cut-based analysis have also been tabulated in Table~\ref{table:2_cut_flow_1a}. In the current case, SR2-A, SR2-B, SR2-C yield a signal significance of $8.0, 8.5, 2.5$ for BP2-A, BP2-B and BP2-C respectively. It is also worthwhile to note the excellent $S/B$ values for the optimixed signal regions. Here, SR2-A, SR2-B and SR2-C results in an exceptional $S/B$ value of $2.89$ (for BP2-A), $3.64$ (for BP2-B) and $1.18$ (for BP2-C), respectively.

The projected $2 \sigma$-exclusion (light blue) and discovery regions (dark blue) derived from direct wino searches in the $Wh$ mediated $1l+2\gamma+jets+\met$ final state at the HL-LHC, have been shown in the $M_{\lspone}-M_{\lsptwo,\chonepm}$ plane, in Figure~\ref{fig:2_exclusion_contour}. The brown dashed line corresponds to the on-shellness condition of $h$ : $M_{\lsptwo,\chonepm} -M_{\lspone} = 125~{\rm GeV}$. It can be observed from Figure~\ref{fig:2_exclusion_contour} that in the presence of $\lambda_{112}^{\prime\prime}u^{c}d^{c}s^{c}$-type of RPV operator, direct wino searches at the HL-LHC are projected to exclude winos up to $\sim 700~{\rm GeV}$ at $2 \sigma$ and the projected wino discovery reach is up to $\sim 600~{\rm GeV}$ for a massless $\lspone$. 

\subsection{Searches in the $WZ$ mediated $3l$~+~$jets (N_{j} \geq 2)$~+~$\met$ channel}
\label{Sec:6j3lmet}

A study by the ATLAS collaboration, which probed the future reach of directly produced winos in the $WZ$ mediated $3l+\met$ final state at the HL-LHC~\cite{ATL-PHYS-PUB-2018-048}, shows a projected exclusion reach up to $M_{\chonepm,\lsptwo} \sim 1150~{\rm GeV}$ for a binolike $M_{\lspone} \sim 100~{\rm GeV}$ at $95\%$ C.L. The future reach of direct wino searches at HL-LHC in the $Wh$ mediated $3l+\met$ channel has also been studied by ATLAS in \cite{ATL-PHYS-PUB-2014-010}. The projected reach excludes winos up to $\sim 650~{\rm GeV}$ for $M_{\lspone} \sim 0~{\rm GeV}$ at $95\%$ C.L. One of the main reasons behind the weaker reach of $Wh$ mediated $3l+\met$ channel compared to the $WZ$ mediated process is the longer cascade decay chain in the former case. This results in a smaller event yield even if one assumes a similar signal region efficiency. In the current subsection, we focus only on the analogous $WZ$ mediated final state reinterpreted in $\lambda_{112}^{\prime\prime}$-type RPV simplified scenario. 

\begin{figure}[htpb!]{\centering
 \subfigure[]{
 \includegraphics[height=2.3in,width=3.2in]{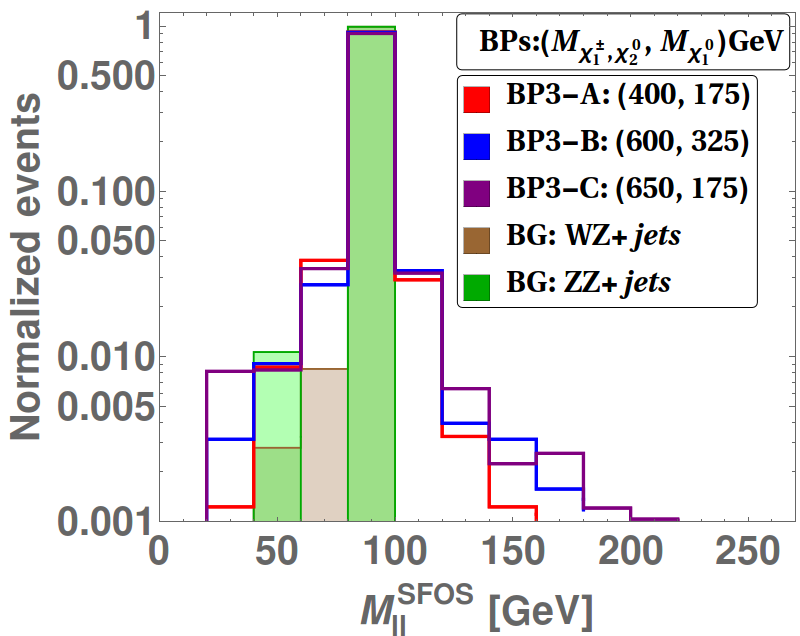}}}
  \subfigure[]{
 \includegraphics[height=2.27in,width=3.2in]{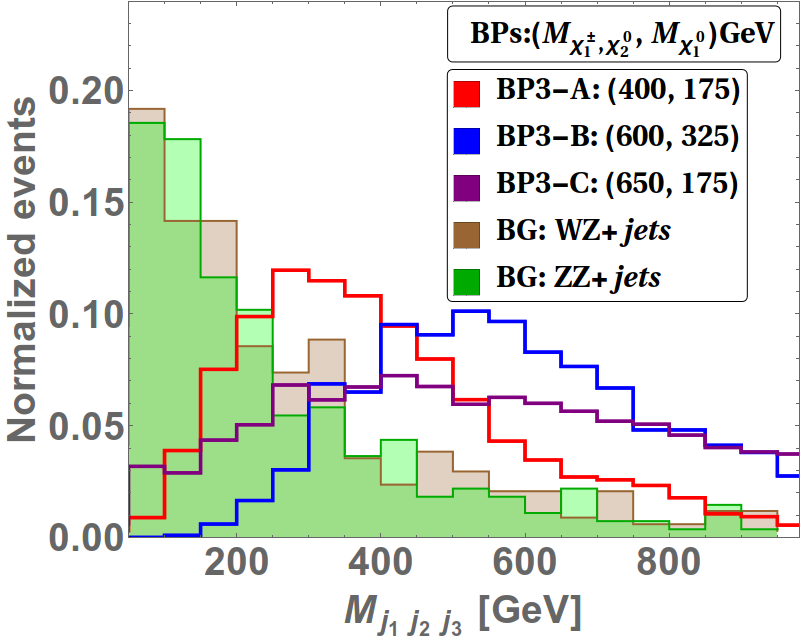}}\\
   \subfigure[]{
 \includegraphics[height=2.3in,width=3.2in]{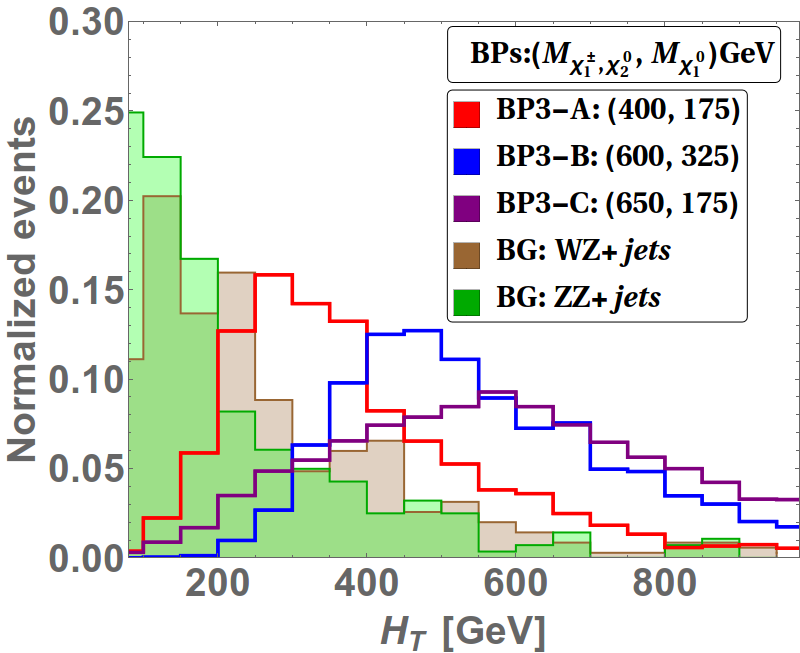}} \\
 \caption{\it \footnotesize Normalized distributions of $M_{ll}^{SFOS}$~(a), $M_{j_1 j_2 j_3}$~(b) and $H_{T}$~(c), for the signal benchmark points, BP3-A~(red solid line), BP3-B~(blue solid line) and BP3-C~(purple solid line), corresponding to the cascade decay process: $ p p \rightarrow \chonepm \lsptwo \to$ $(\chonepm \rightarrow (W^\pm \rightarrow l \nu) (\lspone \rightarrow u d s))$ $(\lsptwo \to (Z \rightarrow ll)$ $(\lspone \rightarrow u d s)) \to 3 l +jets+ \met$, are shown. The brown and green colored distributions represent the most important background processes: $WZ+jets$ and $ZZ+jets$, respectively.}
 \label{fig:DistWZ}
 \end{figure}

In the presence of a $\lambda_{112}^{\prime\prime}$-type RPV coupling, the $\lspone$ would decay into $\lspone \to uds$ resulting in $WZ$ mediated $3l+jets+\met$ final state and the projected exclusions are expected to alter. In the present subsection, we explore this facet and study the projected future reach of $WZ$ mediated $3l+jets+\met$ final state at the HL-LHC within the framework of a simplified $\lambda_{112}^{\prime\prime}$-type RPV scenario. The Feynman diagram of the signal process under consideration is illustrated in Figure~\ref{fig:proc} (c). The decay chain proceed as follows: $pp \to \chonepm\lsptwo$ $\to$ $\left(\chonepm \to W \lspone\right)$ $\left(\lsptwo \to Z \lspone\right)$ $ \to \left(W \to l \nu\right)$ $\left(\lspone \to uds \right)$ $\left(Z \to ll\right)$ $\left(\lspone \to u d s\right)$. The SM value for $Br(Z \to ll)$ ($\sim 6.72\%$~\cite{Tanabashi:2018oca}) has been assumed here. 

An event is required to have exactly three isolated leptons with $p_{T} > 30~{\rm GeV}$ and at least two light jets with $p_{T} > 20~{\rm GeV}$ in the final state. Among the three final state leptons, two are required to form a same flavor opposite charge (SFOS) lepton pair with invariant mass in the range of $|M_{Z} \pm 25~{\rm GeV}|$. In presence of two different SFOS lepton pairs with invariant mass within $|M_{Z}\pm 25~{\rm GeV}|$, the SFOS pair with invariant mass closest to the $Z$ boson mass is considered to be the correct SFOS pair and their invariant mass is represented as $M_{ll}^{SFOS}$. The lepton isolation criteria discussed in Section~\ref{Sec:wh_6jbblnu} is applied here as well. 

The important sources of background are  $WZ+jets$, $ZZ+jets$ and $VVV+jets$ ($V=W,Z$). Potential contribution to background can also arise from $Wh+jets$ and $Zh+jets$ processes, however, their contribution is much lesser when compared to the diboson and triboson backgrounds. Consequently, we ignore the contribution from both, $Wh+jets$ and $Zh+jets$. 

The signal events have been generated for various combinations of  $M_{\lsptwo}~(=M_{\chonepm})$ and $M_{\lspone}$. $M_{\lsptwo,\chonepm}$ has been varied in between $200~{\rm GeV}$ and $1000~{\rm GeV}$ with a step size of $25~{\rm GeV}$ while $M_{\lspone}$ has been varied from $25~{\rm GeV}$ to $M_{\lsptwo,\chonepm} - M_{Z}$ with a step size of $10~{\rm GeV}$. Three representative signal benchmark points with small $\Delta M$ (BP3-A: $M_{\lsptwo}=M_{\chonepm}=400~{\rm GeV}$, $M_{\lspone}=175~{\rm GeV}$), intermediate $\Delta M$ (BP3-B: $M_{\lsptwo}=M_{\chonepm}=600~{\rm GeV}$, $M_{\lspone}=325~{\rm GeV}$) and large $\Delta M$ (BP3-C: $M_{\lsptwo}=M_{\chonepm}=650~{\rm GeV}$, $M_{\lspone}=175~{\rm GeV}$) are chosen. Three optimized signal regions are chosen: SR3-A, SR3-B and SR3-C, with optimized selection cuts which maximize the signal significances of BP3-A, BP3-B and BP3-C, respectively. The Appendix lists the cross section of the background and signal benchmark points.

\begin{table}[htpb!]
\begin{center}\scalebox{0.65}{
\begin{tabular}{||C{2.5cm}||C{4.5cm}|C{2.75cm}|C{2.75cm}|C{2.75cm}|C{2.75cm}|} 
\hline
\hline
\multicolumn{6}{|c|}{SR3-A} \\ \cline{1-6}
\multicolumn{2}{|c|}{Selection cuts for SR3-A} & $M_{ll}^{SFOS}$ & $M_{j_1j_2j_3}$ $>$ 140 GeV & $H_T$ $>$ 220 GeV & - \\ \hline
\multirow{6}{*}{\rotatebox[origin=c]{90}{\centering S and B values}} & \multirow{2}{*}{Signal (BP3-A)} & \multicolumn{4}{c|}{Cut flow of BP3-A: $M_{\lsptwo,\chtwopm}=400~{\rm GeV},~M_{\lspone}=175~{\rm GeV}$} \\ \cline{3-6}
 & & $839$ & $ 808$ & $723$ & - \\ \cline{2-6}
 & \multicolumn{5}{c|}{Cut flow of backgrounds} \\\cline{2-6}
& $WZ+jets$ & $3.64\times 10^{4}$ & $2.47\times 10^{4} $ & $1.57\times 10^{4}$ & - \\
& $ZZ+jets$ & $4146$ & $ 2408$ & $1219$ & - \\
& $VVV+jets$ $(V=W,Z)$  & $3069$ & $2130$ & $1568$ & - \\ \hline 
\multicolumn{2}{|c|}{Total background yield: $1.85\times 10^{4}$} & \multicolumn{2}{c|}{Total signal yield: $723$} & \multicolumn{2}{c|}{Signal significance: $5.3$} \\
\hline
\hline
\multicolumn{6}{|c|}{SR3-B} \\ \cline{1-6}
\multicolumn{2}{|c|}{Selection cuts for SR3-B} & $M_{ll}^{SFOS}$  & $M_{j_1j_2j_3}$ $>$ 110 GeV & $H_T$ $>$ 280 GeV  & -\\ \hline
\multirow{6}{*}{\rotatebox[origin=c]{90}{\centering S and B values}} & \multirow{2}{*}{Signal (BP3-B)} & \multicolumn{4}{c|}{Cut flow of BP3-B: $M_{\lsptwo,\chtwopm}=600~{\rm GeV},~M_{\lspone}=325~{\rm GeV}$} \\ \cline{3-6}
 & & $151$ & 151 & $148$ & - \\ \cline{2-6}
 & \multicolumn{5}{c|}{Cut flow of backgrounds} \\\cline{2-6}
& $WZ+jets$ & $3.64\times 10^{4}$ & $2.74\times 10^{4} $ & $1.20\times 10^{4}$ & - \\
& $ZZ+jets$ & $4146$ & $ 2853$ & $981$ & - \\
& $VVV+jets$ $(V=W,Z)$  & $3069$ & $2359$ & $1320$ & - \\ \hline 
\multicolumn{2}{|c|}{Total background yield: $1.43 \times 10^{4}$} & \multicolumn{2}{c|}{Total signal yield: $148$} & \multicolumn{2}{c|}{Signal significance: $1.2$} \\
\hline
\hline
\multicolumn{6}{|c|}{SR3-C} \\ \cline{1-6}
\multicolumn{2}{|c|}{Selection cuts for SR3-C} & $M_{ll}^{SFOS}$  & $\met$ $>$ 150 GeV & $M_{j_1j_2j_3}$ $>$ 160 GeV & $H_T$ $>$ 160 GeV\\ \hline
\multirow{6}{*}{\rotatebox[origin=c]{90}{\centering S and B values}} & \multirow{2}{*}{Signal (BP3-C)} & \multicolumn{4}{c|}{Cut flow of BP3-C: $M_{\lsptwo,\chtwopm}=650~{\rm GeV},~M_{\lspone}=75~{\rm GeV}$} \\ \cline{3-6}
 & & $235$ & $122$ & $115$ & $114$ \\ \cline{2-6}
 & \multicolumn{5}{c|}{Cut flow of backgrounds} \\\cline{2-6}
& $WZ+jets$ &$3.6 \times 10^{4}$ & $3708$ & $2575$ & $2575$ \\
& $ZZ+jets$ & $4146$ & $45$ & $ 30 $ & $15$  \\
& $VVV+jets$ $(V=W,Z)$  & $3069$ & $542$ & $384$ & $359$ \\ \hline 
\multicolumn{2}{|c|}{Total background yield: $2950$} & \multicolumn{2}{c|}{Total signal yield: $114$} & \multicolumn{2}{c|}{Signal significance: $2.1$} \\
\hline
\hline
\end{tabular}}
\end{center}
\caption{\it \footnotesize The selection cuts on $M_{ll}^{SFOS}$, $\met$, $M_{j_{1}j_{2}j_{3}}$ and $H_{T}$ are listed for the three optimized signal regions: SR3-A, SR3-B and SR3-C, optimized to maximize the signal significance in the $WZ$ mediated $3l+jets+\met$ final state. The cut flow table showing the signal and background yields upon the successive application of selection cuts is also presented. The maximal value of signal significance obtained from the cut-based optimization procedure is also shown.}
\label{table:3_cut_flow_1}
\end{table}

The kinematic variables used to perform the cut-based analysis are: invariant mass of the SFOS pair of leptons ($M_{ll}^{SFOS}$), invariant mass of the three leading $p_{T}$ ordered jets ($M_{j_1 j_2 j_3}$), the scalar sum of the transverse momenta of the three leading $p_{T}$ ordered $jets$ ($H_{T}$) and the missing transverse energy ($\met$).

$M_{j_1 j_2 j_3}$ and $H_{T}$ are observed to be among the most efficient variables in discriminating the signal from the backgrounds. Similar to the observation in Section~\ref{Sec:wh_6jgagalnu}, we observe that the peak of $M_{j_1j_2j_3}$ and $H_{T}$ shift towards higher values and becomes flatter with increasing $\Delta M$ and $M_{\lspone}$. $M_{j_1j_2j_3}$ and $H_{T}$ peak at $\lesssim 100~{\rm GeV}$ for both the $WZ+jets$ and the $ZZ+jets$ background, while all the signal benchmarks peak at values above $\sim 250~{\rm GeV}$. The normalized distributions of $M_{ll}^{SFOS}$, $M_{j_1 j_2 j_3}$ and $H_{T}$ for the signal benchmark points: BP3-A (red solid line), BP3-B (blue solid line) and BP3-C (purple solid line), and the most important background processes: $WZ+jets$ (brown colored region) and $ZZ+jets$ (green colored region), have been illustrated in Figure~\ref{fig:DistWZ} (a), \ref{fig:DistWZ} (b) and \ref{fig:DistWZ} (c), respectively. The selection cuts for the respective signal regions are shown in Table~\ref{table:3_cut_flow_1}. The signal and background yields obtained upon the successive application of the selection cuts have also been listed in Table~\ref{table:3_cut_flow_1} along with the respective values of signal significance.

\begin{figure}[htpb!]
\centering
\includegraphics[scale=0.5]{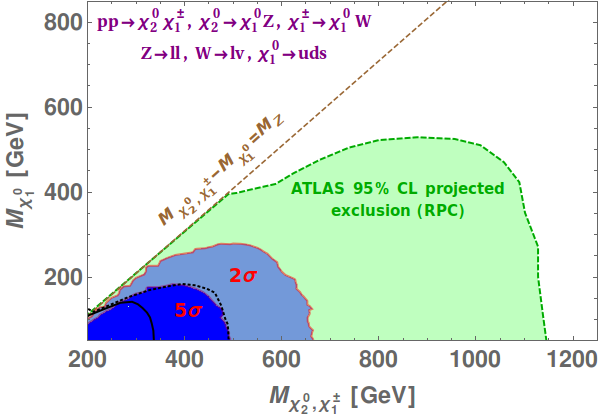}
\caption{\it \footnotesize The projected discovery reach (dark blue) and the projected exclusion reach (light blue) from direct wino searches in the $pp \to \chonepm \lsptwo \to 3l+jets+\met$ final state at the HL-LHC~($\sqrt{s} = 14~{\rm TeV}$, $\mathcal{L} \sim 3~{\rm ab^{-1}}$) is shown in the $M_{\lsptwo,\chonepm}-M_{\lspone}$ plane. The solid black and the dashed black line represents the observed limits (at $95\%$ C.L.) derived by ATLAS ($\sqrt{s}=13~{\rm TeV}$, $\mathcal{L}\sim 139~{\rm fb^{-1}}$)~\cite{Aad:2019vvi} and CMS ($\sqrt{s}=13~{\rm TeV}$, $\mathcal{L}\sim 36~{\rm fb^{-1}}$)~\cite{ATL-PHYS-PUB-2018-048}, respectively, from direct wino searches in the $WZ$ mediated $3l+\met$ channel within a simplified RPC framework. The light green colored region represents the $95\%$ C.L. projected exclusion region derived by ATLAS~\cite{ATL-PHYS-PUB-2018-048} from direct wino searches in the $3l+\met$ final state at the HL-LHC within a simplified RPC framework. The brown dashed line represents the mass correlation: $M_{\chonepm,\lsptwo}-M_{\lspone}=M_{Z}$.}
\label{fig:3_excl}
\end{figure}

The future reach of direct wino searches in the $WZ$ mediated $3l+jets+\met$ final state at HL-LHC is studied. In this context, we evaluate the projected exclusion ($> 2\sigma$) and projected discovery ($>5\sigma$) contours assuming zero systematic uncertainty in the $M_{\lsptwo,\chonepm}$-$M_{\lspone}$ plane (shown in Figure~\ref{fig:3_excl}). The light blue colored region and the dark blue colored regions in Figure~\ref{fig:3_excl} represent the projected exclusion and discovery reach, respectively. The brown dashed line represents the mass correlation: $M_{\chonepm,\lsptwo} - M_{\lspone} = M_{Z}$. The projected exclusion region has a reach up to $M_{\lsptwo,\chonepm} \sim 650~{\rm GeV}$ (wino like) for a binolike $M_{\lspone} \sim 100~{\rm GeV}$, while the projected discovery region has a potential reach up to $M_{\lsptwo,\chonepm} \sim 480~{\rm GeV}$ (wino like) for a binolike $M_{\lspone} \sim 100~{\rm GeV}$. It is to be noted that within the simplified RPC scenario, the projected exclusion contour (at $95\%$ C.L.) of direct wino searches in the $WZ$ mediated $3l+\met$ channel at the HL-LHC reaches up to $\sim 1150~{\rm GeV}$ for a $\lspone$ with mass up to $\sim 100~{\rm GeV}$, as evaluated by ATLAS in \cite{ATL-PHYS-PUB-2018-048} (shown as green dashed line in Figure~\ref{fig:3_excl}). Thus, within the $\lambda_{112}^{\prime\prime} u^{c}d^{c}s^{c}$-type RPV simplified scenario, the projected reach of HL-LHC in direct wino searches in the $3l+jets+\met$ channel is rendered considerably weaker compared to the projected reach of the analogous search in the RPC scenario.

\subsection{Searches in the $WZ$ mediated $3l$~+~$2b$~+~$jets (N_{j} \geq 2)$~+~$\met$ channel}
\label{sec:3l2bjmet}

 In the present subsection, we evaluate the HL-LHC prospects of direct wino searches in the $WZ$ mediated $3l+2b+jets+\met$ final state, produced from the cascade decay of directly produced wino like mass degenerate $\lsptwo\chonepm$ pair. Owing to the presence of $\lambda_{113}^{\prime\prime}u^{c}d^{c}b^{c}$ RPV operator, $\lspone$ decays as: $\lspone \to udb$. Unlike the previous section, the relevant decay chain proceeds as follows: $pp \to \chonepm\lsptwo$ $\to$ $\left(W \lspone\right)$ $\left(Z \lspone\right)$ $ \to \left(l \nu udb \right)$ $\left( ll u d b\right)$, resulting in $WZ$ mediated $3l+2b+jets+\met$ final state. Although the channel containing $3l$ in the final state is substantially analyzed, $3l+2b+jets+\met$ final state is not commonly studied in RPC scenario.

\begin{figure}[h!]{\centering
  \subfigure[]{
  \includegraphics[height=2.3in,width=3.2in]{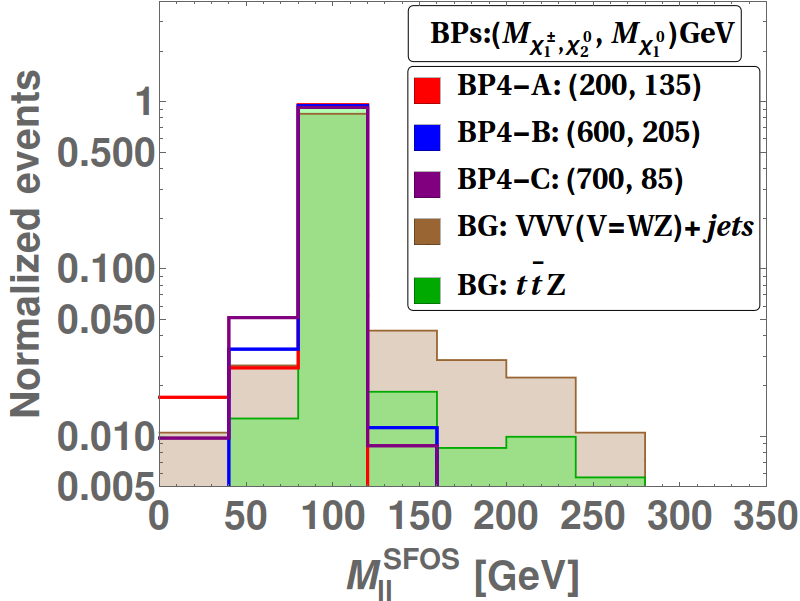}}
  \subfigure[]{
 \includegraphics[height=2.3in,width=3.2in]{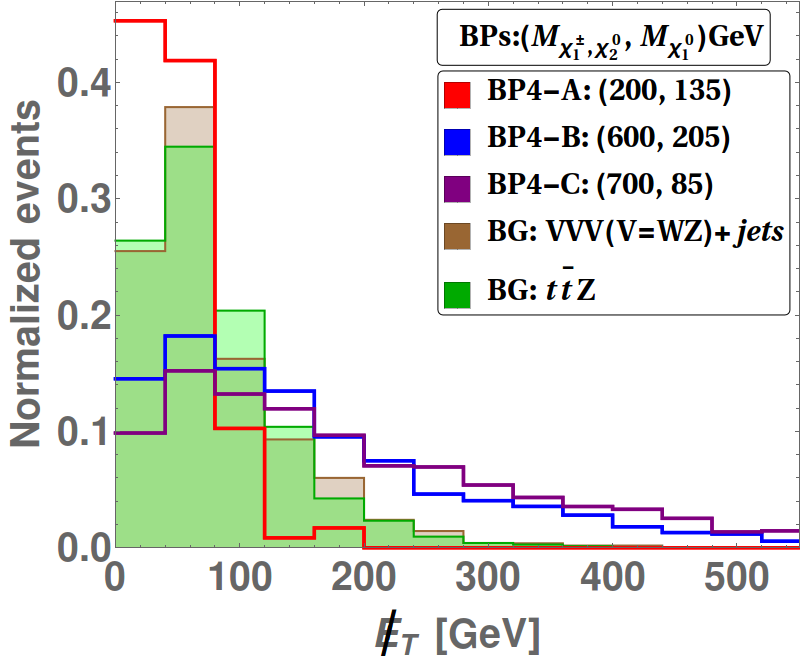}} \\
   \subfigure[]{
 \includegraphics[height=2.3in,width=3.2in]{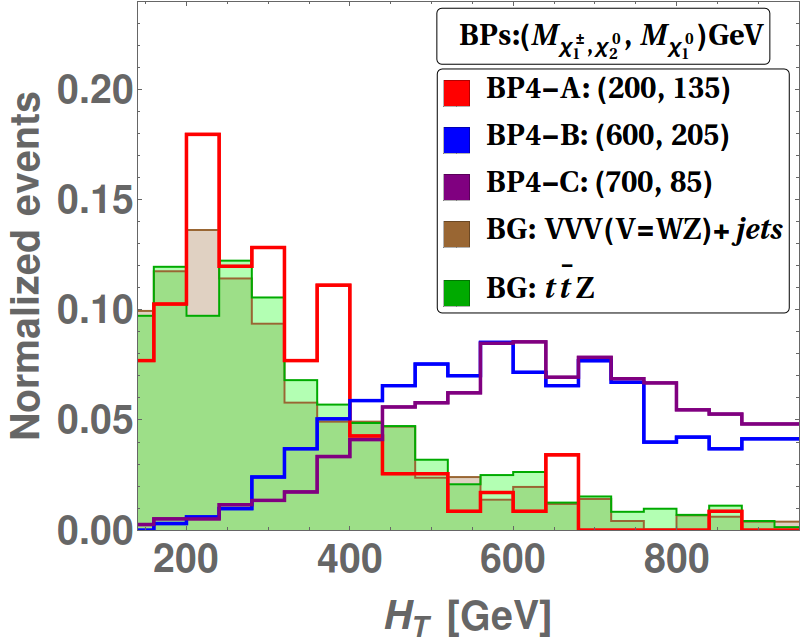}}} 
 \caption{\it \footnotesize Normalized distributions of $M_{ll}^{SFOS}$~(a), $\met$~(b) and $H_{T}$~(c) corresponding to the signal benchmark points, BP4-A~(red solid line), BP4-B~(blue solid line) and BP4-C~(purple solid line), in the $3l+2b+jets+\met$ final state is shown. The respective normalized distributions for the most significant background processes: $t\bar{t}Z$ and $VVV+jets$ are also displayed as brown and green colored regions, respectively.}
 \label{fig:4_DistWZ}
 \end{figure}

The event selection criteria requires the presence of three isolated leptons in the final state along with two $b$ $jets$ and at least two light $jets$. The final state leptons, $b$ $jets$ and light $jets$ are required to have $p_{T} > 30~{\rm GeV}$ and the pseudorapidity must lie within a range of $|\eta| \leq 2.5$. Here as well, we demand the presence of at least one SFOS pair out of the final state leptons with invariant mass $M_{ll}^{SFOS}$ in the range of $M_{Z}\pm 25~{\rm GeV}$. In the presence of two such SFOS pairs, the one with invariant mass closest to the $Z$ boson mass is chosen to be the correct one. The background to the $3l+2b+jets+\met$ final state is constituted by $t\bar{t}Z$, $VVV+jets$, $WZ+jets$ and $ZZ+jets$ processes.

\begin{table}[htpb!]
\begin{center}\scalebox{0.65}{
\begin{tabular}{||C{2.5cm}||C{4.5cm}|C{3.2cm}|C{3.2cm}|C{3.2cm}|} 
\hline
\hline
\multicolumn{5}{|c|}{SR4-A} \\ \cline{1-5}
\multicolumn{2}{|c|}{Selection cuts for SR4-A} & $M_{ll}^{SFOS}$ & $H_{T}$ $>$ 250 GeV & - \\ \hline
\multirow{7}{*}{\rotatebox[origin=c]{90}{\centering S and B values}} & \multirow{2}{*}{Signal (BP4-A)} & \multicolumn{3}{c|}{Cut flow of BP4-A: $M_{\lsptwo,\chtwopm}=250~{\rm GeV},~M_{\lspone}=135~{\rm GeV}$} \\ \cline{3-5}
 & & 186 & $ 128$ & -  \\ \cline{2-5}
 & \multicolumn{4}{c|}{Cut flow of backgrounds} \\ \cline{2-5}
& $t\bar{t}Z$ & 1303 & $1093$ & - \\
& $VVV+jets$ & 1571 & $ 1353$  & - \\
& $WZ+jets$  & 10.4 & $8.2$ & -\\ 
& $ZZ+jets$  & 9.1 & $5.3$ & - \\ \hline 
\multicolumn{2}{|c|}{Total background yield: 2459.5} & \multicolumn{1}{c|}{Total signal yield: 128} & \multicolumn{2}{c|}{Signal significance: 2.6} \\
\hline
\hline
\multicolumn{5}{|c|}{SR4-B} \\ \cline{1-5}
\multicolumn{2}{|c|}{Selection cuts for SR4-B} & $M_{ll}^{SFOS}$ & $\met$ $>$ 160 GeV & - \\ \hline
\multirow{7}{*}{\rotatebox[origin=c]{90}{\centering S and B values}} & \multirow{2}{*}{Signal (BP4-B)} & \multicolumn{3}{c|}{Cut flow of BP4-B: $M_{\lsptwo,\chtwopm}=600~{\rm GeV},~M_{\lspone}=205~{\rm GeV}$} \\ \cline{3-5}
 & & 88 & $34$ & -\\ \cline{2-5}
 & \multicolumn{4}{c|}{Cut flow of backgrounds} \\ \cline{2-5}
& $t\bar{t}Z$ & 1303 & 98 & - \\
& $VVV+jets$ & 1571 & 149 & - \\
& $WZ+jets$  & 10.4 & 3.0 & - \\ 
& $ZZ+jets$ & 9.1 & 0.0 & - \\ \hline 
\multicolumn{2}{|c|}{Total background yield: 250} & \multicolumn{1}{c|}{Total signal yield: 34} & \multicolumn{2}{c|}{Signal significance: 2.1} \\
\hline
\hline
\multicolumn{5}{|c|}{SR4-C} \\ \cline{1-5}
\multicolumn{2}{|c|}{Selection cuts for SR4-C} & $M_{ll}^{SFOS}$ & $\met$ $>$ 160 GeV & $H_{T}$ $>$ 240 GeV \\ \hline
\multirow{7}{*}{\rotatebox[origin=c]{90}{\centering S and B values}} & \multirow{2}{*}{Signal (BP4-C)} & \multicolumn{3}{c|}{Cut flow of BP4-C: $M_{\lsptwo,\chtwopm}=700~{\rm GeV},~M_{\lspone}=85~{\rm GeV}$} \\ \cline{3-5}
 & & 29 & 14.5 & 14.4  \\ \cline{2-5}
 & \multicolumn{4}{c|}{Cut flow of backgrounds} \\ \cline{2-5}
& $t\bar{t}Z$ & 1303 & 98 & 91 \\
& $VVV+jets$ & 1571 & 149 & 134  \\
& $WZ+jets$  & 10.4 & 3.0 & 2.6 \\ 
& $ZZ+jets$ & 9.1 & 0.0 & 0.0 \\ \hline
\multicolumn{2}{|c|}{Total background yield: 227.6} & \multicolumn{1}{c|}{Total signal yield: 14.3} & \multicolumn{2}{c|}{Signal significance: 0.9} \\
\hline
\hline
\end{tabular}}
\caption{\it \footnotesize Selection cuts on $M_{ll}^{SFOS}$, $\met$ and $H_{T}$ are shown along with the cut flow table for the signal and background processes relevant to direct wino searches (in presence of $\lambda_{113}^{\prime\prime} u^{c}d^{c}b^{c}$-type RPV operator) in the $WZ$ mediated $3l+2b+jets+\met$ final state. The signal significance values of BP4-A, BP4-B and BP4-C obtained upon the application of selection cuts corresponding to the signal regions: SR4-A, SR4-B and SR4-C, respectively, are also listed.}
\label{table:4_signal_regions}
\end{center}
\end{table}

Three representative benchmark points: BP4-A ($M_{\lsptwo,\chonepm}=250~{\rm GeV}$, $M_{\lspone}=135~{\rm GeV}$), BP4-B ($M_{\lsptwo,\chonepm}=600~{\rm GeV}$, $M_{\lspone}=205~{\rm GeV}$) and BP4-C ($M_{\lsptwo,\chonepm}=700~{\rm GeV}$, $M_{\lspone}=85~{\rm GeV}$) are chosen according to small, medium and large mass splittings between $\lsptwo (\chonepm)$ and $\lspone$. Cut-based analysis is performed by optimizing the selection cuts on $M_{ll}^{SFOS}$, $\met$ and $H_{T}$ to maximize the signal significance. 
The normalized distribution of $M_{ll}^{SFOS}$, $\met$ and $H_{T}$, both for signal and dominant  backgrounds have been illustrated in Figure~\ref{fig:4_DistWZ} (a), (b) and (c), respectively. The red, blue and purple solid lines represent the normalized distributions of BP4-A, BP4-B and BP4-C respectively, while the most dominant backgrounds: $t\bar{t}Z$ and $VVV+jets$ have been represented by green and brown colored regions respectively. 
Since BP4-A features lower mass difference between $\lsptwo (\chonepm)$ and $\lspone$ than BP4-B and BP4-C, the light $jets$ emanating from the decay of less boosted $\lspone$ in BP4-A carry relatively smaller $p_{T}$ as compared to the $jets$ produced from the decay $\lspone$ in the other two benchmark points. This in turn shifts the peak of the $H_{T}$ distribution towards higher values for BP4-B and BP4-C. 

\begin{figure}[htpb!]
\centering
\includegraphics[scale=0.5]{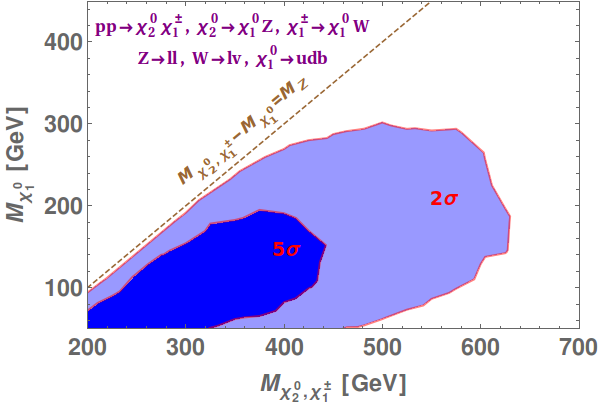}
\caption{\it \footnotesize The projected discovery (dark blue colored) and exclusion (light blue colored) regions for direct wino searches in the $3l+2b+jets+\met$ channel at the HL-LHC. The final state is an implication of $\lambda_{113}^{\prime\prime}u^{c}d^{c}b^{c}$-type RPV operator which implies $\lspone \to udb$. The brown line corresponds to $M_{\lsptwo,\chonepm}-M_{\lspone}=M_{Z}$. }
\label{fig:4b}
\end{figure}

The signal and background cross sections can be seen in Appendix~\ref{Appendix:cs}. The signal significances are optimized for three signal regions: SR4-A, SR4-B and SR4-C. The list of selection cuts on $M_{ll}^{SFOS}$, $\met$ and $H_{T}$ corresponding to the three signal regions have been itemized in Table~\ref{table:4_signal_regions}. The cut flow exhibiting the signal and background yields is also presented in Table~\ref{table:4_signal_regions} along with signal significance.

 We derive the projected exclusion and discovery contour in the context of HL-LHC, which have been illustrated in light blue and dark blue colors, respectively, in Figure~\ref{fig:4b}. The brown dashed line corresponds to the mass correlation: $M_{\chonepm,\lsptwo} - M_{\lspone} = M_{Z}$. The projected exclusion region reaches up to $M_{\chonepm,\lsptwo} \sim 600~{\rm GeV}$ for wino like $\chonepm,\lsptwo$ and binolike $\lspone$ with mass in the range $\sim [150-250]~{\rm GeV}$. 

Thus, we observe that a variety of interesting multiparticle final states can be produced from the cascade decay of direct wino production on account of the introduction of $R$-parity violating operators, many of which display a strong potential to be excluded and even discovered at the HL-LHC. In the present work, we explored the future prospects of two different types of RPV operator: $\lambda_{112}^{\prime\prime}u^{c}d^{c}s^{c}$ (in $Wh$ mediated $1l+2b+jets+\met$ channel, $Wh$ mediated $1l+2\gamma+jets+\met$ channel, $WZ$ mediated $3l+jets+\met$ channel) and $\lambda_{113}^{\prime\prime}u^{c}d^{c}b^{c}$ (in $WZ$ mediated $3l+2b+jets+\met$ channel) by performing a detailed cut-based analysis involving all relevant background processes. We intend to evaluate the implications from various other types of RPV operators on a multitude of search channels in an ongoing work. A more sophisticated analysis of the underlying final state $jets$ and better understanding of the multiparticle backgrounds might help in further improving the future discovery prospects in these channels. Before concluding this work, we briefly discuss the implications from pure-higgsino searches and also analyze the projected sensitivity for a few realistic MSSM benchmark points where the neutralinos and charginos are gaugino-higgsino admixtures.

\section{Benchmark scenarios}
\label{Sec:benchmark}

We begin our discussion in this section by considering two benchmark points with higgsinolike $\lsptwo,\lspthree$, $\chonepm$ and binolike $\lspone$: BP-$\alpha_{\tilde{H}}$ ($M_{\lspthree, \lsptwo,\chonepm} = 450~{\rm GeV}$, $M_{\lspone}=150~{\rm GeV}$) and BP-$\beta_{\tilde{H}}$ ($M_{\lspthree,\lsptwo,\chonepm} = 500~{\rm GeV}$, $M_{\lspone}=200~{\rm GeV}$) and contrast their projected detectability at the HL-LHC with their pure-wino counterparts: BP-$\alpha_{\tilde{W}}$ ($M_{\lsptwo,\chonepm} = 450~{\rm GeV}$, $M_{\lspone}=150~{\rm GeV}$) and BP-$\beta_{\tilde{W}}$ ($M_{\lsptwo,\chonepm} = 500~{\rm GeV}$, $M_{\lspone}=200~{\rm GeV}$), for the four signal channels considered in this work. BP-$\alpha_{\tilde{W}}$ and BP-$\beta_{\tilde{W}}$ fall within the projected exclusion reach of direct wino searches in the $Wh$ mediated $1l+2b+jets+\met$ final state (see Figure~\ref{fig:1_excl_contour}). However, direct higgsino searches in the same search channel results in a signal significance of $\sim 1.56$ and $\sim 0.83$ for their pure-higgsino counterparts BP-$\alpha_{\tilde{H}}$ and BP-$\beta_{\tilde{H}}$, respectively, thereby, putting both these benchmark points outside the projected exclusion region. BP-$\alpha_{\tilde{W}}$ also falls within the projected discovery reach of direct wino searches in the other three search channels (see Figure~\ref{fig:2_exclusion_contour}, \ref{fig:3_excl} and \ref{fig:4b}). However, direct higgsino searches (for BP-$\alpha_{\tilde{H}}$) result in a signal significance of $\sim 2.94$, $\sim 2.30$ and $\sim 1.76$ in $Wh$ mediated $1l+2\gamma+jets+\met$, $WZ$ mediated $3l+jets+\met$ and $WZ$ mediated $3l+2b+jets+\met$ search channels, respectively, and thus, BP-$\alpha_{\tilde{H}}$ falls within (outside) the projected exclusion (discovery) reach of the aforementioned former two channels and even outside the projected exclusion region of the later search channel. BP-$\beta_{\tilde{W}}$ also falls within the projected discovery reach of direct wino searches in the $Wh$ mediated $1l+2\gamma+jets+\met$ channel, and within the projected exclusion reach in the $WZ$ mediated $3l+jets+\met$ and $WZ$ mediated $3l+2b+jets+\met$. On the contrary, in direct higgsino searches, the signal significance of BP-$\beta_{\tilde{H}}$ marginally crosses $2\sigma$ in the $Wh$ mediated $1l+2\gamma+jets+\met$ channel, while registers a value of $\sim 1.66$ and $\sim 1.4$ in the $WZ$ mediated $3l+jets+\met$ and $WZ$ mediated $3l+2b+jets+\met$ channels, respectively. The direct higgsino searches, thus, imply weaker exclusion reach than the analogous wino counterparts, mainly, due to a smaller production cross section.

\begin{table}
\begin{center}\scalebox{0.72}{
\begin{tabular}{||C{2cm}|C{2cm}|C{2.3cm}|C{2.3cm}||C{1.1cm}|C{1cm}|C{1.1cm}||C{1.1cm}|C{1cm}|C{1.1cm}||C{1.1cm}|C{1cm}|C{1.1cm}||C{1.1cm}|C{1cm}|C{1.1cm}||} \hline \hline 
\multirow{4}{*}{$\tan\beta$} & & & & \multicolumn{12}{c||}{Signal significance} \\ \cline{5-16}
 & $\sigma(\chonepm\lsptwo)$ & $Br(\lsptwo \to Z \lspone)$ & $Br(\lsptwo \to h \lspone)$ & \multicolumn{3}{c|}{$Wh$ mediated} & \multicolumn{3}{c|}{$Wh$ mediated} & \multicolumn{3}{c|}{$WZ$ mediated} & \multicolumn{3}{c||}{$WZ$ mediated} \\
 & (fb) & ($\%$) & ($\%$) &  \multicolumn{3}{c|}{$1l+2b+jets+\met$} & \multicolumn{3}{c|}{$1l+2\gamma+jets+\met$} & \multicolumn{3}{c|}{$3l+jets+\met$} & \multicolumn{3}{c||}{$3l+2b+jets+\met$} \\ \cline{5-16} 
 & & & & SR1-A & SR1-B & SR1-C & SR2-A & SR2-B & SR2-C & SR3-A & SR3-B & SR3-C & SR4-A & SR4-B & SR4-C \\ \hline 
 5 & 22.65 & 1.65 & 98.34 & 0.49 & 0.73 & 0.92 & 1.57 & 1.60 & 2.15 & 0.14 & 0.14 & 0.13 & $0.001$ & $0.004$ & $0.004$ \\
 8 & 22.80 & 3.26 & 96.74 & 0.49 & 0.72 & 0.91 & 1.56 & 1.58 & 2.13 & 0.27 & 0.28 & 0.26 & 0.003 & 0.009 & 0.007 \\
 10 & 22.79 & 4.25 & 95.74 & 0.48 & 0.71 & 0.90 & 1.54 & 1.57 & 2.11 & 0.35 & 0.37 & 0.34 & 0.004 & 0.01 & 0.009 \\
 15 & 22.48 & 6.35 & 93.64 & 0.46 & 0.69 & 0.87 & 1.49 & 1.51 & 2.03 & 0.52 & 0.54 & 0.50 & 0.006 & 0.02 & 0.01 \\
 20 & 22.47 & 7.98 & 92.01 & 0.45 & 0.68 & 0.85 & 1.46 & 1.48 & 2.00 & 0.66 & 0.68 & 0.63 & 0.008 & 0.02 & 0.02 \\
 25 & 22.74 & 9.25 & 90.74 & 0.45 & 0.68 & 0.85 & 1.46 & 1.48 & 1.99 & 0.77 & 0.80 & 0.74 & 0.009 & 0.02 & 0.02  \\
 30 & 22.65 & 10.26 & 89.73 & 0.45 & 0.66 & 0.84 & 1.44 & 1.46 & 1.96 & 0.85 & 0.88 & 0.82 & 0.01 & 0.03 & 0.02 \\
 40 & 22.65 & 11.77 & 88.22 & 0.44 & 0.65 & 0.82 & 1.41 & 1.43 & 1.93 & 0.98 & 1.01 & 0.94 & 0.01 & 0.03 & 0.03  \\ \hline
\end{tabular}}
\caption{\it \footnotesize The $\chonepm\lsptwo$ production cross section and $Br(\lsptwo \to Z/h \lspone)$ are shown against the different $\tan\beta$ values for BP-$\beta_{\tilde{W}}$. The respective signal significance values in the $4$ analysis channels ($3$ signal regions in each channel) considered in this work are also listed.}
\label{Tab:tanb_significance}
\end{center}
\end{table}  

In the MSSM, the tree level electroweakino sector is governed by four input parameters: $M_{1}$ (bino mass parameter), $M_{2}$ (wino mass parameter), $\mu$ (higgsino mass parameter) and $\tan\beta$ (ratio of vacuum expectation value of the two Higgs doublets). We first consider the case of BP-$\beta_{\tilde{W}}$ ($M_{1} \sim 200~{\rm GeV}$, $M_{2} \sim 500~{\rm GeV}$, $\mu \sim 2~{\rm TeV}$) and study the collider implications of varying $\tan\beta$. In this respect, we consider $8$ benchmark points with different values of $\tan\beta \sim 5,8,10,15,20,25,30,40$ ($M_{1}$, $M_{2}$ and $\mu$ are kept fixed at the aforesaid values) and compute their signal significance in the $4$ signal channels considered in this work. In these benchmark points, the sleptons and the squarks have been decoupled by fixing their masses at $\sim 3~{\rm TeV}$ and $\sim 1.5~{\rm TeV}$, respectively. In the case of BP-$\beta_{\tilde{W}}$, $\lspthree,\lspfour,\chtwopm$ have a dominant higgsino composition with a mass of $\sim 2~{\rm TeV}$. Being heavier, the direct production cross section of the chargino-neutralino pairs involving any of these higgsinolike inos is negligible compared to the production cross section of the winolike $\chonepm\lsptwo$ pair. Consequently, we only consider the direct production of $\chonepm\lsptwo$ pair while computing the signal significance and ignore the contributions from the other ino pairs. Here, the signal yield is computed in each of the $4$ signal channels by multiplying the $\chonepm \lsptwo$ pair production cross section ($\sigma(\chonepm\lsptwo)$) with the branching rates of the relevant cascade decay modes, the integrated luminosity ($\mathcal{L} = 3000~{\rm fb^{-1}}$) and the efficiency of the respective signal regions. In the case of $Wh$ mediated signal channels ($Wh$ mediated $1l+2b+jets+\met$ and $Wh$ mediated $1l+2\gamma+jets+\met$), the relevant ino branching modes are: $Br(\lsptwo \to h \lspone)$ and $Br(\chonepm \to W \lspone)$, while the relevant ino decay modes in the later two cases ($WZ$ mediated $3l+jets+\met$ and $WZ$ mediated $3l+2b+jets+\met$) are: $Br(\lsptwo \to Z \lspone)$ and $Br(\chonepm \to W \lspone)$. The SM branching rates are considered for the successive decay of $Z,h$ and $W$ bosons. \texttt{Prospino}~\cite{Beenakker:1996ed,Beenakker:1999xh} is used to compute $\sigma(pp \to \chonepm \lsptwo)$ at NLO while \texttt{SUSY-HIT}~\cite{Djouadi:2006bz} is used to compute the ino branching rates. The corresponding ino pair production cross section and the ino branching rates are dependent on $\tan\beta$ and have been shown in Table~\ref{Tab:tanb_significance} against their respective $\tan\beta$ values. We have also listed the respective signal significance values in the $12$ signal regions ($4$ different analysis channels $\times$ $3$ signal regions in each) in Table~\ref{Tab:tanb_significance}.

\begin{figure}
\begin{center}
\includegraphics[scale=0.22]{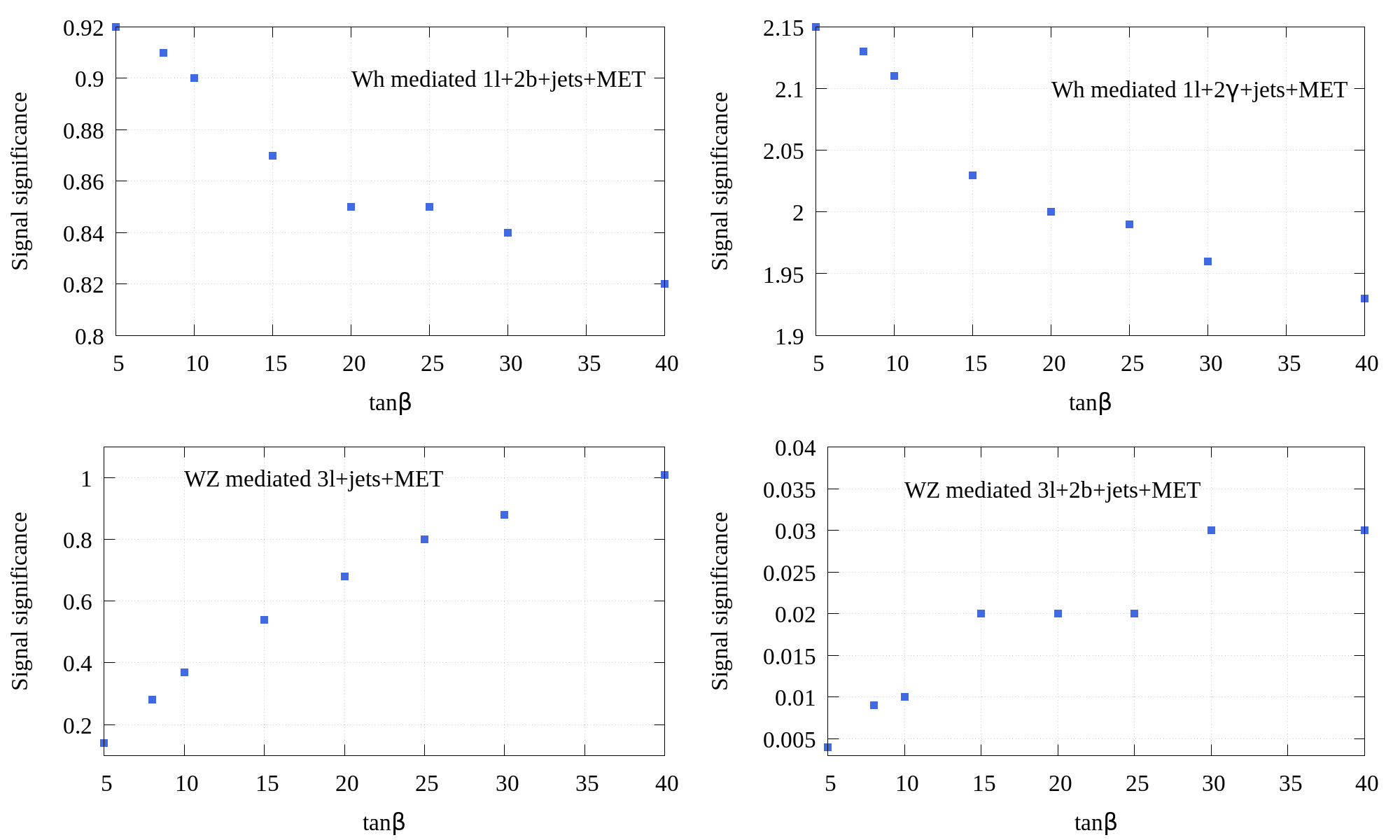}
\caption{\it \footnotesize The largest signal significance value in a particular signal channel and $\tan\beta$ are shown along the y- and x-axis. The $WZ$ mediated and $Wh$ mediated channels exhibit an opposite behavior with variations in $\tan\beta$. The signal significance values fall and rise with increase in $\tan\beta$ for the $Wh$ mediated and $WZ$ mediated channels, respectively. Here, MET refers to $\met$.}
\label{fig:tanb_significance}
\end{center}
\end{figure}

The coupling of the $Z$ boson with a pair of neutralinos ($\lspi\lspj$ ($i,j=1,2,3,4$)) is crucially controlled by the higgsino composition of $\chi_{i,j}^{0}$ while the $h\lspi\lspj$ couplings are proportional to the gaugino-higgsino admixture in $\chi_{i,j}^{0}$. As stated earlier, the tree level neutralino mixing matrix is governed by $M_{1}, M_{2}, \mu$ and $\tan\beta$. Thereby, the $Z\lspi\lspj$ and $h\lspi\lspj$ couplings are also controlled by the same input parameters at the tree level. In the present case, $\lsptwo$ is dominantly wino in nature with a small higgsino component (which varies with $\tan\beta$) in all the $8$ benchmark points. This makes $\lsptwo \to \lspone h$ as the most preferable decay mode of $\lsptwo$ with a branching ratio of $\sim 98.34\%$ for $\tan\beta = 5$. However, $Br(\lsptwo \to \lspone h)$ decreases up to $\sim 88.22\%$ upon increasing $\tan\beta$ to 40. Correspondingly, $Br(\lsptwo \to \lspone Z)$ increases from $\sim 1.65\%$ at $\tan\beta = 5$ up to $\sim 11.77 \%$ at $\tan\beta = 40$, thus, registering a nearly $\sim 7$ times improvement. The increase in the branching rate of $ \lsptwo \to \lspone Z$ is reflected in the signal significance of $WZ$ mediated analysis channels (shown in Table~\ref{Tab:tanb_significance} and bottom panel of Figure~\ref{fig:tanb_significance}). The signal significance of SR3-A, SR3-B and SR3-C (the optimized signal regions corresponding to $WZ$ mediated $3l+jets+\met$ channel) increases from $0.14,~0.14$ and $0.13$ (at $\tan\beta = 5$) to 0.98, 1.01 and 0.94, respectively, at $\tan\beta = 40$. A similar increase is also evident in the signal significance of SR4-A, SR4-B and SR4-C (signal regions corresponding to the $WZ$ mediated $3l+2b+jets+\met$ channel) which has a value of 0.001, 0.004 and 0.004, respectively, at $\tan\beta = 5$ while the respective values at $\tan\beta = 40$ are 0.01, 0.03 and 0.03. Equivalently, the signal significance of the $Wh$ mediated signal channels decrease with increasing $\tan\beta$. In Figure~\ref{fig:tanb_significance}, we have illustrated the variation of the signal significance (shown along y-axis) of the $4$ analysis channels considered in this work with $\tan\beta$ (shown along x-axis) for BP-$\beta_{\tilde{W}}$. For a particular final state, the largest value of signal significance among the respective $3$ signal regions has been considered in Figure~\ref{fig:tanb_significance}.

\begin{table}
\begin{center}\scalebox{0.65}{
\begin{tabular}{||C{3cm}||C{1.6cm}|C{1.6cm}|C{1.6cm}||C{1.6cm}|C{1.6cm}|C{1.6cm}||C{1.6cm}|C{1.6cm}|C{1.6cm}||C{1.6cm}|C{1.6cm}|C{1.6cm}||} \hline
\multicolumn{13}{||c||}{BP-$\beta_{\tilde{W}}^{10}$ ($M_{2}= 515~{\rm GeV}$, $\mu= 670~{\rm GeV}$)}\\ \hline
EW inos & \multicolumn{2}{c|}{$\lspone$} & \multicolumn{2}{c|}{$\lsptwo$} & \multicolumn{2}{c|}{$\lspthree$} & \multicolumn{2}{c|}{$\lspfour$} & \multicolumn{2}{c|}{$\chonepm$} & \multicolumn{2}{c|}{$\chtwopm$} \\ \hline
Mass (in GeV) & \multicolumn{2}{c|}{198.35} & \multicolumn{2}{c|}{496.45} & \multicolumn{2}{c|}{672.90} & \multicolumn{2}{c|}{693.11} & \multicolumn{2}{c|}{496.28} & \multicolumn{2}{c|}{692.73} \\ 
wino $\%$ & \multicolumn{2}{c|}{$10^{-3}$} & \multicolumn{2}{c|}{89.36} & \multicolumn{2}{c|}{0.16} & \multicolumn{2}{c|}{10.43} & \multicolumn{2}{c|}{86.67} & \multicolumn{2}{c|}{13.19} \\
higgsino $\%$ & \multicolumn{2}{c|}{0.65} & \multicolumn{2}{c|}{10.51} & \multicolumn{2}{c|}{99.68} & \multicolumn{2}{c|}{88.99} & \multicolumn{2}{c|}{13.19} & \multicolumn{2}{c|}{86.67} \\ \hline 
\multirow{2}{*}{Cross-section} & \multicolumn{2}{c|}{$\sigma(\chonepm\lsptwo)$} & \multicolumn{2}{c|}{$\sigma(\chonepm\lspthree)$} & \multicolumn{2}{c|}{$\sigma(\chonepm\lspfour)$} & \multicolumn{2}{c|}{$\sigma(\chtwopm\lsptwo)$} & \multicolumn{2}{c|}{$\sigma(\chtwopm\lspthree)$} & \multicolumn{2}{c|}{$\sigma(\chtwopm\lspfour)$} \\ 
 & \multicolumn{2}{c|}{21.94} & \multicolumn{2}{c|}{0.72} & \multicolumn{2}{c|}{0.3} & \multicolumn{2}{c|}{0.02} & \multicolumn{2}{c|}{2.93} & \multicolumn{2}{c|}{2.99} \\ \hline 
 Relevant Brs ($\%$) & \multicolumn{12}{c|}{$\lsptwo \to \lspone Z$ (10.4), $\lsptwo \to \lspone h$ (89.5), $\lspthree \to \lspone Z$ (15.86), $\lspthree \to \lspone h$ (2.97), $\lspfour \to \lspone Z$ (2.74), $\lspfour \to \lspone h$ (12.55), $\chtwopm \to W \lspone$ (14.20) } \\ \hline
 Signal & \multicolumn{3}{c||}{$Wh$ mediated $1l+2b+jets+\met$} & \multicolumn{3}{c||}{$Wh$ mediated $1l+2\gamma+jets+\met$} & \multicolumn{3}{c||}{$WZ$ mediated $3l+jets+\met$} & \multicolumn{3}{c||}{$WZ$ mediated $3l+2b+jets+\met$} \\ \cline{2-13}
significance & SR1-A & SR1-B & SR1-C & SR2-A & SR2-B & SR2-C & SR3-A & SR3-B & SR3-C & SR4-A & SR4-B & SR4-C \\
 & 0.44 & 0.64 & 0.24 & 1.39 & 1.41 & 1.90 & 0.09 & 0.09 & 0.06 & 0.03 & 0.09 & 0.06 \\ \hline \hline 
 \multicolumn{13}{||c||}{BP-$\beta_{\tilde{W}}^{30}$ ($M_{2}= 540~{\rm GeV}$, $\mu= 590~{\rm GeV}$)}\\ \hline
EW inos & \multicolumn{2}{c|}{$\lspone$} & \multicolumn{2}{c|}{$\lsptwo$} & \multicolumn{2}{c|}{$\lspthree$} & \multicolumn{2}{c|}{$\lspfour$} & \multicolumn{2}{c|}{$\chonepm$} & \multicolumn{2}{c|}{$\chtwopm$} \\ \hline
Mass (in GeV) & \multicolumn{2}{c|}{197.92} & \multicolumn{2}{c|}{502.28} & \multicolumn{2}{c|}{593.09} & \multicolumn{2}{c|}{ 632.89} & \multicolumn{2}{c|}{501.79} & \multicolumn{2}{c|}{632.41} \\ 
wino $\%$ & \multicolumn{2}{c|}{0.01} & \multicolumn{2}{c|}{69.33} & \multicolumn{2}{c|}{0.18} & \multicolumn{2}{c|}{30.46} & \multicolumn{2}{c|}{65.09} & \multicolumn{2}{c|}{34.90} \\
higgsino $\%$ & \multicolumn{2}{c|}{0.92} & \multicolumn{2}{c|}{30.31} & \multicolumn{2}{c|}{99.67} & \multicolumn{2}{c|}{69.06} & \multicolumn{2}{c|}{34.90} & \multicolumn{2}{c|}{65.09} \\ \hline 
\multirow{2}{*}{Cross-section} & \multicolumn{2}{c|}{$\sigma(\chonepm\lsptwo)$} & \multicolumn{2}{c|}{$\sigma(\chonepm\lspthree)$} & \multicolumn{2}{c|}{$\sigma(\chonepm\lspfour)$} & \multicolumn{2}{c|}{$\sigma(\chtwopm\lsptwo)$} & \multicolumn{2}{c|}{$\sigma(\chtwopm\lspthree)$} & \multicolumn{2}{c|}{$\sigma(\chtwopm\lspfour)$} \\ 
 & \multicolumn{2}{c|}{18.91} & \multicolumn{2}{c|}{2.86} & \multicolumn{2}{c|}{0.13} & \multicolumn{2}{c|}{0.10} & \multicolumn{2}{c|}{3.77} & \multicolumn{2}{c|}{4.78} \\ \hline 
 Relevant Brs ($\%$) & \multicolumn{12}{c|}{$\lsptwo \to \lspone Z$ (11.62), $\lsptwo \to \lspone h$ (88.37), $\lspthree \to \lspone Z$ (69.17), $\lspthree \to \lspone h$ (10.60), $\lspfour \to \lspone Z$ (4.09), $\lspfour \to \lspone h$ (19.8), $\chtwopm \to W \lspone$ (24.00) } \\ \hline
 Signal & \multicolumn{3}{c||}{$Wh$ mediated $1l+2b+jets+\met$} & \multicolumn{3}{c||}{$Wh$ mediated $1l+2\gamma+jets+\met$} & \multicolumn{3}{c||}{$WZ$ mediated $3l+jets+\met$} & \multicolumn{3}{c||}{$WZ$ mediated $3l+2b+jets+\met$} \\ \cline{2-13}
significance & SR1-A & SR1-B & SR1-C & SR2-A & SR2-B & SR2-C & SR3-A & SR3-B & SR3-C & SR4-A & SR4-B & SR4-C \\
 & 0.39 & 0.55 & 0.69 & 1.24 & 1.25 & 1.68 & 0.21 & 0.11 & 0.09 & 0.29 & 0.78 & 0.39 \\ \hline \hline 
 \multicolumn{13}{||c||}{BP-$\beta_{\tilde{W}}^{50}$ ($M_{2}= 555~{\rm GeV}$, $\mu= 550~{\rm GeV}$)}\\ \hline
EW inos & \multicolumn{2}{c|}{$\lspone$} & \multicolumn{2}{c|}{$\lsptwo$} & \multicolumn{2}{c|}{$\lspthree$} & \multicolumn{2}{c|}{$\lspfour$} & \multicolumn{2}{c|}{$\chonepm$} & \multicolumn{2}{c|}{$\chtwopm$} \\ \hline
Mass (in GeV) & \multicolumn{2}{c|}{197.62} & \multicolumn{2}{c|}{495.15} & \multicolumn{2}{c|}{553.19} & \multicolumn{2}{c|}{615.41} & \multicolumn{2}{c|}{494.27} & \multicolumn{2}{c|}{615.03} \\ 
wino $\%$ & \multicolumn{2}{c|}{0.01} & \multicolumn{2}{c|}{48.29} & \multicolumn{2}{c|}{0.19} & \multicolumn{2}{c|}{51.49} & \multicolumn{2}{c|}{43.53} & \multicolumn{2}{c|}{56.44} \\
higgsino $\%$ & \multicolumn{2}{c|}{1.12} & \multicolumn{2}{c|}{51.04} & \multicolumn{2}{c|}{99.66} & \multicolumn{2}{c|}{48.15} & \multicolumn{2}{c|}{56.44} & \multicolumn{2}{c|}{43.53} \\ \hline 
\multirow{2}{*}{Cross-section} & \multicolumn{2}{c|}{$\sigma(\chonepm\lsptwo)$} & \multicolumn{2}{c|}{$\sigma(\chonepm\lspthree)$} & \multicolumn{2}{c|}{$\sigma(\chonepm\lspfour)$} & \multicolumn{2}{c|}{$\sigma(\chtwopm\lsptwo)$} & \multicolumn{2}{c|}{$\sigma(\chtwopm\lspthree)$} & \multicolumn{2}{c|}{$\sigma(\chtwopm\lspfour)$} \\ 
 & \multicolumn{2}{c|}{18.61} & \multicolumn{2}{c|}{5.90} & \multicolumn{2}{c|}{0.13} & \multicolumn{2}{c|}{0.15} & \multicolumn{2}{c|}{3.26} & \multicolumn{2}{c|}{5.75} \\ \hline 
 Relevant Brs ($\%$) & \multicolumn{12}{c|}{$\lsptwo \to \lspone Z$ (11.91), $\lsptwo \to \lspone h$ (88.08), $\lspthree \to \lspone Z$ (88.12), $\lspthree \to \lspone h$ (11.88), $\lspfour \to \lspone Z$ (3.53), $\lspfour \to \lspone h$ (16.85), $\chtwopm \to W \lspone$ (21.20) } \\ \hline
 Signal & \multicolumn{3}{c||}{$Wh$ mediated $1l+2b+jets+\met$} & \multicolumn{3}{c||}{$Wh$ mediated $1l+2\gamma+jets+\met$} & \multicolumn{3}{c||}{$WZ$ mediated $3l+jets+\met$} & \multicolumn{3}{c||}{$WZ$ mediated $3l+2b+jets+\met$} \\ \cline{2-13}
significance & SR1-A & SR1-B & SR1-C & SR2-A & SR2-B & SR2-C & SR3-A & SR3-B & SR3-C & SR4-A & SR4-B & SR4-C \\
 & 0.39 & 0.54 & 0.68 & 1.24 & 1.26 & 1.68 & 0.34 & 0.13 & 0.11 & 0.55 & 1.43 & 0.65 \\ \hline \hline
 \multicolumn{13}{||c||}{BP-$\beta_{\tilde{W}}^{70}$ ($M_{2}= 600~{\rm GeV}$, $\mu= 535~{\rm GeV}$)}\\ \hline
EW inos & \multicolumn{2}{c|}{$\lspone$} & \multicolumn{2}{c|}{$\lsptwo$} & \multicolumn{2}{c|}{$\lspthree$} & \multicolumn{2}{c|}{$\lspfour$} & \multicolumn{2}{c|}{$\chonepm$} & \multicolumn{2}{c|}{$\chtwopm$} \\ \hline
Mass (in GeV) & \multicolumn{2}{c|}{197.49} & \multicolumn{2}{c|}{502.35} & \multicolumn{2}{c|}{538.15} & \multicolumn{2}{c|}{638.31} & \multicolumn{2}{c|}{501.11} & \multicolumn{2}{c|}{638.11} \\ 
wino $\%$ & \multicolumn{2}{c|}{0.01} & \multicolumn{2}{c|}{26.54} & \multicolumn{2}{c|}{0.18} & \multicolumn{2}{c|}{73.26} & \multicolumn{2}{c|}{22.60} & \multicolumn{2}{c|}{80.56} \\
higgsino $\%$ & \multicolumn{2}{c|}{1.21} & \multicolumn{2}{c|}{72.54} & \multicolumn{2}{c|}{99.66} & \multicolumn{2}{c|}{26.54} & \multicolumn{2}{c|}{80.56} & \multicolumn{2}{c|}{22.60} \\ \hline 
\multirow{2}{*}{Cross-section} & \multicolumn{2}{c|}{$\sigma(\chonepm\lsptwo)$} & \multicolumn{2}{c|}{$\sigma(\chonepm\lspthree)$} & \multicolumn{2}{c|}{$\sigma(\chonepm\lspfour)$} & \multicolumn{2}{c|}{$\sigma(\chtwopm\lsptwo)$} & \multicolumn{2}{c|}{$\sigma(\chtwopm\lspthree)$} & \multicolumn{2}{c|}{$\sigma(\chtwopm\lspfour)$} \\ 
 & \multicolumn{2}{c|}{15.31} & \multicolumn{2}{c|}{6.74} & \multicolumn{2}{c|}{0.08} & \multicolumn{2}{c|}{0.09} & \multicolumn{2}{c|}{1.70} & \multicolumn{2}{c|}{4.67} \\ \hline 
 Relevant Brs ($\%$) & \multicolumn{12}{c|}{$\lsptwo \to \lspone Z$ (12.45), $\lsptwo \to \lspone h$ (87.55), $\lspthree \to \lspone Z$ (88.68), $\lspthree \to \lspone h$ (11.32), $\lspfour \to \lspone Z$ (6.03), $\lspfour \to \lspone h$ (5.44), $\chtwopm \to W \lspone$ (6.78) } \\ \hline
 Signal & \multicolumn{3}{c||}{$Wh$ mediated $1l+2b+jets+\met$} & \multicolumn{3}{c||}{$Wh$ mediated $1l+2\gamma+jets+\met$} & \multicolumn{3}{c||}{$WZ$ mediated $3l+jets+\met$} & \multicolumn{3}{c||}{$WZ$ mediated $3l+2b+jets+\met$} \\ \cline{2-13}
significance & SR1-A & SR1-B & SR1-C & SR2-A & SR2-B & SR2-C & SR3-A & SR3-B & SR3-C & SR4-A & SR4-B & SR4-C \\
 & 0.32 & 0.44 & 0.55 & 1.01 & 1.02 & 1.37 & 0.34 & 0.12 & 0.10 & 0.58 & 1.50 & 0.68 \\ \hline \hline
\end{tabular}}
\caption{\it \footnotesize  The electroweakino mass spectrum of BP-$\beta_{\tilde{W}}^{10}$, BP-$\beta_{\tilde{W}}^{30}$, BP-$\beta_{\tilde{W}}^{50}$ and BP-$\beta_{\tilde{W}}^{70}$ is shown along with the values of $M_{2}$ and $\mu$. The wino and higgsino composition of the electroweakinos are also listed. The NLO cross section of directly produced chargino-neutralino pairs and the electroweakino branching fractions which are relevant to the $4$ signal channels analyzed in this work are shown. Here, $\chonepm$ always decays into $W\lspone$. The signal significance values in the respective analysis channels are also tabulated.}
\label{Tab:M2mu_significance}
\end{center}
\end{table}

We also analyze additional realistic benchmark scenarios where we vary the higgsino and wino admixtures in the neutralinos and charginos, and study the projected reach of HL-LHC in probing them in the $Wh$ mediated $1l+2b+jets+\met$, $Wh$ mediated $1l+2\gamma+jets+\met$, $WZ$ mediated $3l+jets+\met$ and $WZ$ mediated $3l+2b+jets+\met$ final states. We choose four different benchmark points (BP-$\beta_{\tilde{W}}^{10}$, BP-$\beta_{\tilde{W}}^{30}$, BP-$\beta_{\tilde{W}}^{50}$ and BP-$\beta_{\tilde{W}}^{70}$) in such a way that $M_{\lsptwo}$ and $M_{\chonepm}$ are always at roughly $\sim 500~{\rm GeV}$ while $M_{\lspone}$ is at roughly $\sim 200~{\rm GeV}$. We also ensure than $\lspone$ is always binolike by fixing $M_{1}$ at $200~{\rm GeV}$. The sleptons, squarks and gluinos are fixed at $\sim 3~{\rm TeV}$, $\sim 1.5~{\rm TeV}$ and $\sim 4~{\rm TeV}$, respectively, in order to decouple their effects from the processes of our interest. The value of $\tan\beta$ is fixed to $10$. $M_{2}$ and $\mu$ are varied such that the amount of higgsino component in $\lsptwo$ is $\sim 10\%$ (BP-$\beta_{\tilde{W}}^{10}$), $\sim 30\%$ (BP-$\beta_{\tilde{W}}^{30}$), $\sim 50\%$ (BP-$\beta_{\tilde{W}}^{50}$) and $\sim 70\%$ (BP-$\beta_{\tilde{W}}^{70}$). The values of $M_{2}$ and $\mu$ for these $4$ benchmark points along with their electroweakino mass spectrum is shown in Table~\ref{Tab:M2mu_significance}. We have also listed the higgsino and wino composition in Table~\ref{Tab:M2mu_significance}.

In the previous case of BP-$\beta_{\tilde{W}}$, the amount of higgsino admixture in $\lsptwo$ was $\lesssim 1\%$ for $\tan\beta = 10$. Also, $\lspthree, \lspfour$ and $\chtwopm$ were much heavier ($\sim 2~{\rm TeV}$) and therefore, the contributions to the signal yield from $\lsptwo\chonepm$, $\lspthree\chonepm$, $\lspfour \chonepm$, $\lspthree\chtwopm$ and $\lspfour\chtwopm$ production processes could be safely ignored due to their small cross sections. However, in the present case, when we attempt to introduce a finite amount of higgsino admixture in $\lsptwo$ while keeping its mass fixed at $\sim 500~{\rm GeV}$, we are forced to reduce the value of $\mu$. Consequently, $\lspthree, \lspfour$ and $\chtwopm$ are no more in the $\sim O(1)$ TeV range. For example, in the case of BP-$\beta_{\tilde{W}}^{10}$ where the $\lsptwo$ is composed of higgsinos and winos in the proportion of $\sim 10\%$ and $\sim 90\%$, respectively, we are required to choose $\mu \sim 670~{\rm GeV}$. As a result, $\lspthree, \lspfour$ and $\chtwopm$ also become admixtures of winos and higgsinos and have a mass of $\sim 672.90~{\rm GeV}$, $\sim 693.11~{\rm GeV}$ and $\sim 692.73~{\rm GeV}$, respectively. Correspondingly, it would be imperative to take into account the contributions from the heavier chargino-neutralino pairs as well. In the present scenario, therefore, contributions to the signal yield can potentially arise from: $pp \to \chonepm\lsptwo + \chonepm\lspthree +\chonepm\lspfour + \chtwopm\lsptwo + \chtwopm \lspthree + \chtwopm\lspfour$. The production cross section of these chargino-neutralino pairs for BP-$\beta_{\tilde{W}}^{10}$, BP-$\beta_{\tilde{W}}^{30}$, BP-$\beta_{\tilde{W}}^{50}$ and BP-$\beta_{\tilde{W}}^{70}$ are listed in Table~\ref{Tab:M2mu_significance}. Here, we have used \texttt{Prospino-2.1} to compute the cross sections at NLO. The branching ratios of $\lspi \to Z/h \lspone$ ($i=2,3,4$) and $\chi_{j}^{\pm} \to W \lspone$ ($j=1,2$) are also shown in Table~\ref{Tab:M2mu_significance} where \texttt{SUSY-HIT} has been used to compute them. The signal significance of these $4$ benchmark points is computed for (SR1-A, SR1-B, SR1-C), (SR2-A, SR2-B, SR2-C), (SR3-A, SR3-B, SR3-C) and (SR4-A, SR4-B, SR4-C), corresponding to $Wh$ mediated $1l+2b+jets+\met$, $Wh$ mediated $1l+2\gamma+jets+\met$, $WZ$ mediated $3l+jets+\met$ and $WZ$ mediated $3+2b+jets+\met$ channels, respectively, and have been listed in Table~\ref{Tab:M2mu_significance}. It can be observed from Table~\ref{Tab:M2mu_significance} that the signal significance of the $WZ$ mediated channels improve with an increase in the amount of higgsino content in $\lsptwo$. For example, the signal significance of SR3-A and SR4-A increases from $\sim 0.09$ and $\sim 0.03$ for BP-$\beta_{\tilde{W}}^{10}$ to $\sim 0.34$ and $\sim 0.58$ for BP-$\beta_{\tilde{W}}^{70}$.  This increase is mainly an outcome of the combined effect of an increased $\lsptwo/\lspthree \to Z\lspone$ branching rate, an increased $\sigma(\lspthree\chonepm)$ and a lowered $\sigma(\lsptwo\chonepm)$. Upon combining the highest signal significance values in each of the analysis channels in quadrature, we obtain a combined signal significance of $2.01$, $1.98$, $2.33$ and $2.12$ for BP-$\beta_{\tilde{W}}^{10}$, BP-$\beta_{\tilde{W}}^{30}$, BP-$\beta_{\tilde{W}}^{50}$ and BP-$\beta_{\tilde{W}}^{70}$, respectively. Before concluding this section, we would like to note that additional contribution to the signal yield may also arise by considering the cascade decay modes of the heavier charginos and neutralinos. For example, in a generic case, the $\lspthree$ can decay into a $Z/h \lsptwo$ pair and this $\lsptwo$ can further decay into a $Z/h \lspone$ pair, resulting in an additional $Z/h$ bosons in the final state. A multifarious number of such possibilities are potentially feasible, and are outside the scope of this present work. We intend to explore such scenarios on a case by case basis in a future work. 

\begin{figure}
\begin{center}
\includegraphics[scale=0.2]{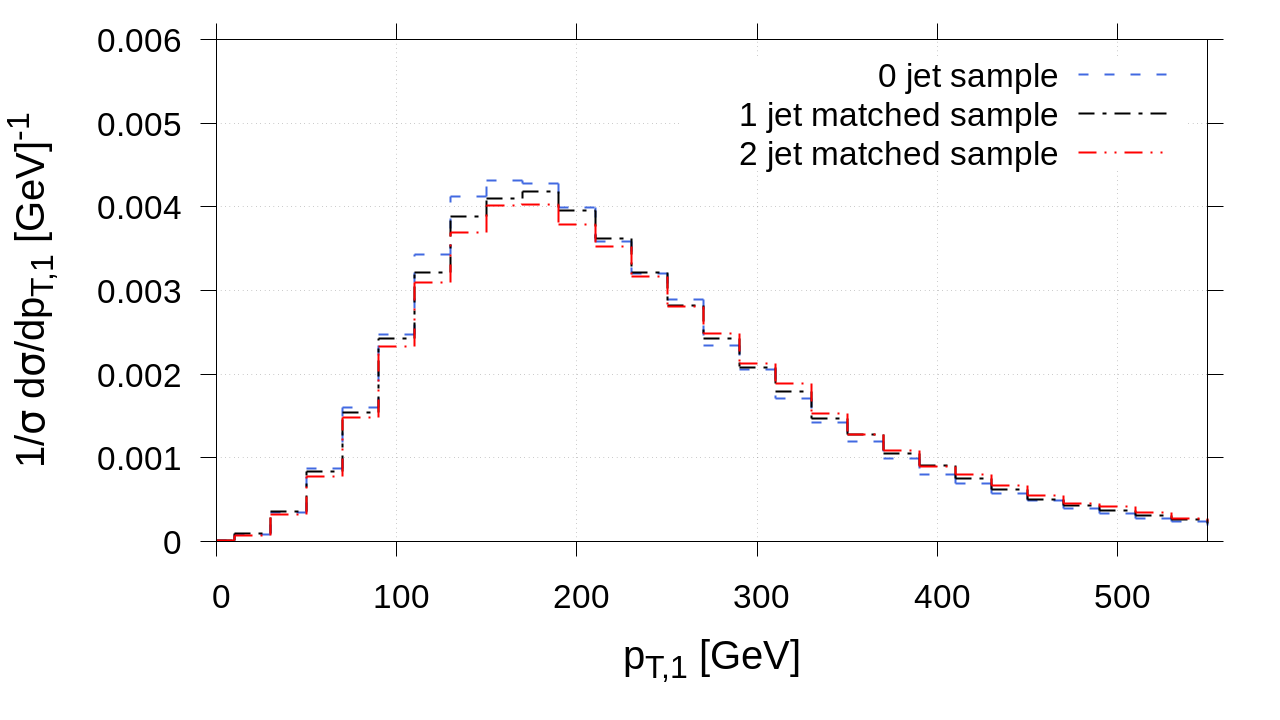}\includegraphics[scale=0.2]{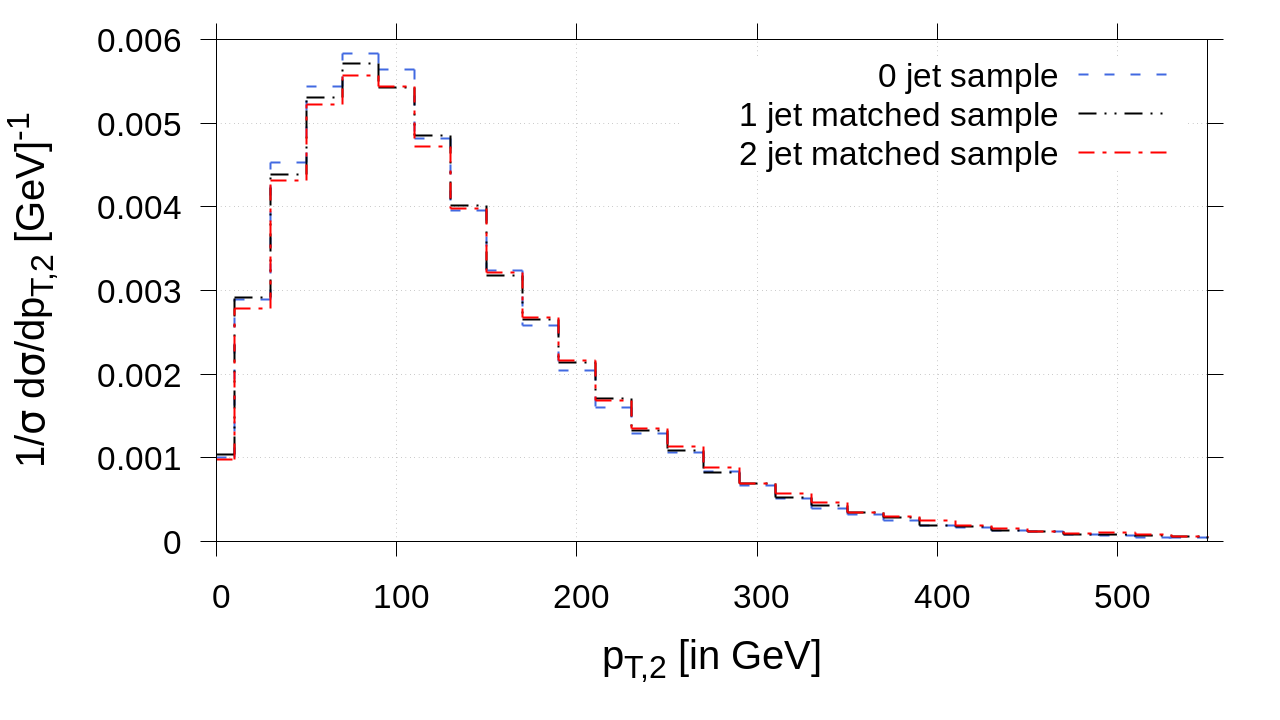}
\caption{Normalized distributions illustrating the $p_{T}$ of the $\lspone$'s produced via $pp \to$ $\chonepm \lsptwo \to$ $ (W^{\pm} \lspone)~(Z\lspone)$ at the HL-LHC. $p_{T,1}$~(left panel) and $p_{T,2}$~(right panel) represents the transverse momentum of the $\lspone$ with the highest and the second highest $p_{T}$. The distributions correspond to $2$ $jet$~(red dot-dot-dashed line), $1$ $jet$~(black dot-dashed line) and $0$ $jet$ matched~(blue dashed line) samples.}
\label{fig:pT_dist_jetmatching}
\end{center}
\end{figure}

\begin{figure}[!htb]
\begin{center}
\includegraphics[scale=0.2]{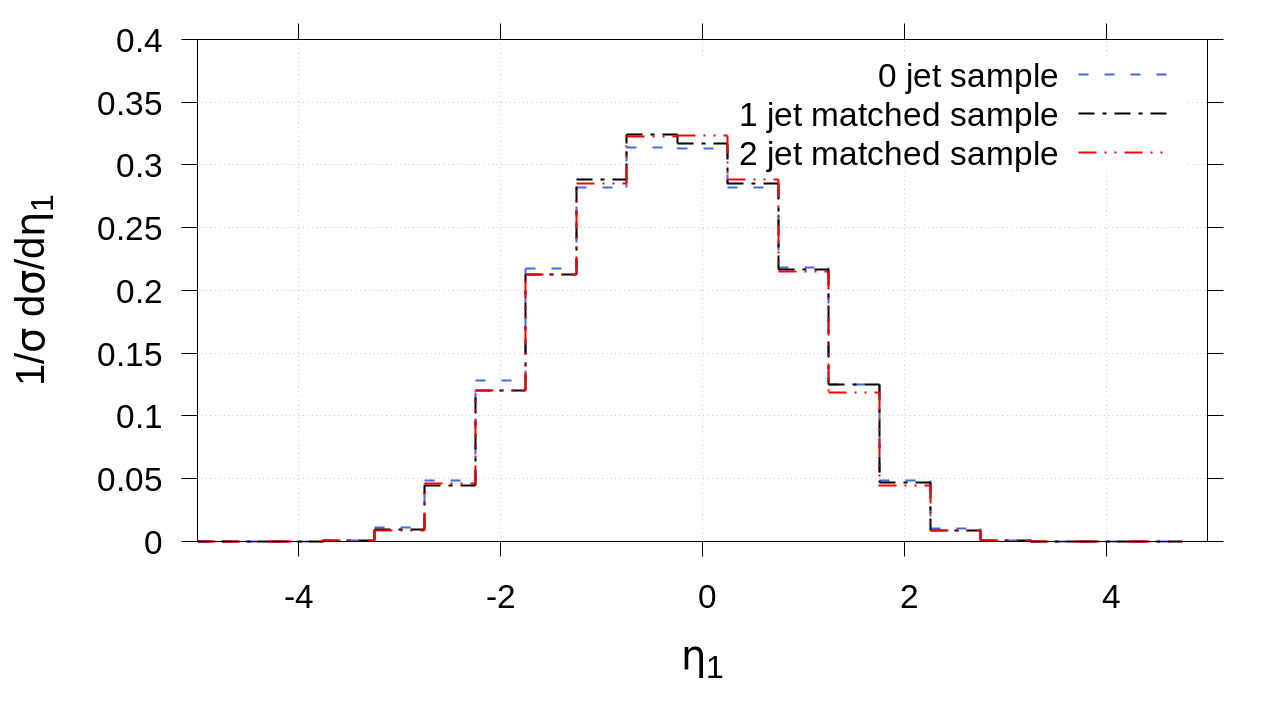}\includegraphics[scale=0.2]{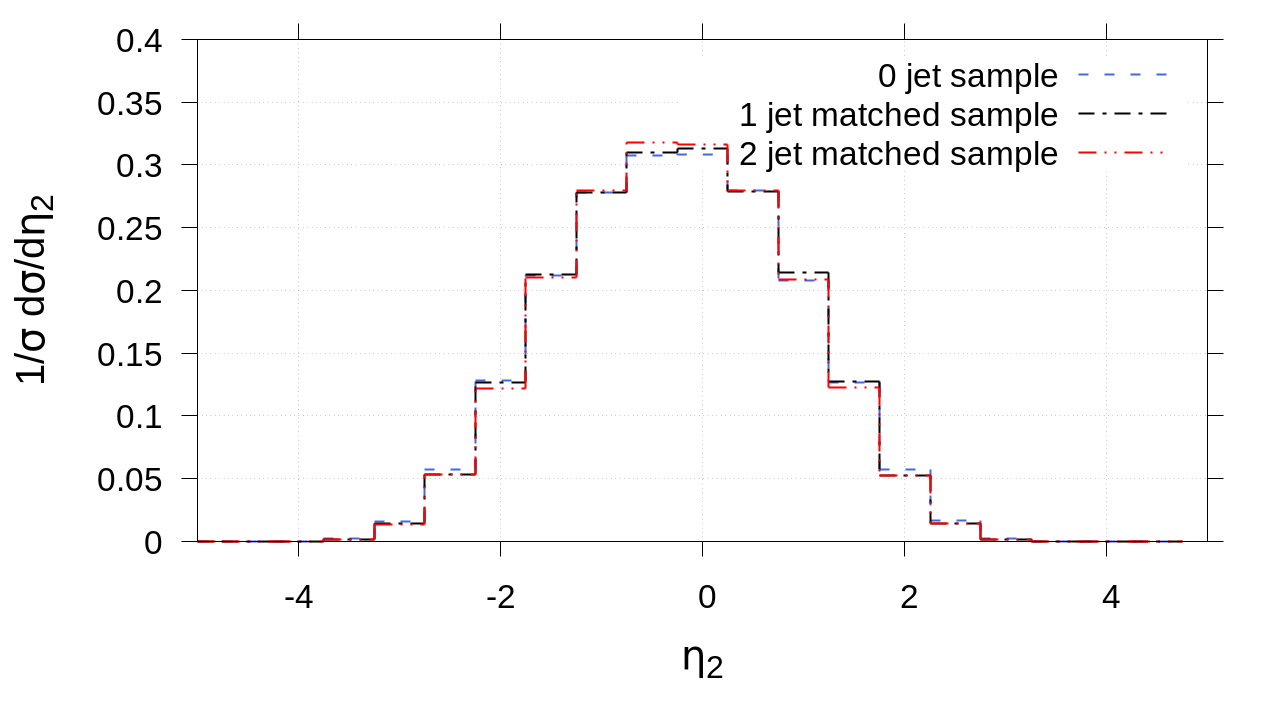}
\caption{Normalized distributions illustrating the $\eta$ of the $\lspone$'s produced via $pp \to \chonepm \lsptwo \to$ $(W^{\pm} \lspone)~(Z\lspone)$ at the HL-LHC. $\eta_{1}$~(left panel) and $\eta_{2}$~(right panel) represents the pseudorapidity of the $\lspone$ with the highest and second highest $p_{T}$. The distributions correspond to $2$ $jet$~(red dot-dot-dashed line), $1$ $jet$~(black dot-dashed line) and $0$ $jet$ matched~(blue dashed line) samples.}
\label{fig:eta_dist_jetmatching}
\end{center}
\end{figure}

We also briefly discuss the collider implications of $jet$ matched signal samples. In the analysis presented in Section~\ref{sec-3}, we used the $jet$ matched background samples while the signal samples were generated without any $jet$ matching. The underlying reason is the negligible difference in the kinematic features of the final state particles of the signal produced with and without $jet$ matching. In Figure~\ref{fig:pT_dist_jetmatching} and \ref{fig:eta_dist_jetmatching}, we illustrate the $p_{T}$ and $\eta$ distributions, respectively, of the $\lspone$'s with the highest and second highest $p_{T}$ produced via $pp \to (\lsptwo \to \lspone W^{\pm}) (\chonepm \to \lspone Z)$, at the HL-LHC. The blue dashed line corresponds to the $0~jet$ matched sample while the black dot-dashed and red-dot-dot-dashed lines represent the distributions for $1~jet$ and $2~jet$ matched samples. These distributions indicate that $jet$ matching has almost a negligible effect on the final state kinematics and it is safe to consider the $0~jet$ matched signal samples only. 

\begin{figure}[!htb]
\begin{center}
\includegraphics[scale=0.2]{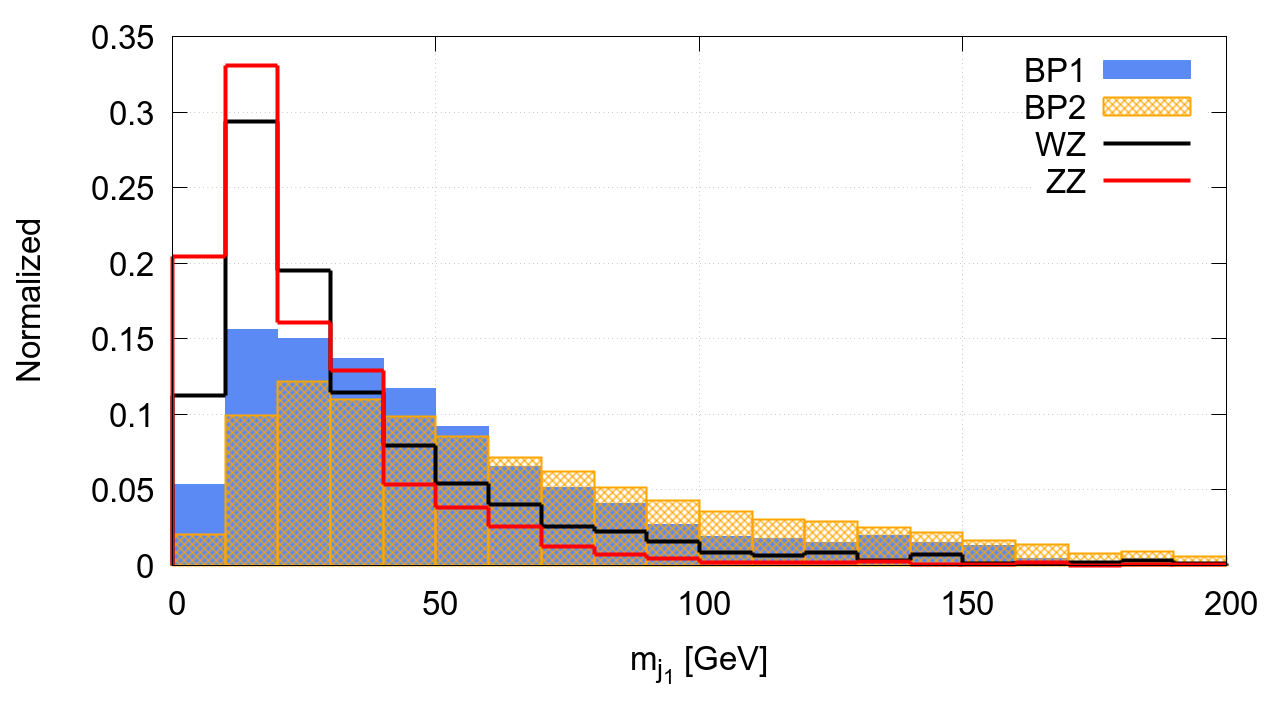}\includegraphics[scale=0.2]{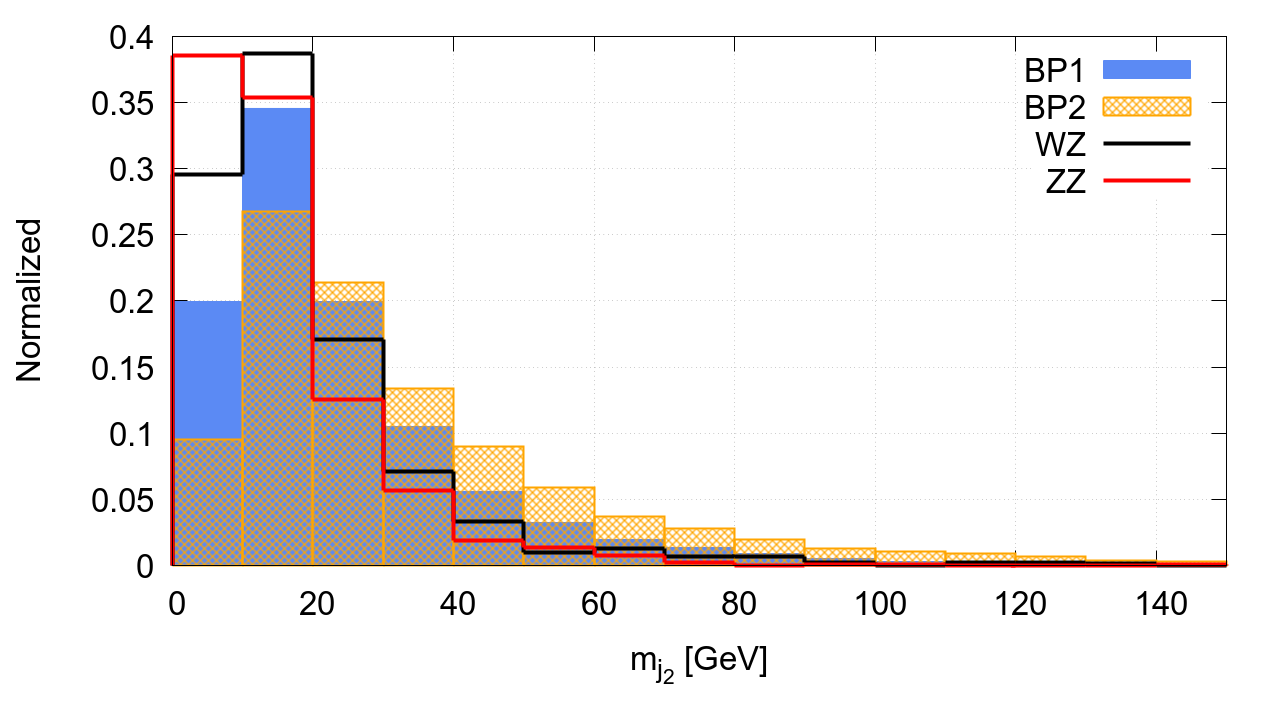}
\includegraphics[scale=0.2]{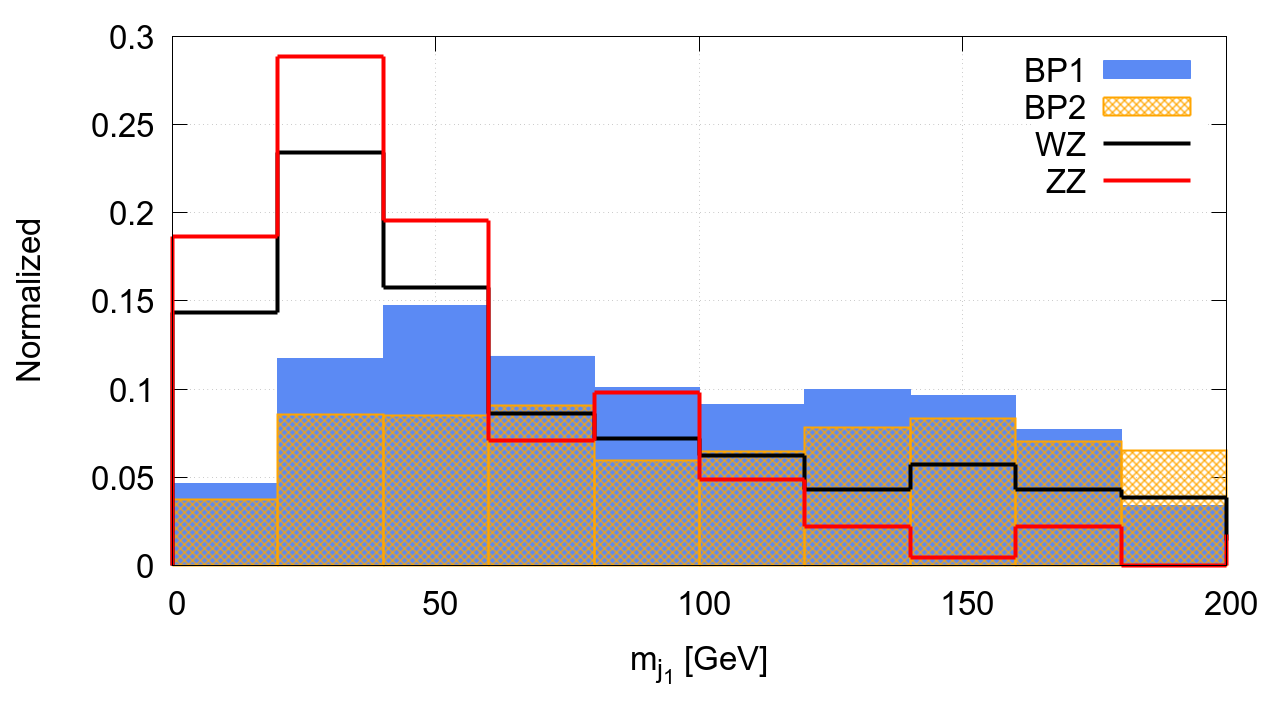}\includegraphics[scale=0.2]{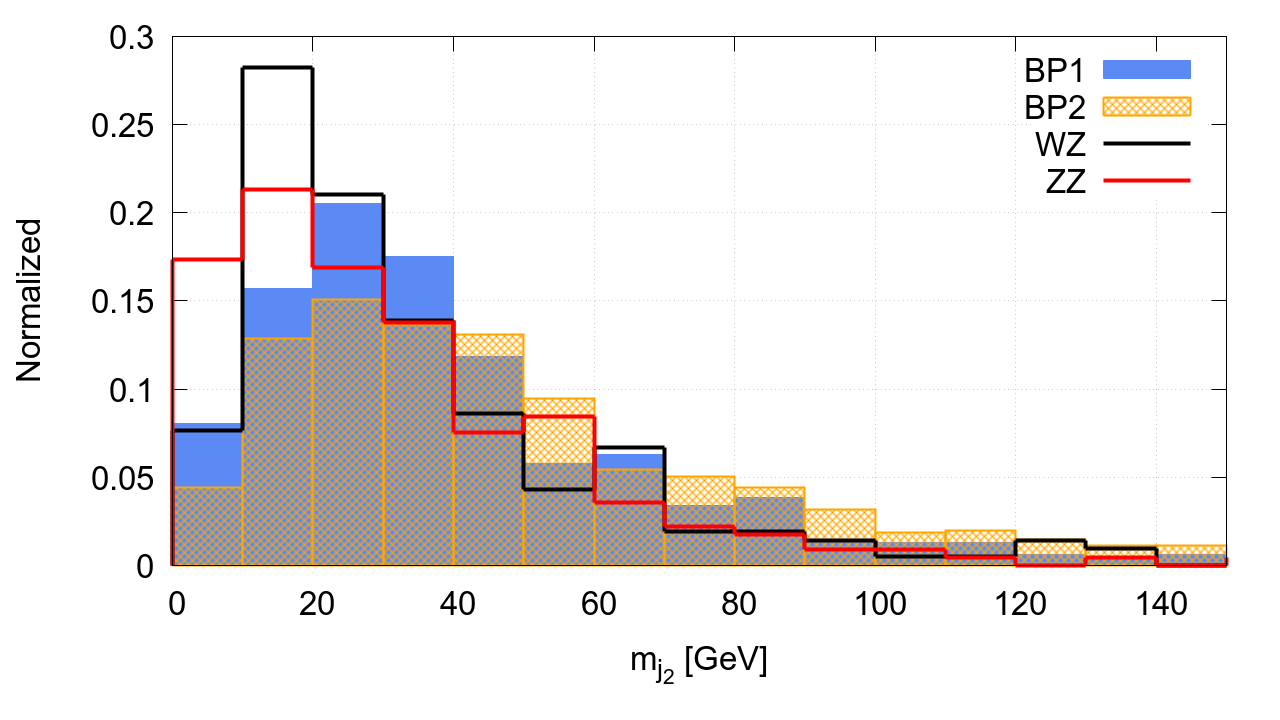}
\caption{\textit{Top panel}: Normalized distributions illustrating the invariant mass of $j_{1}$~($m_{j_1}$) and $j_{2}$~($m_{j_2}$), reconstructed with $R = 0.8$, for BP-J1~(blue shaded region), BP-J2~(orange shaded region), $WZ +jets$~(black solid line) and $ZZ+jets$~(red solid line), in the left and right panels, respectively. \textit{Bottom panel}:  Same as the top panel, but with $jets$ reconstructed with $R = 1.5$.}
\label{fig:jet_sub_mj}
\end{center}
\end{figure}

In this work, we have used the traditional cut-based analysis to discriminate the signal from the background. It would be important and interesting to gauge the benefits of employing advanced analysis techniques like $jet$ substructure to perform the signal-background discrimination. Before concluding this paper, we intend to briefly discuss this aspect as well. In order to study the study the prospects of $jet$ substructure techniques, we consider two signal benchmark points: BP-J1~($M_{\chonepm,\lsptwo}=350~{\rm GeV}$,~$M_{\lspone} = 150~{\rm GeV}$) and BP-J2~($M_{\chonepm,\lsptwo}=550~{\rm GeV}$,~$M_{\lspone} = 300~{\rm GeV}$), and simulate the signal process: $pp \to \chonepm\lsptwo \to (W^{\pm}\lspone)(Z \lspone) \to (W \to l^{\prime}\nu)(\lspone \to jets)(Z \to l^{\prime}l^{\prime})(\lspone \to jets) \to 3l + jets + \met$. The relevant background processes are $WZ+jets$ and $ZZ+jets$. In order to study the prospects of $jet$ substructure techniques, we modify the anti-$k_{t}$ $jet$ reconstruction definition by choosing a relatively large $jet$ radius parameter~($R$). In this regard, we choose two different values of $R$, 0.8 and 1.5. We intend to investigate the internal structure of the reconstructed $jets$ and their $p_{T}$ distributions for both, signal and background, in order to understand the merit of using the $jet$ substructure for signal-background discrimination. Our basic event selection criteria demands at least $2$ $jets$ and $3$ leptons~(electrons or muons) in the final state. The invariant mass distribution of the leading and subleading $p_{T}$ ordered $jets$~(represented as $j_{1}$ and $j_{2}$, respectively) would provide a window to the internal structure of the $jets$. We illustrate the invariant mass of $j_{1}$ and $j_{2}$, reconstructed with $R = 0.8$, in Figure~\ref{fig:jet_sub_mj}~(top left panel) and (top right panel), respectively. The invariant mass distribution for $j_{1}$ and $j_{2}$ reconstructed with $R = 1.5$ are also shown in the bottom left panel and  the bottom right panel, respectively, of the same figure.

It is observed from Figure~\ref{fig:jet_sub_mj} that the signal benchmark points do not display any significant difference or any distinct kinematic features from the background processes for both $R= 0.8$ and $R=1.5$. Choosing a larger $jet$ radius parameter makes the $jet$ invariant distributions flatter and the peak slightly shifts to a higher value. We also identified the fraction of events with a large $p_{T,j_{1,2}}$ $i.e.$ above $2M_{\lspone}$, $2M_{Z}$ and $2M_{W}$. The results have been displayed in Table~\ref{tab:frac1}. We observe that the fraction of events with $p_{T,j_{1}}$ above the $2M_{\lspone}$ threshold is $\sim 16\%$ and $\sim 12\%$ for BP-J1 with $R=0.8$ and $R=1.5$, respectively. In the case of BP-J2, where the $\lspone$ is much heavier, the corresponding fraction of events are further small, $\sim 6\%$ and $\sim 3\%$, respectively. It can be concluded from these discussions that $jet$ substructure techniques might help in improving the signal-background discrimination to some extent, however, gaining a thorough understanding of the improvement warrants a detailed analysis of its own which is beyond the scope of this paper.

\begin{table}[!htb]
\begin{center}
\begin{tabular}{C{4cm}||C{1.5cm}C{1.5cm} C{1.5cm} C{1.5cm}|C{1.5cm}C{1.5cm}C{1.5cm}}
\multirow{2}{*}{($M_{\lspone},~M_{\lsptwo,\chonepm}$)} & & \multicolumn{3}{c|}{$R = 0.8$} & \multicolumn{3}{c}{$R = 1.5$} \\  \cline{2-8}
& $p_{T}$ & $> 2 M_{\lspone}$ & $> 2M_{Z}$ & $> 2M_{W}$ & $> 2 M_{\lspone}$ & $> 2M_{Z}$ & $> 2M_{W}$ \\ \hline \hline
BP1 & \multicolumn{1}{c|}{$j_{1}$} & 0.16 & 0.46 & 0.56 & 0.12 & 0.46 & 0.54 \\
(150~GeV, 350~GeV) & \multicolumn{1}{c|}{$j_{2}$} & 0.02 & 0.11 & 0.15 & $6 \times 10^{-3}$ & 0.06 & 0.07 \\ \hline
BP1 & \multicolumn{1}{c|}{$j_{1}$} & 0.06 & 0.8 & 0.86 & 0.03 & 0.75 & 0.8\\
(300~GeV, 550~GeV) & \multicolumn{1}{c|}{$j_{2}$} & $5\times 10^{-3}$ & 0.35 & 0.45 & $8 \times 10^{-4}$ & 0.19 & 0.26\\ \hline
\end{tabular}
\caption{Table illustrating the fraction of events~(after passing the selection cuts: $N_{j} \geq 2$, $N_{l} = 3$, $N_{b} = 0$) with $p_{T,j_{i}}$~($j=1,2$) above $2M_{\lspone}$, $2M_{Z}$ and  $2M_{W^{\pm}}$.}
\label{tab:frac1}
\end{center}
\end{table}

We conclude our discussions here and proceed to provide a detailed summary and conclusion to our results.

\section{Summary and Conclusions}
\label{Sec:Conclusion}

In this work, we studied the sensitivity of the future HL-LHC to direct wino searches in a simplified scenario with $\lambda^{\prime\prime}_{112} u^{c}d^{c}s^{c}$ and $\lambda^{\prime\prime}_{113}u^{c}d^{c}b^{c}$-type RPV operators. The collider implications of $\lambda^{\prime\prime}_{112} u^{c}d^{c}s^{c}$-type RPV coupling on direct wino searches in the $Wh$ mediated $1l+2b+jets(N_{j} \gtrsim 2)+\met$, $Wh$ mediated $1l+2\gamma+jets(N_{j} \gtrsim 2)+\met$ and $WZ$ mediated $3l+jets(N_{j} \gtrsim 2)+\met$ has been studied in Section~\ref{Sec:wh_6jbblnu}, \ref{Sec:wh_6jgagalnu} and \ref{Sec:6j3lmet}, respectively, while Section~\ref{sec:3l2bjmet} examined the projected reach of direct wino searches in the $WZ$ mediated $3l+2b+jets(N_{j}\gtrsim 2)+\met$ final state, by the virtue of $\lambda^{\prime\prime}_{113}u^{c}d^{c}b^{c}$-type RPV operator. The direct production of mass degenerate wino-type $\chonepm\lsptwo$ was considered which eventually underwent cascade decay into bino-type $\lspone$ along with the $h, W$ and/or $Z$ bosons. In the presence of $\lambda^{\prime\prime}_{u^{c}d^{c}s^{c}(b^{c})}$-type RPV operator, the $\lspone$ decays into $\lspone \to uds$ ($\lspone \to udb$) resulting in $jets$ in the final state.

A detailed collider analysis was performed in the aforesaid channels by taking into account all relevant background samples and by considering a multitude of important kinematic variables. Direct wino searches in the $Wh$ mediated $1l+2b+jets+\met$ final state in the context of $\lambda^{\prime\prime}u^{c}d^{c}s^{c}$-type RPV scenario exhibited a projected $2\sigma$ exclusion ($5\sigma$ discovery) reach up to $M_{\chonepm,\lsptwo} \sim 680~{\rm GeV}$ ($\sim 450~{\rm GeV}$) for a binolike $\lspone$ with mass up to $\sim 0~{\rm GeV}$. It is to be noted that the respective reinterpretation within a RPC scenario (studied in \cite{ATL-PHYS-PUB-2018-048}) furnishes considerably stringent projections, and the respective $95\%$ C.L. projected exclusion contour reaches up to $M_{\chonepm,\lsptwo}$ up to $\sim 1100~{\rm GeV}$ for binolike $\lspone$ with mass $M_{\lspone} \sim 0~{\rm GeV}$. The same simplified RPV scenario was also interpreted in terms of searches in the $Wh$ mediated $1l+2\gamma+jets+\met$ final state, and a relatively stronger potential reach was observed. Here, the projected exclusion and discovery contours had reach up to $\sim 700~{\rm GeV}$ and $\sim 600~{\rm GeV}$, respectively. 

As discussed previously in Section~\ref{Sec:6j3lmet}, the future reach of direct wino searches in the $WZ$ mediated $3l+\met$ final state within a RPC framework has been studied in \cite{ATL-PHYS-PUB-2018-048}, and the projected $95\%$ C.L. exclusion contour reaches up to $\sim 1150~{\rm GeV}$ for a binolike $\lspone$ with mass up to $\sim 0~{\rm GeV}$. We performed a collider study to derive the projected reach of direct wino searches in the $WZ$ mediated $3l+jets+\met$ final state and reinterpreted the projected reach within a $\lambda^{\prime\prime}_{112}u^{c}d^{c}s^{c}$-type RPV scenario. The projected exclusion and discovery contours displayed a considerably weaker reach as compared to the RPC scenario \cite{ATL-PHYS-PUB-2018-048}. The projected $2\sigma$ exclusion contour reached up to $\sim 660~{\rm GeV}$ while the projected $5\sigma$ discovery contour reached up to $\sim 490~{\rm GeV}$. Similarly, in Section~\ref{sec:3l2bjmet}, the projected reach of direct wino searches in the $WZ$ mediated $3l+2b+jets+\met$ final state was reinterpreted to simplified scenario with $\lambda^{\prime\prime}_{113} u^{c}d^{c}b^{c}$-type RPV coupling. The projected exclusion reach of this search channel reaches up to $\sim 600~{\rm GeV}$ for $M_{\lspone}$ in between $\sim 150~{\rm GeV}$ and $\sim 250~{\rm GeV}$.

A few benchmark scenarios have been explored in Section~\ref{Sec:benchmark}. The future reach of direct higgsino production at the HL-LHC in the aforesaid channels was analyzed for BP-$\alpha_{\tilde{H}}$ and BP-$\beta_{\tilde{H}}$. It is observed that the direct higgsino searches furnish weaker projection contours compared to the wino counterparts due to a smaller production rate. Furthermore, the sensitivity of the analysis channels to $\tan\beta$ is also studied. The $WZ$ mediated channels displayed an improvement in signal significance with an increase in $\tan\beta$, will all other MSSM input parameters kept fixed. The implications of a finite wino-higgsino mixing in the heavier ino states on the projected reach of the search channels considered in this work is also analyzed for the case of BP-$\beta_{\tilde{W}}^{10}$, BP-$\beta_{\tilde{W}}^{30}$, BP-$\beta_{\tilde{W}}^{50}$ and BP-$\beta_{\tilde{W}}^{70}$. These benchmark points resulted in a combined signal significance of $2.01, 1.98, 2.33$ and 2.12, respectively, thereby, marginally falling within the projected exclusion reach (except for BP-$\beta_{\tilde{W}}^{30}$) of HL-LHC. Towards the end, we also briefly discuss the prospects of using $jet$ substructure to improve the signal-background discrimination.

\section{Acknowledgements}
Work of B. Bhattacherjee was supported by Department of Science and Technology, Government of India under the Grant Agreement numbers IFA13-PH-75 (INSPIRE Faculty Award).
The work of Najimuddin Khan was supported by the Department of Science and Technology,
Government of INDIA under the SERB-Grant PDF/2017/00372. IC acknowledges support from DST, India, under grant number IFA18-PH214 (INSPIRE Faculty Award).


\providecommand{\href}[2]{#2}\begingroup\raggedright\endgroup

\newpage
\appendix
\section{Signal and background cross sections}
\label{Appendix:cs}
\begin{table}[htpb!]
\begin{center}\scalebox{0.70}{
\begin{tabular}{|C{4cm}|C{4cm}|C{9cm}|C{4cm}|}
\hline
\hline
Final state & \multicolumn{2}{c|}{Process} & Cross section (fb)\\
\hline \hline
& Signal & BP1-A ($M_{\lsptwo} =  200~{\rm GeV}$, $M_{\lspone} = 55~{\rm GeV}$) & 381 \\
& benchmark points & BP1-B ($M_{\lsptwo} =  350~{\rm GeV}$, $M_{\lspone} = 165~{\rm GeV}$) & 46 \\
$Wh$ mediated &  & BP1-C ($M_{\lsptwo} =  500~{\rm GeV}$, $M_{\lspone} = 25~{\rm GeV}$) & 10 \\  \cline{2-4}
$1l+2b+jets+\met$ & \multirow{6}{*}{Background} & $t\bar{t}+jets$ & $9.15\times 10^{5}$  \\ 
(Section~\ref{Sec:wh_6jbblnu})& & $WW+jets$ & $8.92\times 10^{4}$  \\ 
& & $WZ+jets$ & $4.01\times 10^{4}$  \\ 
& & $ZZ+jets$ & $1.17\times 10^{4}$  \\ 
& & $Wh+jets$ $(h\rightarrow b\bar{b},W\rightarrow l^{\prime} \nu)$ & 334   \\
& & $Zh+jets$ $(h\rightarrow b\bar{b},Z\rightarrow l^{\prime} l^{\prime})$ & 54 \\
\hline \hline
& Signal & BP2-A ($M_{\lsptwo} =  250~{\rm GeV}$, $M_{\lspone} = 100~{\rm GeV}$) &  0.67 \\
$Wh$ mediated & benchmark points & BP2-B ($M_{\lsptwo} = 425~{\rm GeV}$, $M_{\lspone} = 100~{\rm GeV})$ &  0.08 \\
$1l+2\gamma + jets$ & & BP2-C ($M_{\lsptwo} = 600~{\rm GeV}$, $M_{\lspone} = 150~{\rm GeV})$ &  0.02 \\\cline{2-4}
(Section~\ref{Sec:wh_6jgagalnu}) & \multirow{3}{*}{Background} & $ t\bar{t}h+jets~(h\rightarrow \gamma\gamma)$ & 0.82 \\
 & & $Wh+jets~(h\rightarrow \gamma\gamma,W\rightarrow l^{\prime} \nu)$ & 0.61 \\
& & $Zh+jets~(h\rightarrow \gamma\gamma,Z\rightarrow l^{\prime} l^{\prime})$ & 0.10 \\
\hline \hline
& Signal & BP3-A $( M_{\lsptwo}=400~{\rm GeV}$, $M_{\lspone}=175~{\rm GeV})$ &  4.71 \\
& benchmark points & BP3-B $( M_{\lsptwo}=600~{\rm GeV}$, $M_{\lspone}=325~{\rm GeV})$ &  0.82 \\
$WZ$ mediated & & BP3-C $( M_{\lsptwo}=650~{\rm GeV}$, $M_{\lspone}=175~{\rm GeV})$ &  0.56 \\ \cline{2-4}
$3l+jets+\met$ & \multirow{2}{*}{Background} & $WZ+jets$ & 40080 \\
(Section~\ref{Sec:6j3lmet})& & $ZZ+jets$ & 11690 \\
& & $VVV+jets$ & 799 \\
\hline
\hline
& Signal & BP4-A $( M_{\lsptwo}=250~{\rm GeV}$, $M_{\lspone}=135~{\rm GeV})$ & 29  \\
& benchmark points & BP4-B $( M_{\lsptwo}=600~{\rm GeV}$, $M_{\lspone}=205~{\rm GeV})$ & 0.8  \\
$WZ$ mediated  & & BP4-C $( M_{\lsptwo}=700~{\rm GeV}$, $M_{\lspone}=85~{\rm GeV})$ & 0.4  \\ \cline{2-4}
$3l+2b+jets+\met$ & \multirow{4}{*}{Background} & $t\bar{t}Z$ & 762  \\
(Section~\ref{sec:3l2bjmet}) & & $VVV+jets$ & 1037 \\
& & $WZ+jets$ ($W \to l^{\prime} \nu$, $Z \to l^{\prime}l^{\prime}$) & 248 \\
& & $ZZ+jets$ ($Z \to l^{\prime} l^{\prime}$, $Z \to l^{\prime}l^{\prime}$) & 319 \\
\hline
\hline
\end{tabular}
}
\end{center}
\caption{\it \footnotesize The cross section of the signal benchmark points and the background processes are shown. The signal cross sections are at NLO-NLL order (taken from \cite{Fuks:2012qx,Fuks:2013vua}). For the background processes, the LO cross section values computed by \texttt{MadGraph$\_$aMC@NLO} have been considered with the exception of $t\bar{t}+jets$ for which the NLO cross section is considered (the NLO cross section has been obtained by multiplying the LO cross section with the $k$ factor ($k\simeq 1.5$).) }
\label{tab:Appendix_cs}
\end{table}

\end{document}